\documentclass[preprint]{aastex}

\usepackage{amsmath}	
\usepackage{amssymb}	
\usepackage{graphicx,color}

\usepackage{natbib}

\newcommand{\rhoi}{\rho_i}
\newcommand{\rhoe}{\rho_e}
\newcommand{\vai}{v_{\rm Ai}}
\newcommand{\vae}{v_{\rm Ae}}

\newcommand{\ki}{k_i}
\newcommand{\ke}{k_e}

\newcommand{\tauAi}{\tau_{Ai}}

\newcommand{\xir}{\hat\xi_r}
\newcommand{\xiphi}{\hat\xi_\varphi}

\newcommand{\cph}{c_{ph}}
\newcommand{\cg}{c_g}
\newcommand{\ck}{c_k}

\shorttitle{Dispersion of fast kink waves in magnetic flux tubes}
\shortauthors{Oliver et al.}

\begin{document}

\title{PROPAGATION AND DISPERSION OF TRANSVERSE WAVE TRAINS IN~MAGNETIC FLUX TUBES}

\author{R. Oliver}
\affil{Departament de F\'\i sica, Universitat de les Illes Balears, E-07122 Palma de Mallorca, Spain}
\email{ramon.oliver@uib.es}

\and
\author{M. S. Ruderman}
\affil{School of Mathematics and Statistics, University of Sheffield, Hicks Building, Hounsfield Road, Sheffield S3 7RH, UK \\ Space Research Institute (IKI), Russian Academy of Sciences, Moscow 117997, Russia}

\and
\author{J. Terradas}
\affil{Departament de F\'\i sica, Universitat de les Illes Balears, 07122 Palma de Mallorca, Spain}

\begin{abstract}
The dispersion of small amplitude, impulsively excited wave trains propagating along a magnetic flux tube is investigated. The initial disturbance is a localized transverse displacement of the tube that excites a fast kink wave packet. The spatial and temporal evolution of the perturbed variables (density, plasma displacement, velocity, \ldots) is given by an analytical expression containing an integral that is computed numerically. We find that the dispersion of fast kink wave trains is more important for shorter initial disturbances (i.e. more concentrated in the longitudinal direction) and for larger density ratios (i.e. for larger contrasts of the tube density with respect to the environment density). This type of excitation generates a wave train whose signature at a fixed position along a coronal loop is a short event (duration $\simeq 20$~s) in which the velocity and density oscillate very rapidly with typical periods of the order of a few seconds. The oscillatory period is not constant but gradually declines during the course of this event. Peak values of the velocity are of the order of 10~km~s$^{-1}$ and are accompanied by maximum density variations of the order of 10--15\% the unperturbed loop density.
\end{abstract}


\section{INTRODUCTION}

In the last two decades abundant evidence about waves and oscillations in the solar atmosphere has been gathered. Events of various nature (standing and propagating waves) and in various environments (chromosphere, prominences, active regions, coronal holes) have been detected. Here we present some examples, while emphasizing that the following list is not exhaustive. In the chromosphere, propagating and standing transverse waves have been detected in spicules \citep{zaqarashvili2007,okamoto2011}, in mottles \citep{kuridze2012}, and in active region fibrils \citep{pietarila2011}. Regarding solar prominences, \citet{lin2007} used high-resolution H$\alpha$ filtergrams and observed traveling transverse waves in thin filament threads. \citet{okamoto2007} found transverse oscillations of flowing active region filament threads observed with {\em Hinode} SOT. Oscillatory events of different nature have been observed in coronal loops: transverse oscillations of active region loops triggered by a disturbance that propagates from the central flare site \citep{aschwanden1999,nakariakov1999}; high-frequency, compressible waves traveling along an active region coronal loop \citep{williams2001,williams2002}; Doppler shift oscillations caused by waves propagating in the upper part of coronal loops \citep{tian2012}; ubiquitous waves in the solar corona propagating upwards along magnetic field lines \citep{tomczyk2007,tomczyk2009}; etc. Reviews about waves and oscillations in spicules, prominences, and coronal structures can be found in \citet{zaqarashvili_erdelyi2009}, \citet{arregui2012}, \citet{nakariakov_verwichte2005}, and \citet{demoortel_nakariakov2012}.

\begin{figure}[ht!]
\centerline{(a)\includegraphics[width=0.3\textwidth,angle=0]{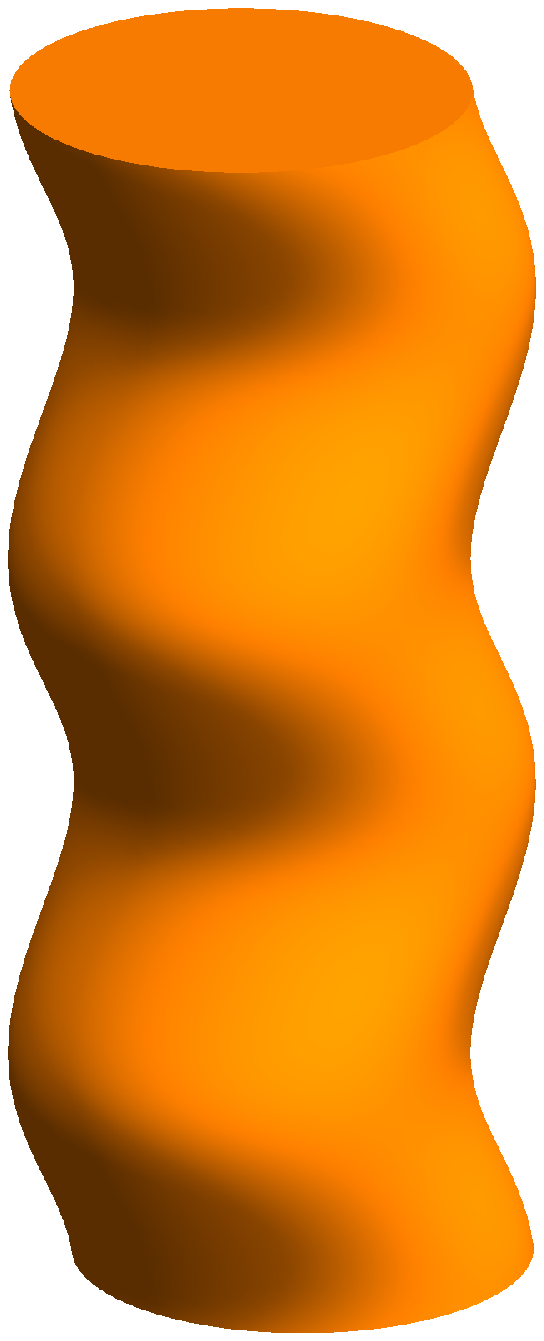}\hspace{1cm}(b)\includegraphics[width=0.3\textwidth,angle=0]{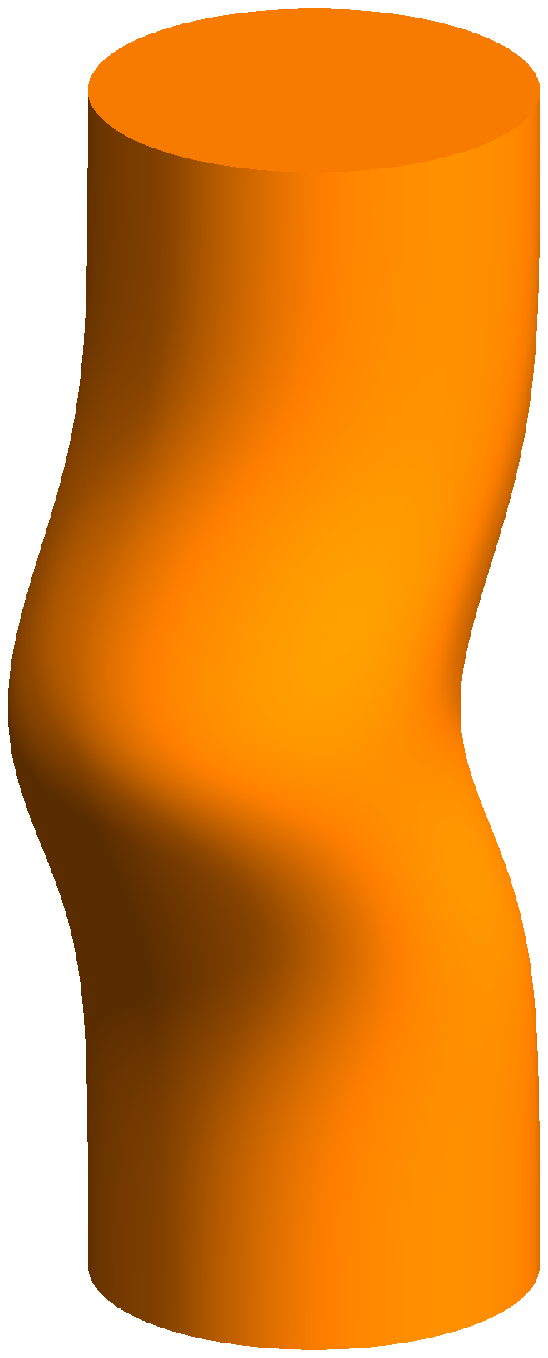}}
  \caption{Uniform cylindrical magnetic flux tube subject to (a) a fast kink mode (i.e. transverse, periodic along the tube) perturbation and (b) a localized transverse perturbation. (A color version of this figure is available in the online journal.)}
  \label{fig_cylinder_kink}
\end{figure}

Many of these events are examples of magnetic flux tubes being perturbed by an external agent. The transverse loop oscillations described by \citet{aschwanden1999} and \citet{nakariakov1999} are a remarkable case because time series of EUV images allow to see the lateral swaying of a coronal loop. This particular phenomenon has been interpreted as a standing fast kink mode oscillation. Fast kink (i.e. transverse) waves propagating along magnetic flux tubes are more frequent in the literature than their standing oscillation counterparts. They have been observed not only in coronal loops \citep[e.g.][]{tian2012}, but also in spicules \citep{zaqarashvili2007}, in mottles \citep{kuridze2012}, in active region fibrils \citep{pietarila2011}, and in filament threads \citep{lin2007,okamoto2007}. What is common to all these phenomena is that an external excitation causes the transverse displacement of a magnetic tube and that this perturbation propagates along the tube, where it is usually detected as a time variation of the Doppler velocity or the magnetic tube position. These propagating waves can be excited, e.g., by a periodic driver acting at a fixed position of the magnetic tube. This has been the mechanism invoked by \citet{VTG2010} and \citet{TGV2010} to explain the waves propagating along the coronal magnetic field observed by \citet{tomczyk2007} and \citet{tomczyk2009}. A fast kink wave propagating along a uniform, cylindrical magnetic tube produces a periodic transverse motion of the tube such as that in Figure~\ref{fig_cylinder_kink}(a).

Propagating transverse waves can also be generated by an impulsive, concentrated transverse displacement of the magnetic tube (see Figure~\ref{fig_cylinder_kink}(b)). Such a perturbation of the magnetic tube is a superposition of fast kink eigenmodes, each with its own amplitude. In the absence of dispersion, the initial wave form would keep its original shape during its propagation, that is, the bulge of Figure~\ref{fig_cylinder_kink}(b) would simply propagate unaltered along the cylinder. Fast kink eigenmodes of a magnetic tube are dispersive, however, and so as time evolves the initial hump transforms into an oscillatory train containing several maxima and minima. In other words, away from the excitation point the magnetic tube can suffer not one, but several lateral oscillations about its equilibrium position as the wave train passes by. Hence, transverse oscillations of magnetic tubes do not necessarily require the presence of a continuous driver.

The purpose of this work is to study the dispersion of linear fast kink wave trains propagating along uniform, cylindrical magnetic flux tubes. These wave trains are excited by an impulsive, localized initial disturbance (Figure~\ref{fig_cylinder_kink}(b)). To solve this initial value problem we use a technique based on the method of Fourier integrals \citep[cf.][Section~11.2]{whitham1974}, that consists of expressing the initial perturbation as a sum of eigenmodes. This means that the properties of the magnetic tube eigenmodes need to be well known in advance, and for this reason their main features are summarized in Sections~\ref{sect_normal_modes} and \ref{sect_cont_modes}. Next the method of Fourier integrals is applied to the cylindrical magnetic tube, and 
an analytical expression describing the spatial and temporal variation of  perturbed variables is obtained (Section~\ref{sect_fourier}). This formula is expressed in terms of an integral that contains contributions from all eigenmodes, with amplitudes that depend exclusively on the initial conditions. Accurate approximations to this integral are computed numerically. The propagation of a concentrated transverse disturbance is considered in Section~\ref{subsect_numerics}, paying special attention to the dispersion of the wave train during its propagation. An application to coronal loop transverse oscillations is carried out in Section~\ref{sect_loop_application} and a discussion of the results and our conclusions are presented in Section~\ref{sect_conclusions}.

\section{GOVERNING EQUATIONS}\label{sect_govern}

In our analysis we use the linear ideal magnetohydrodynamic (MHD) equations in the cold plasma approximation,

\begin{equation}\label{eq_mass}
\rho = -\nabla\cdot(\rho_0\boldsymbol{\xi}), 
\end{equation}

\begin{equation}\label{eq_momen}
\rho_0\frac{\partial^2\boldsymbol{\xi}}{\partial t^2} =
   \frac1{\mu_0}(\nabla\times\boldsymbol{b})\times\boldsymbol{B} , 
\end{equation}

\begin{equation}\label{eq_induc}
\boldsymbol{b} = \nabla\times(\boldsymbol{\xi}\times\boldsymbol{B}). 
\end{equation}

\noindent
Here $\rho_0$ and $\boldsymbol{B}$ are the equilibrium density and magnetic field, $\rho$ and $\boldsymbol{b}$ are the density and magnetic field perturbations, $\boldsymbol{\xi}$ is the plasma displacement, and $\mu_0$ is the magnetic permeability of free space. In what follows we use cylindrical coordinates $(r,\varphi,z)$. In these coordinates $\boldsymbol{\xi} = (\xi_r,\xi_\varphi,0)$ and $\boldsymbol{b} = (b_r,b_\varphi,b_z)$. We assume that the equilibrium magnetic field is uniform and directed along the $z$\/-axis. The equilibrium consists of a magnetic tube of radius $a$ and plasma density $\rhoi$ in a magnetic environment with density $\rhoe$, both quantities being constants. The corresponding Alfv\'en speeds are determined by

\begin{equation}\label{eq_alfven}
{\vai^2} =\frac{B^2}{\mu_0\rhoi}, \quad \vae^2 = \frac{B^2} {\mu_0\rhoe} .
\end{equation}

\noindent
Here we consider an overdense magnetic tube, that is, $\rhoi>\rhoe$ and $\vai<\vae$.

To restrict our analysis to kink waves we assume that perturbations of all quantities are proportional to $e^{i\varphi}$\/. Then the system of Equations~(\ref{eq_mass})--(\ref{eq_induc}) reduces to

\begin{equation}\label{eq_xi_r}
\frac{\partial^2\xi_r}{\partial t^2} -
   v_A^2\frac{\partial^2\xi_r}{\partial z^2} = 
   -\frac1{\rho_0}\frac{\partial P}{\partial r} , 
\end{equation}

\begin{equation}\label{eq_xi_phi}
\frac{\partial^2\xi_\varphi}{\partial t^2} -
   v_A^2\frac{\partial^2\xi_\varphi}{\partial z^2} = -\frac{iP}{r\rho_0}, 
\end{equation}

\begin{equation}\label{eq_induc_1}
b_r = B\frac{\partial\xi_r}{\partial r} , \quad
b_\varphi = B\frac{\partial\xi_\varphi}{\partial r}, 
\end{equation} 

\begin{equation}\label{eq_press}
P = -\frac{\rho_0 v_A^2}r\left[\frac{\partial(r\xi_r)}{\partial r} + i \xi_\varphi\right], 
\quad \rho = \frac P{v_A^2} , 
\end{equation}

\noindent
where $P = Bb_z/\mu_0$ is the magnetic pressure and $v_A$ stands for the Alfv\'en speed at any position. These equations have to be supplemented with the conditions that $\xi_r$ and $P$ are continuous at the tube boundary ($r = a$).

The system of Equations~(\ref{eq_xi_r})--(\ref{eq_press}) is fourth order in time. Hence, to set up an initial value problem we need to impose four initial conditions,

\begin{equation}\label{init_cond}
\begin{split}
&\xi_r(t=0,r,z) = f_r(r,z), \quad \xi_\varphi(t=0,r,z) = f_\varphi(r,z), \quad \\
&\frac{\partial\xi_r}{\partial t} (t=0,r,z)= g_r(r,z), \quad
\frac{\partial\xi_\varphi}{\partial t}(t=0,r,z) = g_\varphi(r,z).
\end{split}
\end{equation}

\section{NORMAL (OR PROPER) MODES}\label{sect_normal_modes}

\subsection{Dispersion Equation and Eigenfrequencies}\label{subsect_disp_eq}

We next take perturbations of all variables in the form

\begin{equation}\label{rzt_dependence}
\xi_r(t,r,z) = \hat{\xi_r}(r)\exp[i(-\omega t + k z)],
\end{equation}

\noindent
and similarly for $\xi_\varphi$, $b_r$, $b_\varphi$, $P$, and $\rho$. The $\varphi$-dependence of perturbed variables is omitted in what follows. After substituting these expressions in Equations~(\ref{eq_xi_r})--(\ref{eq_press}) and in the boundary conditions, which are the conditions of continuity of $\xi_r$ and $P$ at $r = a$\/, we obtain an eigenvalue problem. The eigenvalues are the zeros of the dispersion equation, that defines $\omega$ for a given longitudinal wavenumber, $k$. This equation has been derived by many authors \citep[e.g.][]{edwinroberts1983} and can be expressed as

\begin{equation}\label{dr_mod}
D(\omega) \equiv \frac{J_1'(\ki a)}{\ki J_1(\ki a)} + 
   \frac{K_1'(\kappa_e a)}{\kappa_e K_1(\kappa_e a)} = 0.
\end{equation}

\noindent 
where $J_1$ is the Bessel function of the first kind and first order, $K_1$ is the modified Bessel function of the second kind (McDonald function), and a prime indicates a derivative. The radial wavenumbers, $\ki$ and $\kappa_e$, are given by

\begin{equation}\label{ks}
\ki^2 = \frac{\omega^2-k^2\vai^2}{\vai^2}, \quad 
\kappa_e^2 = -\frac{\omega^2-k^2\vae^2}{\vae^2},
\end{equation}

\noindent
where we take $\ki > 0$ when $\ki^2 > 0$, and $\kappa_e > 0$ when $\kappa_e^2 > 0$.

A detailed description of the procedure that leads to Equation~(\ref{dr_mod}) can be found in \cite{ruderman2006}, who also discuss the general properties of the dispersion equation. A short summary now follows. We consider proper eigenmodes that decay exponentially with the distance from the tube. The eigenfrequencies of these eigenmodes are the zeros of the function $D(\omega)$. If $\omega$ is an eigenfrequency of linear ideal MHD equations describing perturbations about a static equilibrium, then $\omega^2$ is real \citep[e.g.][]{Goedbloed2004,priest1984}, and, consequently, $\omega$ is either real or purely imaginary. Purely imaginary frequencies correspond to unstable solutions and since a magnetic tube with straight field lines is stable, all eigenfrequencies must be real. The condition that the eigenmodes must decay exponentially with the distance from the tube implies that $\kappa_e^2 > 0$, so $\omega^2 < k^2\vae^2$ \citep[see][for more details]{ruderman2006}. Since $D(\omega)$ depends on $\omega^2$ only, it follows that if $\omega$ is an eigenfrequency then $-\omega$ is also an eigenvalue. The two eigenfrequencies with opposite signs correspond to two waves propagating in opposite directions. Given  that the sign of $\omega$ is irrelevant we have only considered $\omega\geq 0$ in this work.

The sign of $k$ can also be ignored because the longitudinal wavenumber only appears as $k^2$ in the dispersion equation. In this section we restrict ourselves to solutions with $k\geq 0$, although both positive and negative $k$ are required to obtain the solution of the initial value problem (cf. Section~\ref{sect_fourier}). For any fixed $k$ there is a finite number of eigenvalues (see Figure~\ref{fig_dr_over}(a)). For $k < k_{c1}$ (cf. Equation~(\ref{eq_kcn}) below) there is only one eigenvalue corresponding to the eigenmode fundamental in the radial direction. This eigenmode does not have nodes in the radial direction. When $k \to 0$, its frequency tends to~$\ck k$ (see Figure~\ref{fig_dr_over}(b)), where the kink speed $\ck$ is defined by

\begin{equation}\label{eq_ck}
\ck^2 = \frac{2}{1+\rhoe/\rhoi} \vai^2.
\end{equation}

\noindent
Here we follow \cite{ruderman2006} and refer to this mode as the global kink mode to distinguish it from other fast kink modes. According to \citet{goossens2009}, in the long wavelength approximation the global kink mode has an Alfv\'enic character.

When $k_{c1} < k < k_{c2}$\/, there are two eigenmodes, one fundamental and one first overtone in the radial direction. The first overtone has exactly one node in the radial direction, and this node is inside the tube. When $k_{c2} < k < k_{c3}$\/, there are three eigenmodes, one fundamental, one first overtone, and one second overtone in the radial direction. The second overtone has exactly two nodes in the radial direction, and these nodes are inside the tube. And so on, with the number of eigenmodes tending to infinity when $k \to \infty$\/. When $k = k_{cn}$\/, the frequency of the $n$\/th overtone is defined by the condition $\kappa_e = 0$, which gives for its frequency $\omega = k\vae$\/. Then it is straightforward to obtain the expression for $k_{cn}$\/,

\begin{equation}\label{eq_kcn}
k_{cn} = \frac{j_{1n}\vai}{a\sqrt{\vae^2 - \vai^2}} ,
\end{equation}

\noindent
where $j_{1n}$ is the $n$\/th zero (in increasing order) of $J_1(x)$. 

When solving the dispersion equation it is convenient to use the dimensionless variables $ka$ and $\omega\tauAi$, with $\tauAi=a/\vai$ the internal Alfv\'en travel time. Then, one only needs to fix the density ratio $\rhoi/\rhoe$ (or the Alfv\'en speed ratio $\vae/\vai$) to compute $\omega\tauAi$ for a given $ka$ from Equation~(\ref{dr_mod}). Some features of the global kink mode and its first radial overtones are displayed in the dispersion diagram of Figure~\ref{fig_dr_over}(a), computed for the density ratio $\rhoi/\rhoe=4$. Such as described above, for $0\leq ka<k_{c1}a$ there is only the global kink mode, that exists for all wavenumbers. In the range $k_{c1}a \leq ka<k_{c2}a$ the global kink mode and its first radial overtone are the only eigenmodes, etc.

\begin{figure}[ht!]
  \centerline{
    \scriptsize{(a)}
    \includegraphics[width=0.33\textwidth,angle=-90]{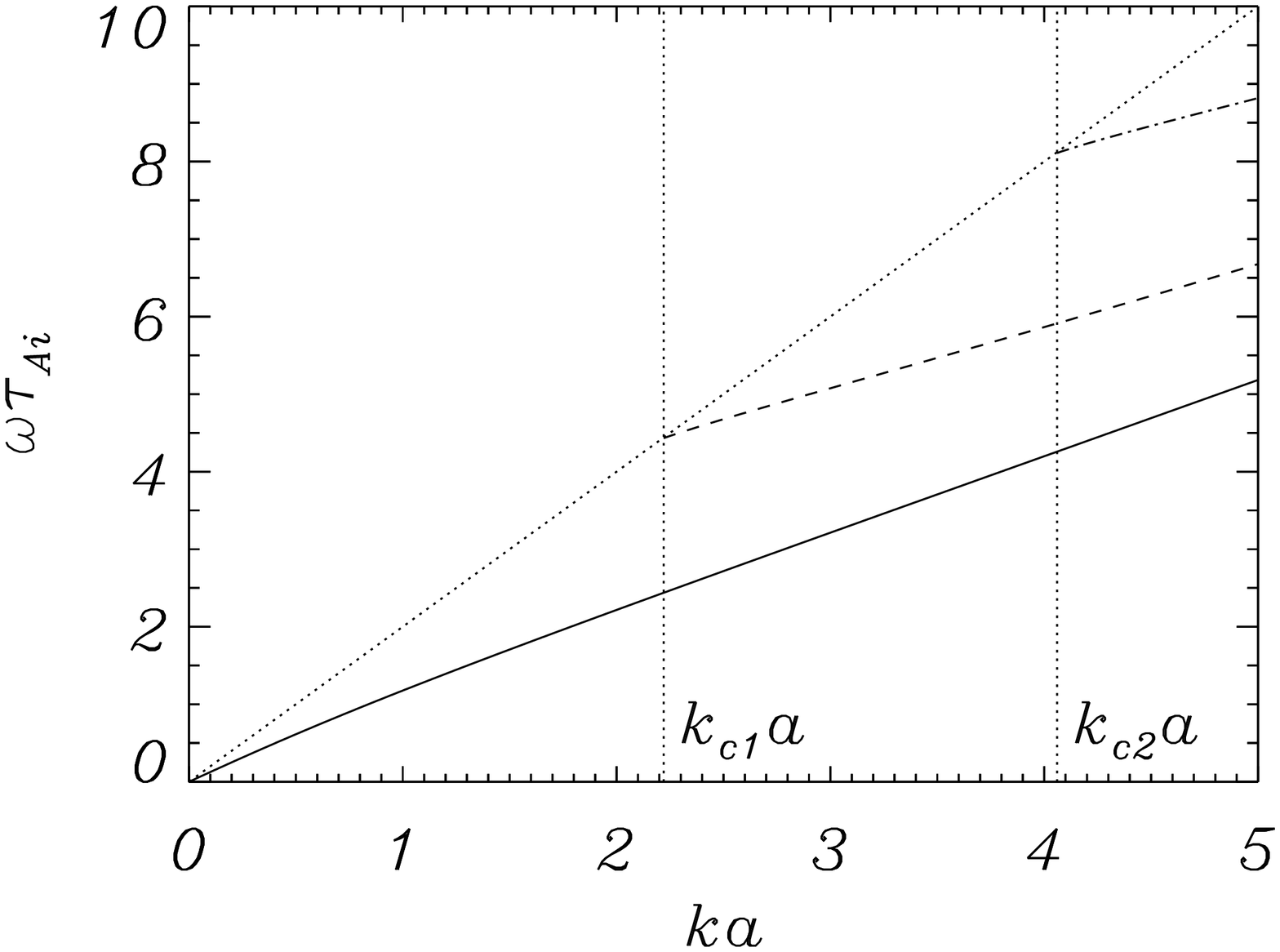} \\
    \scriptsize{(b)}
    \includegraphics[width=0.33\textwidth,angle=-90]{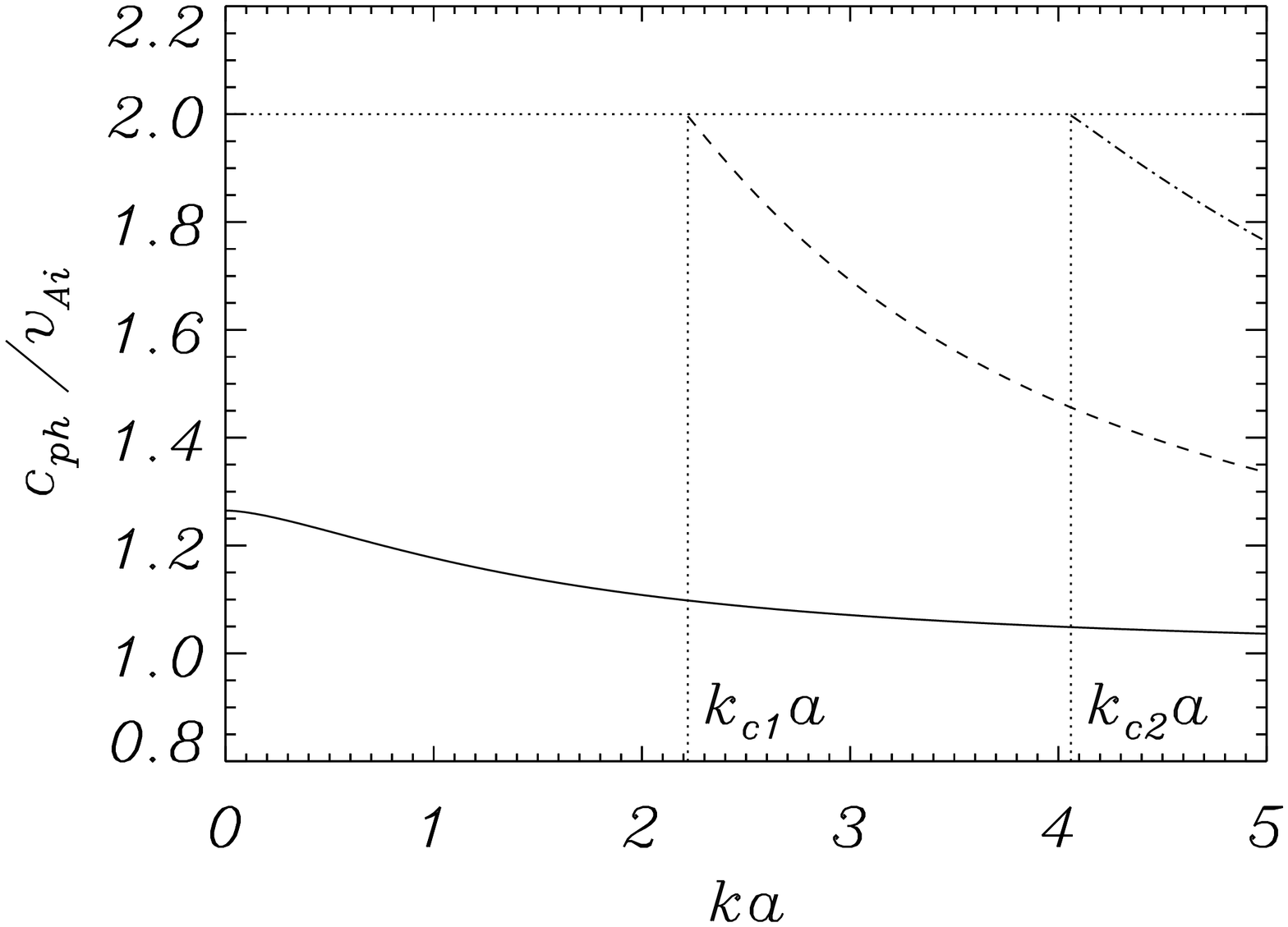} \\
  }
  \centerline{
    \scriptsize{(c)}
    \includegraphics[width=0.33\textwidth,angle=-90]{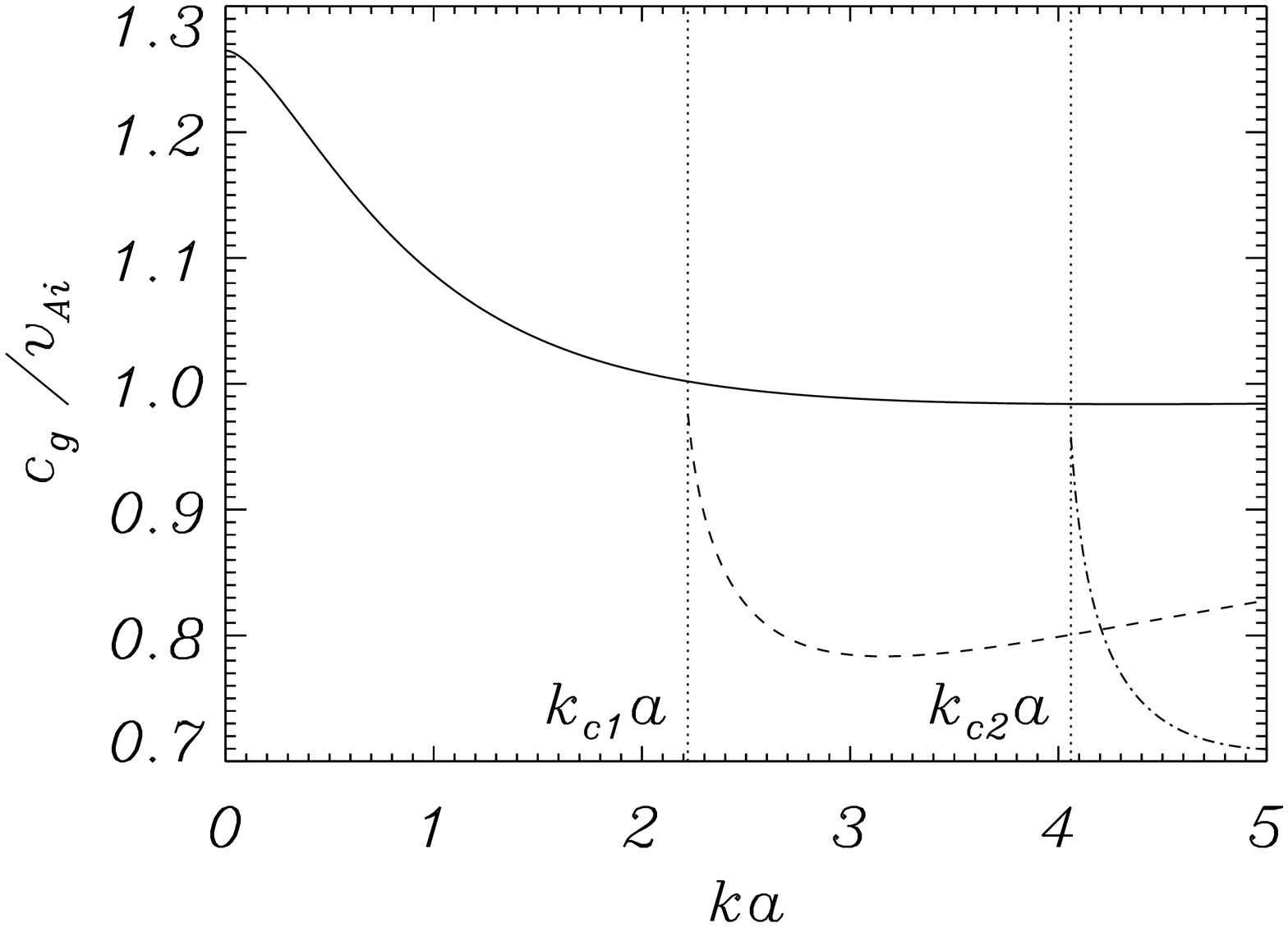} \\
  }
  \caption{(a) Frequency, (b) phase speed, and (c) group velocity vs. the longitudinal wavenumber for the global (i.e. fundamental) fast kink mode (solid line) and its first two overtones in the radial direction (dashed and dash-dotted lines). The inclined and horizontal dotted lines in (a) and (b), respectively, correspond to the frequency $\omega=k\vae$, whereas the vertical dotted lines give the cut-off wavenumbers $k_{c1}$ and $k_{c2}$ of the first and second overtones, respectively. This figure corresponds to $\rhoi/\rhoe=4$, i.e.\ to the Alfv\'en speed ratio $\vae/\vai=2$. To obtain the dimensionless quantities, the magnetic cylinder radius ($a$), Alfv\'en speed ($\vai$), and transit time ($\tauAi=a/\vai$) are used.}
  \label{fig_dr_over}
\end{figure}

\subsection{Phase Speed and Group Velocity}
\label{sect_cph_cg}

The phase speed and group velocity of a particular eigenmode are defined as

\begin{equation}\label{cph_cg}
\cph = \frac\omega k, \quad \cg=\frac{\partial\omega}{\partial k}.
\end{equation}

\noindent 
The phase speed gives the propagation velocity of the eigenmode with wave\-number $k$ along the magnetic tube. The group velocity is relevant when one considers the propagation of a wave packet, as we do in this work. If the medium is dispersive, i.e.\ if $\cg \neq \cph$\/, then the initial shape of the wave packet will become distorted in time as it propagates. The dependence of $\cg$ on $k$ then determines the shape of the packet.

We first turn our attention to the global kink mode. In the long wavelength approximation the dispersion relation for the global kink mode propagating in the positive $z$\/-direction is $\omega = \ck k$\/, so that $\cph=\cg=\ck$ in this limit. It is also straightforward to derive the approximate dispersion relation valid for small wavelengths ($ka \gg 1$) from Equation~(\ref{dr_mod}),

\begin{equation}\label{dr_gg_ak}
\omega \approx \vai\left(k + \frac{j'_1}{2ka^2}\right) ,
\end{equation}

\noindent
where $j'_1$ is the smallest root of $J'_1(x)$. This relation is derived for the fundamental mode in the radial direction. From the approximate dispersion relation~(\ref{dr_gg_ak}) we obtain approximate expressions for the phase speed and group velocity valid for $ka \gg 1$,

\begin{equation}\label{phg_gg_ak}
\frac{\cph}{\vai} \approx 1 + \frac{j'_1}{2k^2 a^2}, \quad 
\frac{\cg}{\vai} \approx 1 - \frac{j'_1}{2k^2 a^2}.
\end{equation}

\noindent
It follows from these expressions that both $\cph$ and $\cg$ tend to $\vai$ as $ka \to \infty$\/. We thus see that the phase speed (group speed) is a monotonically decreasing (monotonically increasing) function of $ka$ for $ka \gg 1$. 

After numerically solving the dispersion equation one can compute the phase speed and the group velocity. The results of these calculations are presented in Figures~\ref{fig_dr_over}(b) and (c) for  the density ratio $\rhoi/\rhoe=4$. It follows from these results and the asymptotic analysis for $ka \gg 1$ that the global kink mode phase speed, $\cph$, is a monotonically decreasing function of $k$\/. On the other hand, the dependence of the group velocity, $\cg$, on $k$ is non-monotonic. When $k$ varies from 0 to $\infty$, $\cg$ first monotonically decreases, takes its minimum value, which is smaller than $\vai$\/, at an intermediate value of $ka = k_m a$\/, and then starts to grow monotonically tending to $\vai$ as $ka \to \infty$\/. It is important to mention that this minimum of $\cg$ is rather shallow.


Regarding the kink mode radial overtones, we have already mentioned that the $n$th overtone only exists for $k\geq k_{cn}$. For this particular wavenumber, the phase and group velocities are equal to $\vae$ and $\vai$, respectively (Figures~\ref{fig_dr_over}(b), (c)). In addition, $\cph$ presents a monotonically decreasing behavior whereas $\cg$ shows a non-monotonic dependence with $k$, with a minimum at an intermediate wavenumber.

We next apply the information about the group velocity to explore the behavior of a localized wave packet. Such a feature, initially concentrated in space, is the sum of eigenmodes distributed over a wavenumber range. We can consider a wave packet with a wide range of wavenumbers as a composition of many wave subpackets with wavenumbers confined to small intervals, so the group velocity is approximately constant for any subpacket. Then the subpackets with higher $\cg$ travel at the wave train front, while those with smaller $\cg$ remain at its back. According to Figure~\ref{fig_dr_over}(c), a concentrated perturbation made of a sum of global kink modes will disperse as it travels along a magnetic tube, with long (short) wavelengths occupying its leading (trailing) edge. Therefore, the various wavelengths that make an initial perturbation also arrive at different times at a given point along the tube. If this point is a distance $z_0$ from the source, then long wavelengths (i.e.\ low frequencies) are the first to arrive, after $t = z_0/\ck$. Then progressively shorter waves with $ka < k_m a$ as well as waves with $ka > k_m a$ arrive. The wavelengths with $k \approx k_m$ will be the last reaching the detection point at a time slightly larger than $z_0/\vai$\/. Regarding the $n$th overtone (dashed and dash-dotted lines of Figure~\ref{fig_dr_over}(c)), the maximum propagation speed is attained either for wavenumbers close to $k_{cn}$ or for very large wavenumbers, with intermediate values of $k$ having smaller propagation speeds. This information determines the dispersion properties of a wave packet made exclusively of a kink mode overtone.

\begin{figure}[ht!]
  \centerline{
    \scriptsize{(a)}
    \includegraphics[width=0.33\textwidth,angle=-90]{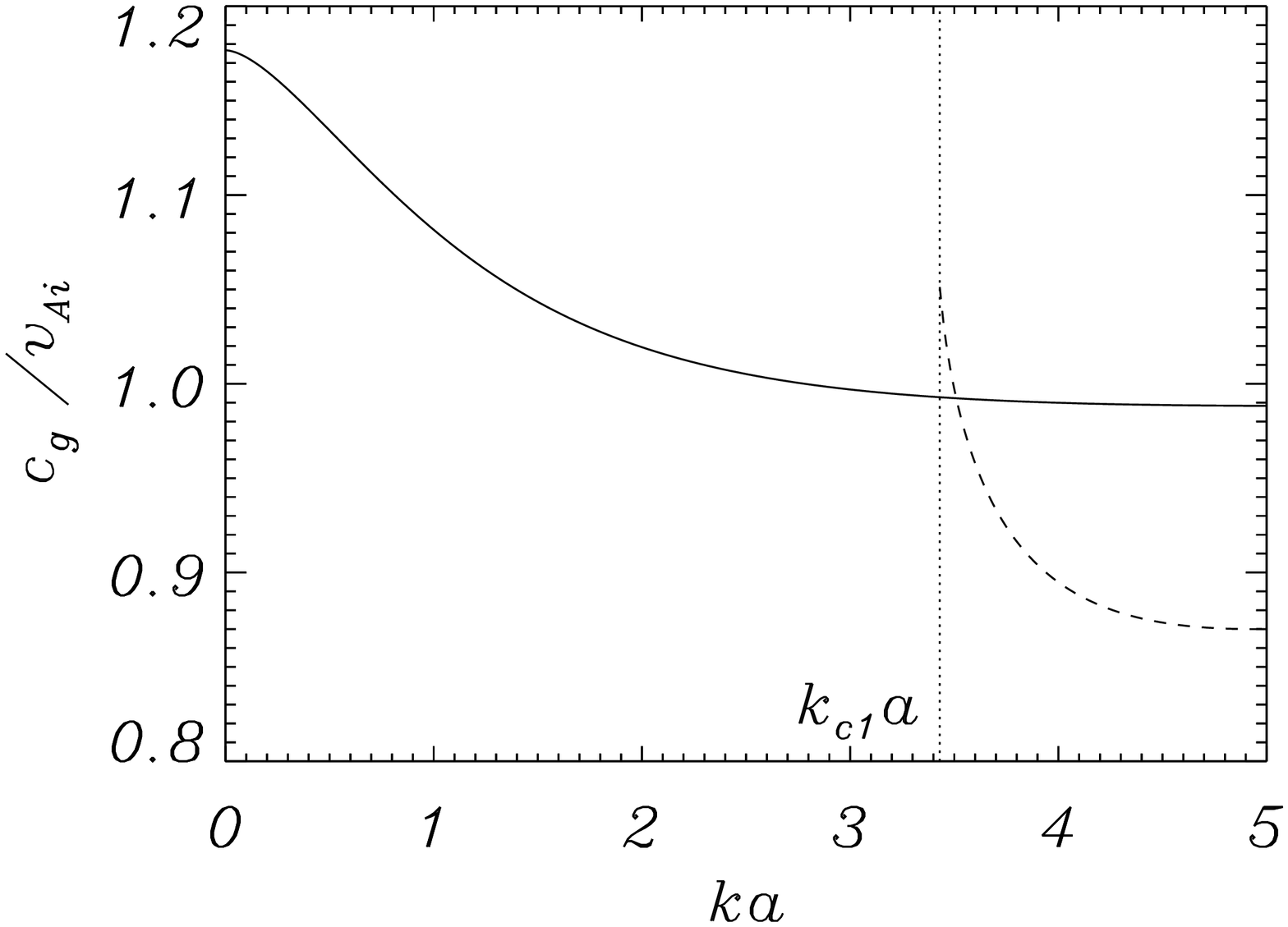} \\
    \scriptsize{(b)}
    \includegraphics[width=0.33\textwidth,angle=-90]{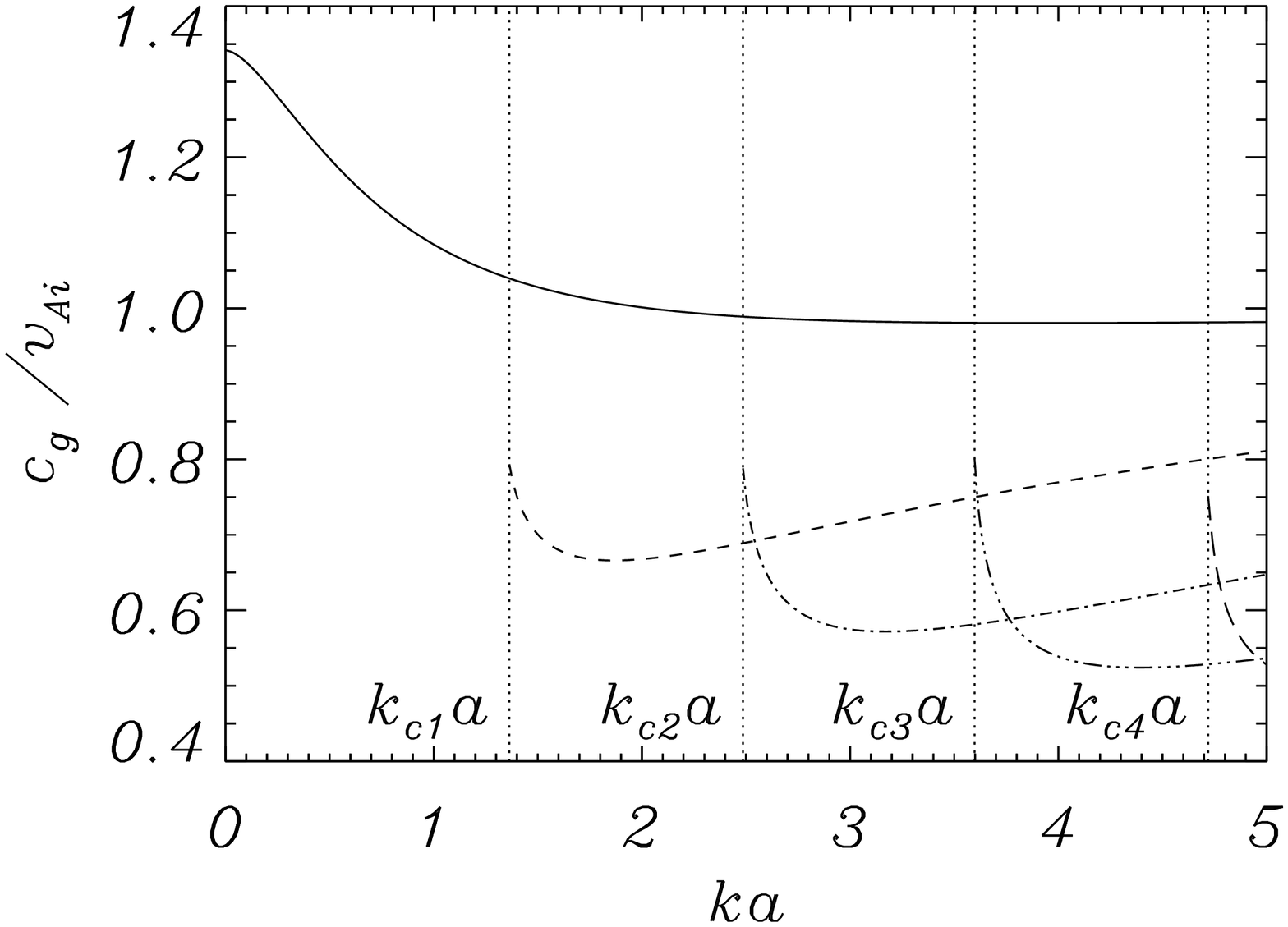} \\
  }
  \caption{Group velocity vs. the longitudinal wavenumber for the global (i.e. fundamental) fast kink mode (solid line) and its first overtones in the radial direction (dashed, dash-dotted, dash--triple-dot, long-dashed lines). The vertical dotted lines give the cut-off wavenumber, $k_{cn}$, of the $n$th overtone. The density ratio is (a) $\rhoi/\rhoe=2.25$ and (b) $\rhoi/\rhoe=9$. To obtain the dimensionless quantities, the magnetic cylinder radius ($a$), Alfv\'en speed ($\vai$), and transit time ($\tauAi=a/\vai$) are used.}
  \label{fig_cg_vae}
\end{figure}

We next explore the influence of the equilibrium density ratio, $\rhoi/\rhoe$, on the fast kink modes group velocity. The density ratio has very little influence on the shape of the global kink mode group velocity (solid line of Figure~\ref{fig_cg_vae}) and this implies that the description of a wave packet dispersion given above does not change when $\rhoi/\rhoe$ is varied. Nevertheless, $\ck$ becomes smaller when the density ratio is reduced and so the range of variation of $\cg$ is also reduced. This in turn implies that a wave train dispersion becomes less important for smaller values of $\rhoi/\rhoe$. As for the radial overtones, Equation~(\ref{eq_kcn}) shows that a decrease (increase) of the density ratio leads to larger (smaller) values of the cut-off frequencies, such as can be appreciated in Figure~\ref{fig_cg_vae}. This directly affects the longer wavelengths that can have a wave train made of a single overtone. The maximum and minimum values of $\cg$ for a given kink overtone are also influenced by the density ratio, but the overall dependence of the group velocity on $k$ remains qualitatively the same.

\subsection{Radial Dependence of Perturbed Variables}\label{sect_radial}

The spatial and temporal behavior of eigenmodes is described by Equation~(\ref{rzt_dependence}) for $\xir(r)$ and similar ones for all other perturbed variables. The radial and azimuthal plasma displacements can be written in the form

\begin{equation}\label{xir_of_r}
\xir(r) = 
   \begin{cases}
    \hphantom{-}\ki^{-1} K_1(\kappa_e a) J'_1(\ki r), & r < a, \vspace{1mm}\\
    -\kappa_e^{-1}  J_1(\ki a) K'_1(\kappa_e r), & r > a,
  \end{cases}
\end{equation}

\begin{equation}\label{xiphi_of_r}
\xiphi(r) = \frac ir
  \begin{cases}
    \hphantom{-}\ki^{-2} K_1(\kappa_e a) J_1(\ki r), & r < a, \vspace{1mm}\\
    -\kappa_e^{-2} J_1(\ki a) K_1(\kappa_e r), & r > a.
  \end{cases}
\end{equation}

\noindent 
Expressions for $\hat{b}_r(r)$, $\hat{b}_\varphi(r)$, $\hat{P}(r)$, and $\hat{\rho}(r)$ can be obtained using Equations~(\ref{eq_induc_1}) and (\ref{eq_press}). Note that Equations~(\ref{xir_of_r}) and (\ref{xiphi_of_r}) are independent of the sign of $\omega$\/, so they are the same for the eigenmode propagating in the positive $z$\/-direction and the eigenmode propagating in the negative $z$\/-direction.

It follows from Equations~(\ref{xir_of_r}) and (\ref{dr_mod}) that the plasma displacement in the radial direction is continuous at $r = a$ and, with the aid of Equations~(\ref{eq_press}), (\ref{xir_of_r}), and (\ref{xiphi_of_r}), it is straightforward to show that the magnetic pressure perturbation, $P$, is also continuous at the tube boundary.

\section{IMPROPER MODES}
\label{sect_cont_modes}

The eigenmodes corresponding to the zeros of $D(\omega)$ are called \textit{proper}\/. The corresponding eigenfrequencies constitute the point spectrum of the eigenvalue problem. The distinctive property of the proper eigenmodes is that they are square integrable with respect to $r$ on the interval $(0,\infty)$. In addition to the point spectrum there is a continuous spectrum that consists of the union of two intervals, $(-\infty,-|k|\vae]$ and $[|k|\vae,\infty)$. Eigenfunctions corresponding to frequencies from the continuous spectrum are called \textit{improper} because they are not square integrable with respect to $r$ \citep[e.g.][]{andries2007}.

We explained above that there is a finite number of proper eigenmodes for a given longitudinal wavenumber, $k$, and that their frequencies lie in the range $0<|\omega|<|k|\vae$. On the other hand, for a fixed $k$ improper modes can have any frequency $|\omega| > |k|\vae$.

\subsection{Radial Dependence of Perturbed Variables}\label{sect_radial_improper}

Now we obtain expressions similar to Equations~(\ref{xir_of_r}) and (\ref{xiphi_of_r}) for improper eigenmodes. Taking $\xi_r$\/, $\xi_\varphi$, and $P$ in Equations~(\ref{eq_xi_r}), (\ref{eq_xi_phi}), and (\ref{eq_press}) in the form given by Equation~(\ref{rzt_dependence}), and eliminating $\hat{\xi}_r(r)$ and $\hat{\xi}_\varphi(r)$ from the obtained expressions, we obtain the following equation for $\hat{P}(r)$\/,

\begin{equation}\label{equ_for_hatP}
\frac{d^2\hat{P}}{d r^2} + \frac1r\frac{d\hat{P}}{d r} +
   \bigg(k_r^2 - \frac1{r^2}\bigg)\hat{P} = 0,
\end{equation}

\noindent 
where $k_r = \ki$ for $r < a$\/, and $k_r = \ke$ for $r > a$\/. Here the external radial wavenumber, $\ke$, is given by

\begin{equation}\label{ke}
\ke^2 = \frac{\omega^2-k^2\vae^2}{\vae^2},
\end{equation}

\noindent
where we take $\ke \geq 0$.

In addition we obtain expressions for $\hat{\xi}_r$ and $\hat{\xi}_\varphi$ in terms of $\hat{P}$\/,

\begin{equation}\label{hatxir_in_hatP}
\hat{\xi}_r = \frac1{\rho_0 v_A^2 k_r^2}\frac{d\hat{P}}{d r}, \quad
\hat{\xi}_\varphi = \frac{i\hat{P}}{r\rho_0 v_A^2 k_r^2} .
\end{equation}

The solution to Equation~(\ref{equ_for_hatP}) inside the tube regular at $r = 0$ is $\hat{P}(r) = \rhoi \vai^2 a^2 k_e^2 J_1(\ki r)$ (up to a multiplicative constant). The reason for introducing the multiplier $k_e^2$ will be explained later. The factor $a^2$ is needed for $\hat P$ to have the right units. The general solution to Equation~(\ref{equ_for_hatP}) outside the tube is $\hat{P}(r) = \rhoe \vae^2 a^2 k_e^2 [C_J J_1(\ke r) + C_Y Y_1(\ke r)]$, where $Y_1$ is the Bessel function of the second kind and first order, and $C_J$ and $C_Y$ are constants to be determined. Using the conditions that $\hat{P}$ and $\hat{\xi}_r$ are continuous at $r = a$\/, and the identity \citep{abramowitz1964}

\begin{equation}\label{determ}
J_1(z) Y'_1(z) - J'_1(z) Y_1(z) = \frac2{\pi z} ,
\end{equation}

\noindent 
we obtain the expressions for $C_J$ and $C_Y$,

\begin{equation}\label{C_J_C_Y}
  \begin{array}{l}\displaystyle
    C_J = \frac{\pi a\ke}{2\ki}\left[\ki J_1(\ki a) Y'_1(\ke a) - \ke J'_1(\ki a) Y_1(\ke a)\right],
       \vspace{2mm}\\ \displaystyle
    C_Y = \frac{\pi a\ke}{2\ki}\left[\ke J'_1(\ki a) J_1(\ke a) - \ki J_1(\ki a) J'_1(\ke a)\right] . 
  \end{array}
\end{equation} 

\noindent 
Using Equation~(\ref{hatxir_in_hatP}) and the expression for $\hat{P}(r)$ inside and outside the tube we obtain the expressions for $\hat{\xi}_r(r)$ and $\hat{\xi}_\varphi(r)$ for an improper eigenmode,

\begin{equation}\label{xir_of_r_im}
\xir(r) = 
   \begin{cases}
    \hphantom{xxxx}a^2 \ke^2\ki^{-1} J'_1(\ki r), & r < a, \vspace{1mm}\\
    a^2\ke[C_J J'_1(\ke r) + C_Y Y'_1(\ke r)], & r > a,
  \end{cases}
\end{equation}

\begin{equation}\label{xiphi_of_r_im}
\xiphi(r) = \frac ir
  \begin{cases}
    \hphantom{xxxx}a^2\ke^2\ki^{-2} J_1(\ki r), & r < a, \vspace{1mm}\\
    a^2\left[\right.C_J J_1(\ke r) + C_Y Y_1(\ke r)\left.\right], & r > a.
  \end{cases}
\end{equation}

Note that we have $\ke \to 0$ as $\omega \to |k|\vae$\/. Using the asymptotic expressions valid for small argument \citep{abramowitz1964}

\begin{equation}\label{asymp_J_Y}
J_1(z) \sim \frac z2, \quad Y_1(z) \sim -\frac 2{\pi z},
\end{equation}

\noindent
we obtain that both $\xir(r)$ and $\xiphi(r)$ tend to finite limits as $\ke \to 0$. If we had not introduced the multiplier $\ke^2$ in the expression for $\hat{P}(r)$ inside the tube, these perturbed variables would have singularities at $\ke = 0$.

It is straightforward to check that both the plasma displacement in the radial direction and the magnetic pressure perturbation are continuous at $r = a$ for an improper eigenmode.

\section{SOLUTION BY FOURIER INTEGRALS}\label{sect_fourier}

Now we assume that the magnetic flux tube suffers a localized, small amplitude disturbance at a given time ($t=0$). We only consider plane polarized kink waves that  cause a lateral displacement of the tube such as that of Figure~\ref{fig_cylinder_kink}(b). We do not require the perturbation to be periodic along the tube, as is the case when a normal mode is excited.

\subsection{Initial Conditions}\label{sect_ics}

We introduce Cartesian coordinates ($x$\/, $y$\/, $z$) with the $z$\/-axis along the tube axis and the $x$\/-axis in the direction of the wave polarization. We impose the initial conditions

\begin{equation}\label{init_xi_x}
\xi_x(t=0,r,z) = \xi_0 \psi(r)\exp(-z^2/\Delta^2), \quad \xi_y(t=0,r,z) = 0,
\end{equation}

\noindent
where

\begin{equation}\label{init_psi}
\psi(r) = \left\{\begin{array}{cl} 1, & r \leq a, \vspace*{2mm}\\ \exp(-(r-a)^2/l^2) , & r \geq a.
\end{array}\right. 
\end{equation}

\noindent
Thus at the initial time the $z=0$ cross-section of the tube is displaced a distance $\xi_0$ from its equilibrium position. Tube cross-sections with increasing $|z|$ suffer smaller initial displacements. This is accounted for by the Gaussian factor. Hence, these initial conditions correspond to a transverse hump of the tube localized about $z=0$. The shape of the magnetic cylinder of Figure~\ref{fig_cylinder_kink}(b) has been obtained with these initial conditions and $\xi_0=0.5a$, $\Delta=0.5a$. The length of the initial hump is of the order of $4\Delta$.

We next derive expressions for $\xi_r$ and $\xi_\varphi$ at $t=0$. Without loss of generality we can assume that the angle $\varphi$ is counted from the $x$\/-axis. In Section~\ref{sect_govern} we assumed that the perturbed variables have an azimuthal dependence proportional to $e^{i\varphi}$\/. Then, the real expressions for the $r$- and $\varphi$-components of the displacement are $\Re\left(\xi_r e^{i\varphi}\right)$ and $\Re\left(\xi_\varphi e^{i\varphi}\right)$, where $\Re$ indicates the real part of a quantity. Now, we have the relations

\begin{equation}\label{Car_cylin}
\begin{array}{l}
\xi_x = \cos\varphi\,\Re\left(\xi_r e^{i\varphi}\right) - 
   \sin\varphi\,\Re\left(\xi_\varphi e^{i\varphi}\right), \vspace*{2mm}\\
\xi_y = \sin\varphi\,\Re\left(\xi_r e^{i\varphi}\right) + 
   \cos\varphi\,\Re\left(\xi_\varphi e^{i\varphi}\right) .
\end{array}
\end{equation}

\noindent
Taking into account that $\xi_x$ is independent of $\varphi$ and $\xi_y = 0$, we immediately obtain from these relations that

\begin{equation}\label{cylin_Car}
\xi_r = \xi_x, \quad \xi_\varphi = i\xi_x .
\end{equation}

\noindent
It follows from Equations~(\ref{init_cond}), (\ref{init_xi_x}), and (\ref{cylin_Car}) that

\begin{equation}\label{f_xi_x0}
f_r(r,z) = \xi_0\psi(r)\exp(-z^2/\Delta^2), \quad f_\varphi(r,z) = i\xi_0\psi(r)\exp(-z^2/\Delta^2) = i f_r(r,z).
\end{equation}

As mentioned in Section~\ref{sect_govern}, the initial conditions must be completed by supplying the time derivative of $\xi_r$ and $\xi_\varphi$ at $t=0$, that is, the functions $g_r$ and $g_\varphi$ in Equation~(\ref{init_cond}). We assume that the magnetic tube is initially at rest, and so the time derivative of $\xi_x$ and $\xi_y$ are initially zero. Therefore, the same applies to $\xi_r$ and $\xi_\varphi$, which means that

\begin{equation}\label{g_init_1}
g_r(r,z) = g_\varphi(r,z) = 0 .
\end{equation}

\subsection{Solution to Initial Value Problem} \label{sect_fourier_general}

Our aim now is to solve Equations~(\ref{eq_xi_r})--(\ref{init_cond}). Since the initial conditions~(\ref{init_cond}) are imposed on $\xi_r$ and $\xi_\varphi$ and their time derivatives, the temporal dependence of the perturbed variables is obtained using the radial and azimuthal displacements only. For this reason it is convenient to use the variable $\boldsymbol{\xi}(t,r,z)=(\xi_r,\xi_\varphi)$ in what follows. The complete derivation of the solution for $\boldsymbol{\xi}(t,r,z)$ is presented in Appendix~\ref{app-fourier} and only the main expressions are shown here. The method is based on introducing the Fourier transform of $\boldsymbol{\xi}(t,r,z)$ with respect to $z$,

\begin{equation}\label{Four_trans-main}
\tilde{\boldsymbol{\xi}}(t,r,k) = \int_{-\infty}^\infty
   \boldsymbol{\xi}(t,r,z) e^{-ikz}\,dz.
\end{equation}

\noindent
Note that $\tilde{\boldsymbol{\xi}}$ is also a vector and so it must be interpreted as $\tilde{\boldsymbol{\xi}}=(\tilde\xi_r,\tilde\xi_\varphi)$, with $\tilde\xi_r(t,r,k)$ and $\tilde\xi_\varphi(t,r,k)$ the Fourier transforms of $\xi_r(t,r,z)$ and $\xi_\varphi(t,r,z)$ with respect to $z$.

Now, for a fixed $k$ the displacement vector $\tilde{\boldsymbol{\xi}}(t,r,k)$ can be expressed as a linear combination of the eigenfunctions for this value of $k$. According to Sections~\ref{sect_normal_modes} and \ref{sect_cont_modes}, these eigenfunctions are those of proper eigenmodes (with discrete $\omega$; $|\omega|<|k|\vae$) and of improper modes (with a continuous range of frequencies; $|\omega|>|k|\vae$). Thus,

\begin{align}\label{expand_eigen-main}
\tilde{\boldsymbol{\xi}}(t,r,k) &= \left[A_0^+(k)e^{-i\omega_0(k)t} + 
   A_0^-(k)e^{i\omega_0(k)t}\right]\hat{\boldsymbol{\xi}}_0(r,k) \nonumber\\
&+ \sum_{j=1}^N \left[A_j^+(k)e^{-i\omega_j(k)t} + 
   A_j^-(k)e^{i\omega_j(k)t}\right]\hat{\boldsymbol{\xi}}_j(r,k) \nonumber\\
&+ \int_{|k|\vae}^\infty\left[A_\omega^+(k)e^{-i\omega t} + 
   A_\omega^-(k)e^{i\omega t}\right]\hat{\boldsymbol{\xi}}_\omega(r,k)\,d\omega .  
\end{align}

\noindent
In this expression, the global fast kink mode, whose frequency is denoted by $\omega_0(k)$, contributes with the terms with amplitudes $A_0^\pm(k)$ and its vector displacement $\hat{\boldsymbol{\xi}}_0(r,k)$ has components $\xir(r,k)$ and $\xiphi(r,k)$ given by Equations~(\ref{xir_of_r}) and (\ref{xiphi_of_r}). In Equation~(\ref{expand_eigen-main}) $N$ denotes the number of radial overtones for the longitudinal wavenumber $k$. The $j$th overtone has frequency $\omega_j(k)$ and amplitude $A_j^\pm(k)$. Its vector displacement, $\hat{\boldsymbol{\xi}}_j(r,k)$, also has components given by Equations~(\ref{xir_of_r}) and (\ref{xiphi_of_r}). One must bear in mind that for $|k|<k_{c1}$ no overtones exist and so the terms with amplitude $A_j^\pm(k)$ must be omitted. Finally, the integral in Equation~(\ref{expand_eigen-main}) comes from improper modes with amplitudes $A_\omega^\pm(k)$; their vector displacement is denoted by $\hat{\boldsymbol{\xi}}_\omega(r,k)$ and corresponds to $\xir(r,k)$ and $\xiphi(r,k)$ in Equations~(\ref{xir_of_r_im}) and (\ref{xiphi_of_r_im}). In what follows we do not show the dependence of eigenmode frequencies on $k$ explicitly and write $\omega_j$ instead of $\omega_j(k)$.

The coefficients $A_0^\pm(k)$, $A_j^\pm(k)$, and $A_\omega^\pm(k)$ can be obtained from the initial conditions. Their expressions are

\begin{equation}\label{A_pm-main}
A_j^\pm(k) = \frac{\mathcal{N}_j(k)}{2\omega_j(k)\mathcal{D}_j(k)} , \qquad (j = 0,1,\dots,N),
\end{equation}

\noindent
with

\begin{align}\label{A_numer-main}
\mathcal{N}_j(k) &=
   \frac{\rhoi K_1(\kappa_e a)}{\ki^2} \int_0^a \left[(\omega_j \tilde{f}_r \pm 
   i\tilde{g}_r) \ki r J'_1(\ki r) \right. \nonumber\\
&\hspace{18ex}-\left. (i\omega_j \tilde{f}_\varphi \mp \tilde{g}_\varphi)J_1(\ki r)\right]\,dr \nonumber\\ 
&- \frac{\rhoe J_1(\ki a)}{\kappa_e^2} \int_a^\infty \left[(\omega_j \tilde{f}_r\pm
   i\tilde{g}_r) \kappa_e r K'_1 (\kappa_e r) \right. \nonumber\\ 
&\hspace{18ex}-\left. (i\omega_j \tilde{f}_\varphi \mp \tilde{g}_\varphi) K_1(\kappa_e r)\right]\,dr ,
\end{align}

\begin{align}\label{A_denom-main}
\mathcal{D}_j(k) &= \frac{\rhoi K_1^2(\kappa_e a)}{\ki^4} 
   \int_0^a \left[(r\ki)^2 {J'_1}^2(\ki r) + J_1^2(\ki r)\right]\frac{dr}r \nonumber\\
&+ \frac{\rhoe J_1^2(\ki a)}{\kappa_e^4} \int_a^\infty 
   \left[(r\kappa_e)^2 {K'_1}^2(\kappa_e r) + K_1^2(\kappa_e r)\right]\frac{dr}r  .
\end{align}

\noindent
Here $\tilde f_r(r,k)$, $\tilde f_\varphi(r,k)$, $\tilde g_r(r,k)$, $\tilde g_\varphi(r,k)$ are the Fourier transforms with respect to $z$ of the initial conditions. It follows from Equation~(\ref{A_pm-main}) that any initial perturbation with $g_r(r,z)=g_\varphi(r,z)=0$ has $A_j^-(k) = A_j^+(k)$, which implies that the initial disturbance generates two identical wave trains that propagate in opposite directions along the magnetic tube. This is the case with our particular choice of Equations~(\ref{f_xi_x0}) and (\ref{g_init_1}), for which

\begin{align}\label{hat_f_xi_x0}
& \tilde{f}_r(r,k) = \xi_0\sqrt\pi\Delta\psi(r)\exp(-\Delta^2 k^2/4), \quad 
\tilde{f}_\varphi(r,k) = i \tilde{f}_r(r,k) , \nonumber\\
& \tilde{g}_r(r,k) = \tilde{g}_\varphi(r,k) = 0.
\end{align}

\noindent
Hence, $\omega_j \tilde{f}_r \pm i\tilde{g}_r = -(i\omega_j \tilde{f}_\varphi \mp \tilde{g}_\varphi) = \omega_j \tilde{f}_r$.

The amplitudes of improper modes are 

\begin{equation}\label{A_pm_om-main}
A_\omega^\pm(k) = \frac{\mathcal{N}_\omega(k)}{2\omega q(\omega)} ,
\end{equation}

\noindent
where

\begin{align}\label{A_numer_om-main}
\mathcal{N}_\omega(k) &=
   \frac{\rhoi a^2k_e^2}{\ki^2} \int_0^a \left[(\omega \tilde{f}_r \pm 
   i\tilde{g}_r) \ki r J'_1(\ki r) - (i\omega\tilde{f}_\varphi \mp \tilde{g}_\varphi)J_1(\ki r)\right]\,dr  \nonumber\\
&+ \frac{\rhoe a^2k_e^2}{\ke^2} \int_a^\infty \left\{(\omega\tilde{f}_r \pm i\tilde{g}_r) \ke r 
    \left[C_J J'_1(\ke r) + C_Y Y'_1(\ke r)\right] \right. \nonumber\\
&\hspace{12ex} -\left. (i\omega\tilde{f}_\varphi \mp \tilde{g}_\varphi)
   \left[C_J J_1(\ke r) + C_Y Y_1(\ke r)\right]\right\}\,dr ,
\end{align}

\noindent
and

\begin{equation}\label{q_omega-main}
q(\omega) = \frac{\rho_e\vae^2k_e^2}{\omega}\left(C_J^2 + C_Y^2\right) .
\end{equation}

\noindent Again, an initial perturbation such that $g_r(r,z)=g_\varphi(r,z)=0$ has $A_\omega^-(k) = A_\omega^+(k)$ and so the contribution of improper modes is also made of two identical wave trains propagating in opposite directions.

The last expression we need to introduce is the inverse Fourier transform of $\tilde{\boldsymbol{\xi}}(t,r,k)$,

\begin{equation}\label{Four_trans_inv-main1}
\boldsymbol{\xi}(t,r,z) = \frac1{2\pi}\int_{-\infty}^\infty
   \tilde{\boldsymbol{\xi}}(t,r,k) e^{ikz}\,dk.
\end{equation}

\noindent
Taking into account that $\tilde{\boldsymbol{\xi}}(t,r,-k)=\tilde{\boldsymbol{\xi}}(t,r,k)$, we can reduce this expression to

\begin{equation}\label{Four_trans_inv-main}
\boldsymbol{\xi}(t,r,z) = \frac1{\pi}\int_0^\infty
   \tilde{\boldsymbol{\xi}}(t,r,k) e^{ikz}\,dk.
\end{equation}

\noindent
This formula gives the solution to our initial value problem. Note that one can separate the contributions to $\boldsymbol{\xi}(t,r,z)$ coming from proper and improper modes only by including their respective terms when computing $\tilde{\boldsymbol{\xi}}(t,r,k)$ from Equation~(\ref{expand_eigen-main}). That is, to evaluate the contribution of proper modes the integral in Equation~(\ref{expand_eigen-main}) is eliminated, whereas this integral is the only term retained in this expression to determine the contribution of improper modes.

Having computed the amplitudes $A_j^\pm(k)$ and $A_\omega^\pm(k)$ one can also compute all other perturbed variables, and not just the radial and azimuthal displacement. For example, to determine $v_r(t,r,z)$ one needs to write an expression analogous to Equation~(\ref{expand_eigen-main}) in which the eigenfunctions $\hat{\boldsymbol{\xi}}_j(r,k)$ and $\hat{\boldsymbol{\xi}}_\omega(r,k)$ are replaced by those of the radial velocity components.

\subsection{Asymptotic Behavior for Large $t$}
\label{sect_asymptotic}

We now study the contribution of improper modes in Equation~(\ref{expand_eigen-main}), i.e. the term with the integral. Using integration by parts yields

\begin{eqnarray}\label{interfer}
&& \int_{|k|\vae}^\infty\left[A_\omega^+(k)e^{-i\omega t} + 
   A_\omega^-(k)e^{i\omega t}\right]\hat{\boldsymbol{\xi}}_\omega(r,k)\,d\omega
   \nonumber\\
&& = \frac it\left[A_{|k|\vae}^-(k)e^{i|k|\vae t} - 
   A_{|k|\vae}^+(k)e^{-i|k|\vae t}\right]\hat{\boldsymbol{\xi}}_{|k|\vae}(r,k) \nonumber\\
&& \hphantom{xx} +\, \frac it\int_{|k|\vae}^\infty
   \left\{e^{i\omega t}\frac\partial{\partial\omega} 
   \left[A_\omega^-(k)\hat{\boldsymbol{\xi}}_\omega(r,k)\right]\right.\nonumber\\ 
&& \left. \hphantom{xxxxx} -\, e^{-i\omega t}\frac\partial{\partial\omega}\left[A_\omega^+(k)
   \hat{\boldsymbol{\xi}}_\omega(r,k)\right]\right\}\,d\omega . 
\end{eqnarray} 

\noindent
This result shows that the third term in Equation~(\ref{expand_eigen-main}) decays as $1/t$ with time. This is similar to the asymptotic behavior in time of the Alfv\'en continuum modes studied in \citet{tataronis1975} and also in \citet{poedts2004}; see Chapter~10. Hence, the asymptotic expression for $\tilde{\boldsymbol{\xi}}(t,r,k)$ for large time is given by the first two terms on the right-hand side of Equation~(\ref{expand_eigen-main}). For $|k| < k_{c1}$ there is no sum on the right-hand side of Equation~(\ref{expand_eigen-main}), and the large time asymptotic behavior of $\tilde{\boldsymbol{\xi}}(t,r,k)$ is described by the first term. The large time asymptotic behavior of $\boldsymbol{\xi}(t,r,z)$ is described by the inverse Fourier transform of the asymptotic expression for $\tilde{\boldsymbol{\xi}}(t,r,k)$. If $\tilde{\boldsymbol{f}}(r,z) = \tilde{\boldsymbol{g}}(r,z) = 0$ for $|k| > k_{c1}$\/, then the asymptotic behavior of $\boldsymbol{\xi}(t,r,z)$ is described by the inverse Fourier transform of the first term on the right-hand side of Equation~(\ref{expand_eigen-main}). In other words, the temporal evolution of the magnetic tube is described by the global kink mode.

\subsection{Numerical Procedure} \label{numerical_procedure}

Here we describe the steps to obtain a solution to the initial value problem from a numerical solution to Equation~(\ref{Four_trans_inv-main}). Normal and improper modes are treated separately.

For normal modes we consider a grid of wavenumbers and solve the dispersion relation on this grid. Then, we have a discrete approximation to $\omega_j(k)$. For a fixed $k$ in the grid, numerical approximations to the integrals in Equations~(\ref{A_numer-main}) and (\ref{A_denom-main}) are computed. In these expressions one must substitute $\ki$ and $\kappa_e$ from Equation~(\ref{ks}) and the Fourier transforms of the initial conditions from Equation~(\ref{hat_f_xi_x0}). By repeating these numerical integrations for all $k$ in the grid, we end up with numerical approximations to $\mathcal{N}_j(k)$ and $\mathcal{D}_j(k)$, so that the amplitudes $A_j^\pm(k)$ can be obtained with the help of Equation~(\ref{A_pm-main}).

Concerning improper modes, we consider a two-dimensional grid of wavenumbers and frequencies and since these modes only exist for $|\omega|\geq|k|\vae$ we ignore the frequencies that do not satisfy this condition. Now, numerical approximations to $A_\omega^\pm(k)$ can be computed in our two-dimensional grid following an analogous procedure to that of proper modes. Here we need to make use of Equation~(\ref{ke}) for $\ke$ and Equation~(\ref{C_J_C_Y}) for $C_J$ and $C_Y$.

We are ready to use Equation~(\ref{Four_trans_inv-main}). At this point we select the time ($t$) and position ($r,z$) where the displacement vector will be computed. A wavenumber is selected and $\tilde{\boldsymbol{\xi}}(t,r,k)$ is calculated using Equation~(\ref{expand_eigen-main}). A numerical approximation to the integral in this formula can be computed using the approximations to $A_\omega^\pm(k)$ on the two-dimensional $(k,\omega)$ grid. Using Equation~(\ref{expand_eigen-main}) for all the wavenumbers in the grid results in a discrete approximation to $\tilde{\boldsymbol{\xi}}(t,r,k)$, that can be used to obtain $\boldsymbol{\xi}(t,r,z)$ numerically from Equation~(\ref{Four_trans_inv-main}).

\section{NUMERICAL RESULTS}\label{subsect_numerics}

In this section we present the results of numerical calculations by examining the evolution of the wave train in time. In the calculations presented in this work we always consider a point on the magnetic tube boundary, that is, we set $r=a$. In addition, in Sections~\ref{subsubsect_t=0} and \ref{subsubsect_spatial} the amplitude of the initial disturbance is taken as $\xi_0=1$ in arbitrary units.

\subsection{Accuracy of the Numerical Solution and Contribution of Proper and Improper Modes}\label{subsubsect_t=0}

\begin{figure}[ht!]
  \centerline{
    \scriptsize{(a)}
    \includegraphics[width=0.33\textwidth,angle=-90]{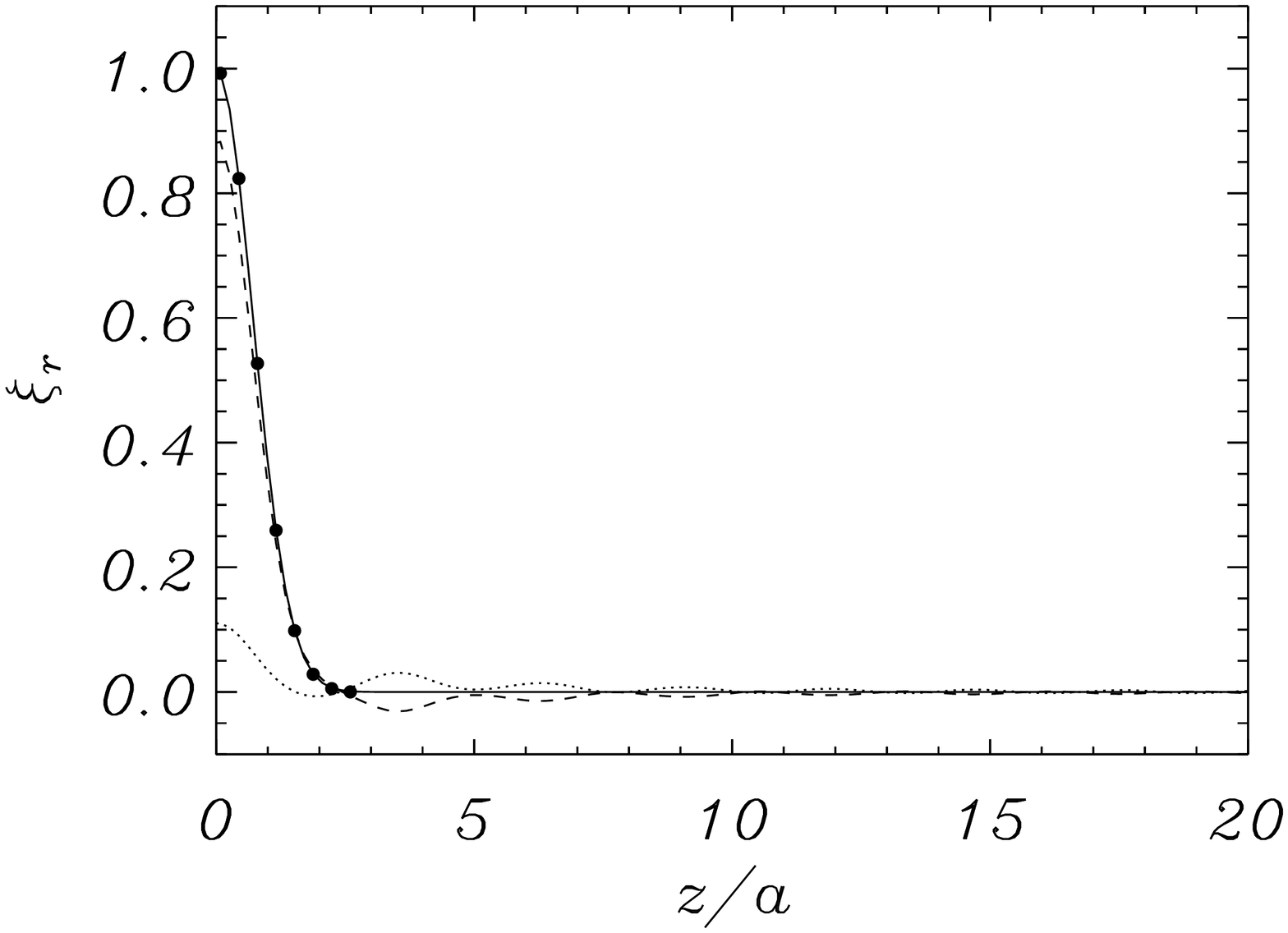} \\
    \scriptsize{(b)}
    \includegraphics[width=0.33\textwidth,angle=-90]{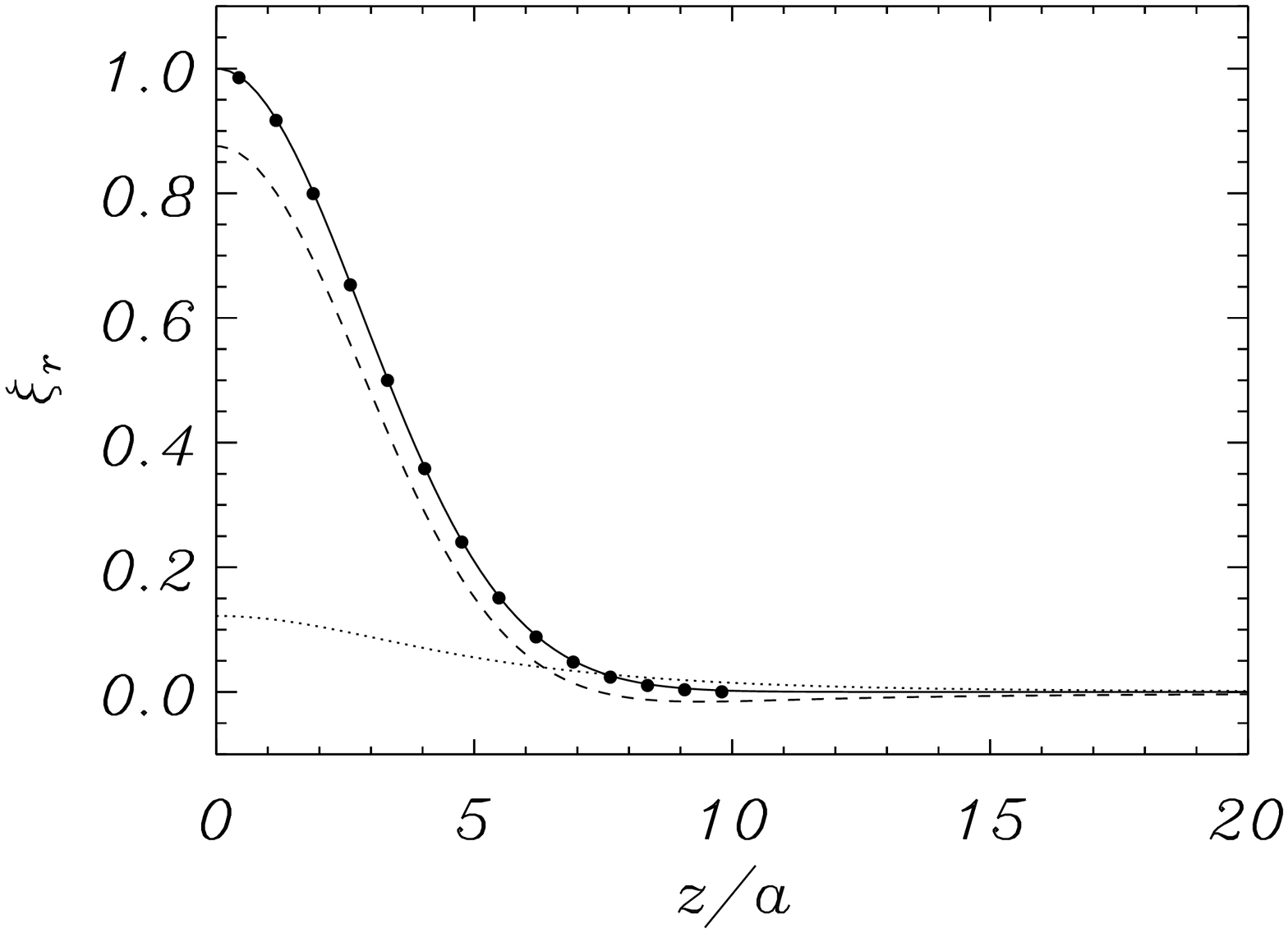} \\
  }
  \centerline{
    \scriptsize{(c)}
    \includegraphics[width=0.33\textwidth,angle=-90]{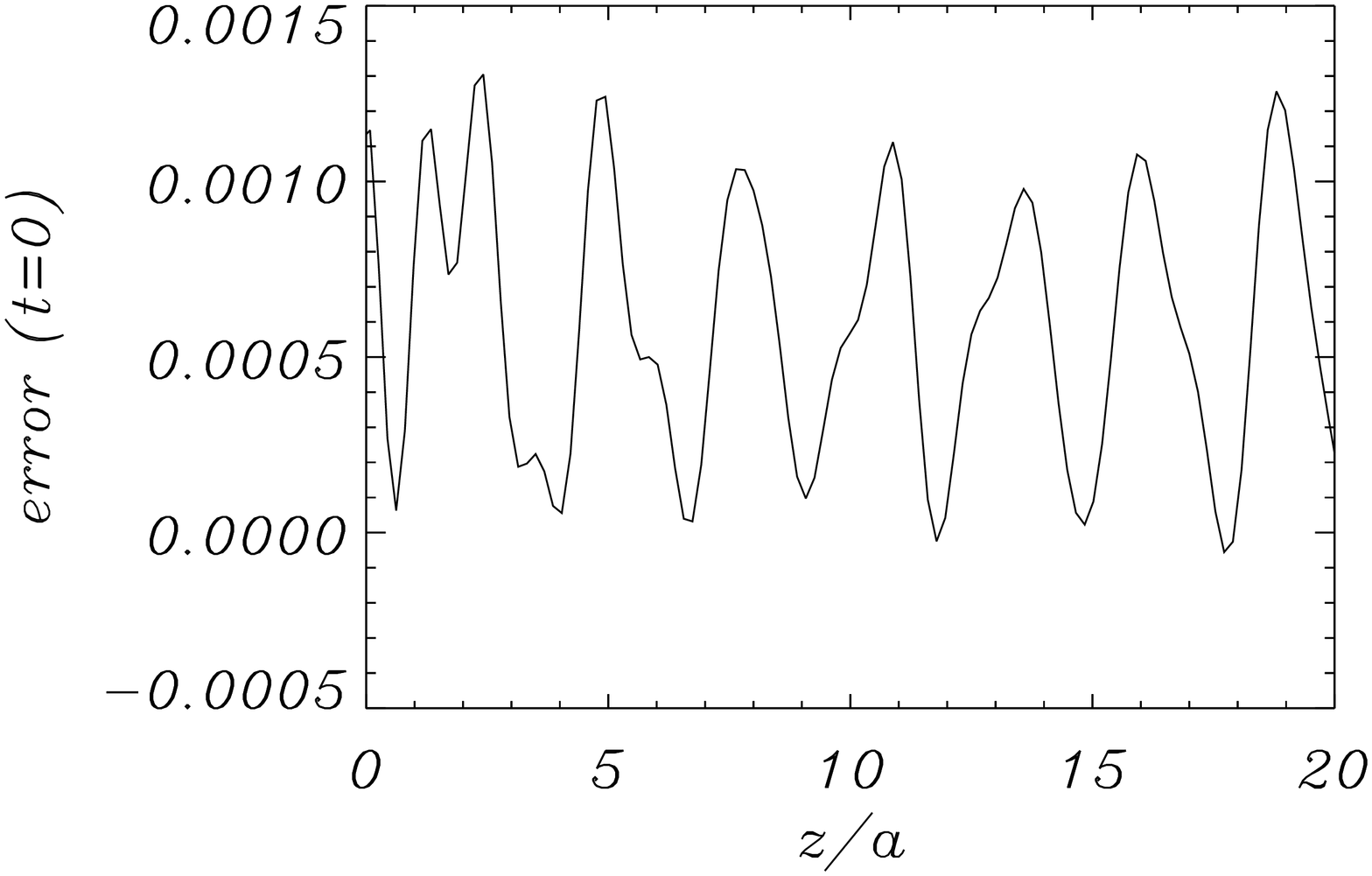} \\
    \scriptsize{(d)}
    \includegraphics[width=0.33\textwidth,angle=-90]{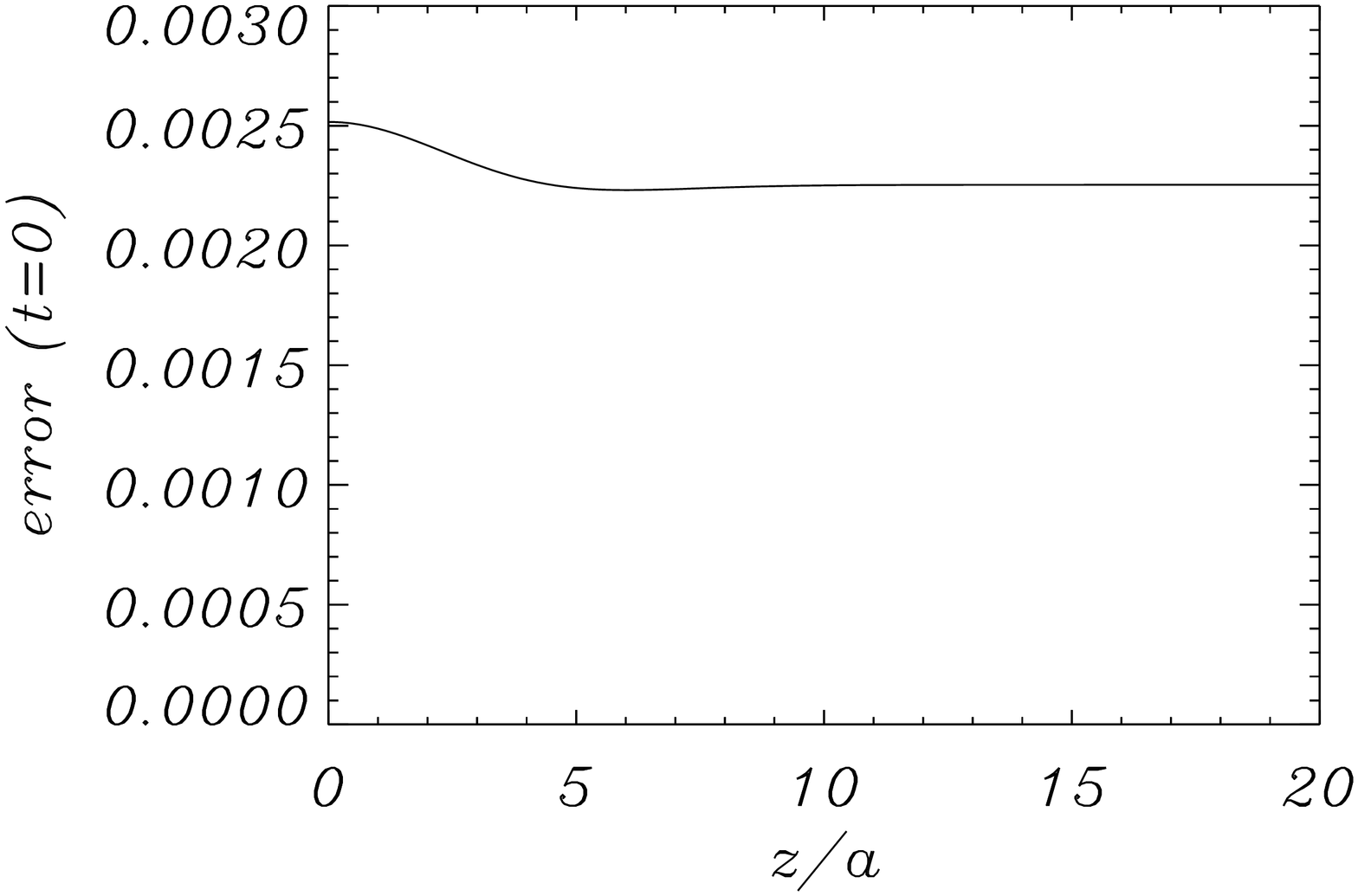} \\
  }
  \caption{Top: initial condition for $\xi_r$ along the magnetic tube (solid line) and numerically obtained solution for $t=0$ (circles). The contributions of proper and improper modes are shown as dashed and dotted lines, respectively. Bottom: difference between the initial condition for $\xi_r$ and the numerically obtained solution for $t=0$. In all panels the density ratio is $\rhoi/\rhoe=4$, while $\Delta=a$ (left) and $\Delta=4a$ (right), where $a$ denotes the magnetic tube radius. It is worth recalling that the length of the initial disturbance is of the order of $4\Delta$.}
  \label{fig_t=0}
\end{figure}

The procedure outlined in Section~\ref{numerical_procedure} requires the evaluation of many approximations to integrals over finite and semi-infinite intervals and so the numerical errors might become so important that the solution is unreliable. The accuracy of the numerical solution is first judged by computing the radial displacement of the tube boundary at $t=0$ with Equation~(\ref{Four_trans_inv-main}) and comparing it with the imposed initial condition ($f_r$ of Equation~(\ref{f_xi_x0})). The results are presented in Figure~\ref{fig_t=0} for two lengths of the initial tube displacement and the density ratio $\rhoi/\rhoe=4$; other values of this parameter yield a similar outcome. Figures~\ref{fig_t=0}(a) and (b) show an extremely good agreement between the initial displacement of the tube (solid line) and the solution computed with Equation~(\ref{Four_trans_inv-main}) for $t=0$ (circles). To have a better idea about the error of the numerical approximation to $\xi_r$, the difference between these exact and numerical initial conditions is presented in Figures~\ref{fig_t=0}(c) and (d). This error is everywhere smaller than 0.3\%, which proves that the Fourier-based method and its numerical implementation work satisfactorily. Note that this consistency of our calculations reveals that the contributions of both proper and improper modes (solid and dotted lines, respectively, in Figures~\ref{fig_t=0}(a) and (b)) have been calculated adequately or else their sum would not accurately reproduce the initial solution. In Figure~\ref{fig_t=0} we have restricted ourselves to positive $z$ because the numerical solutions computed with Equation~(\ref{Four_trans_inv-main}) are symmetric with respect to $z=0$, as expected. This is not only true for $t=0$, but also for $t\neq 0$, as can be seen in Figure~\ref{fig_vae2_delta1_short_t}.

\begin{figure}[ht!]
  \centerline{
    \scriptsize{(a)}
    \includegraphics[width=0.33\textwidth,angle=-90]{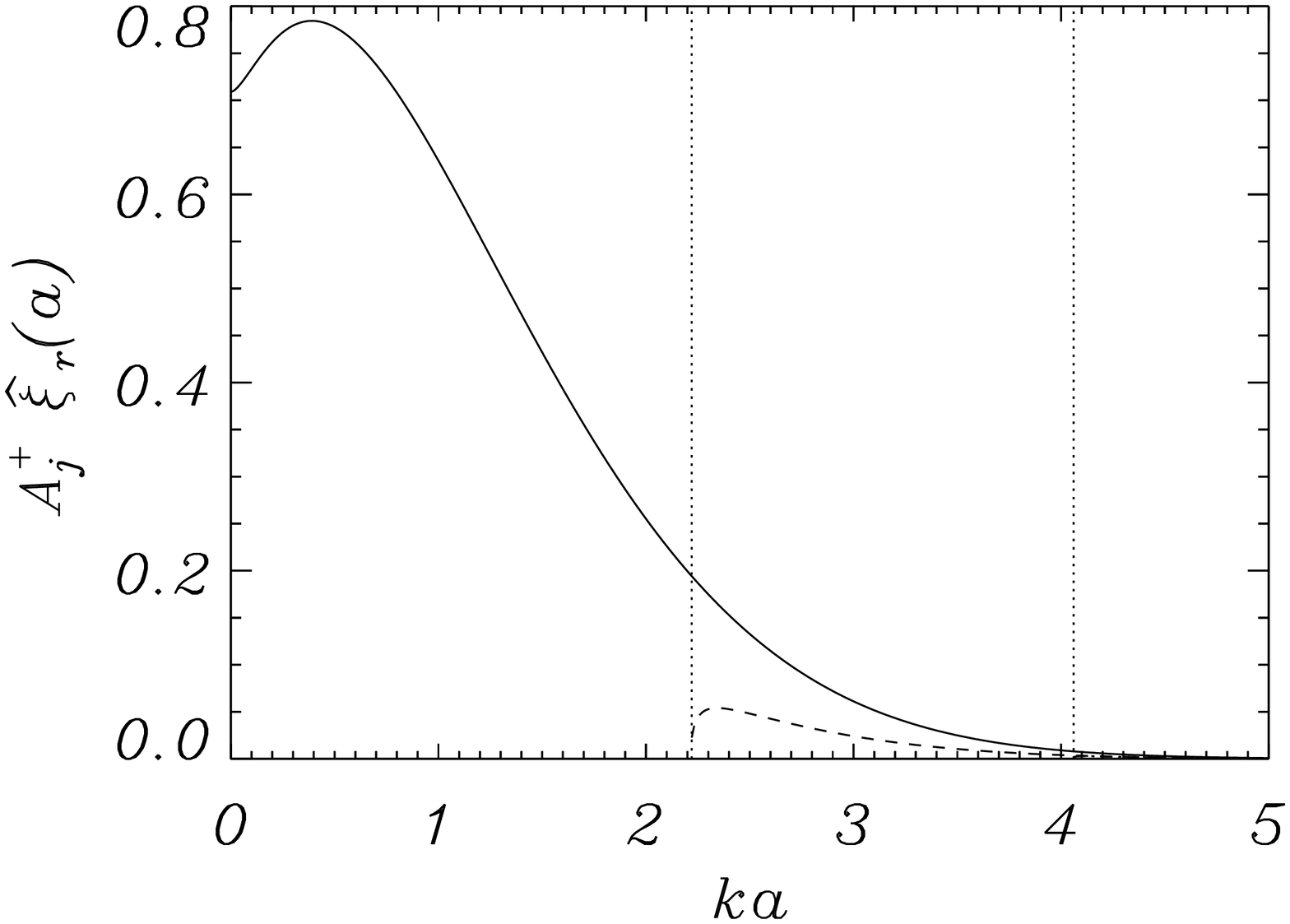} \\
    \scriptsize{(b)}
    \includegraphics[width=0.33\textwidth,angle=-90]{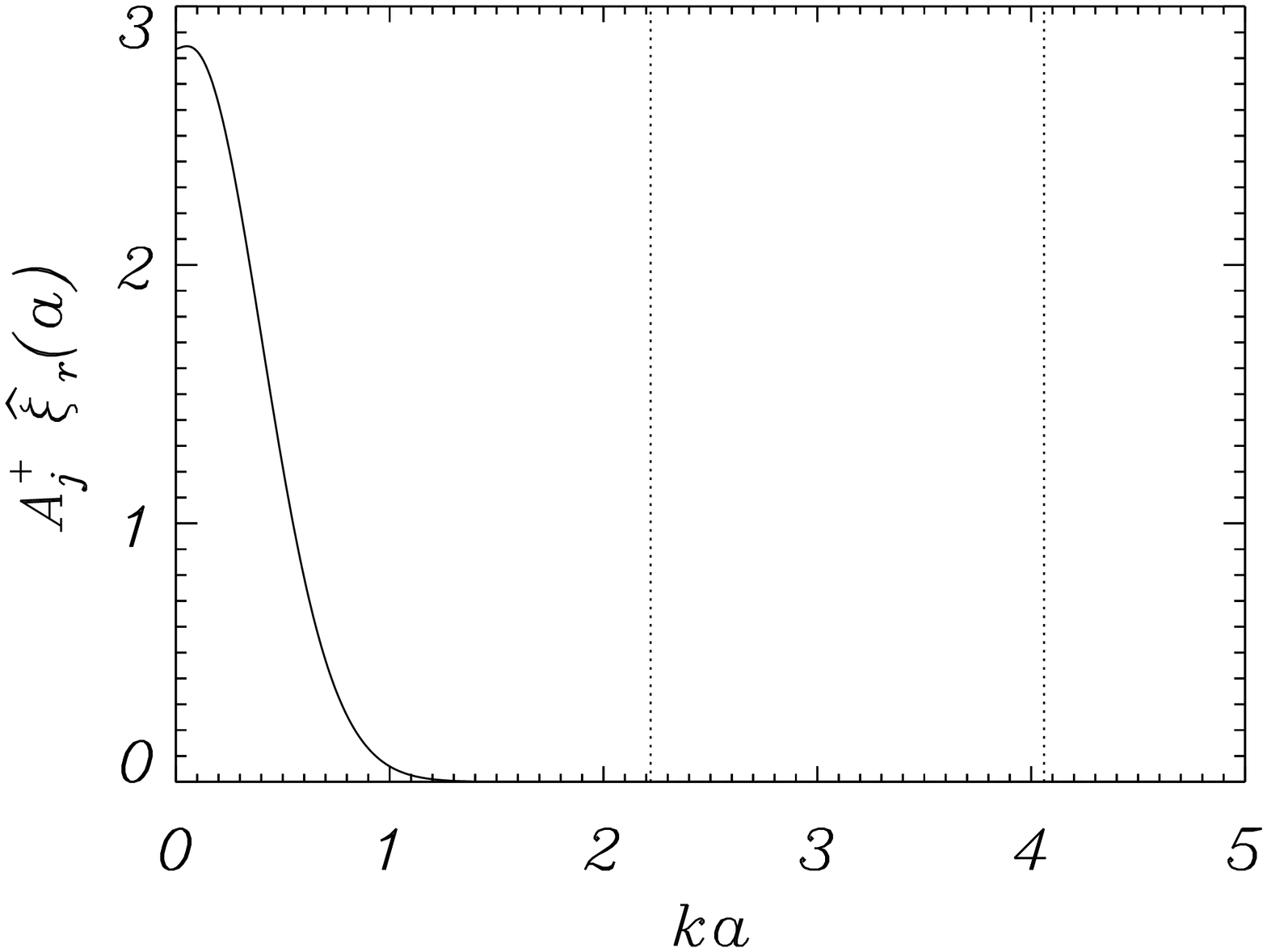} \\
  }
  \centerline{
    \scriptsize{(c)}
    \includegraphics[width=0.33\textwidth,angle=-90]{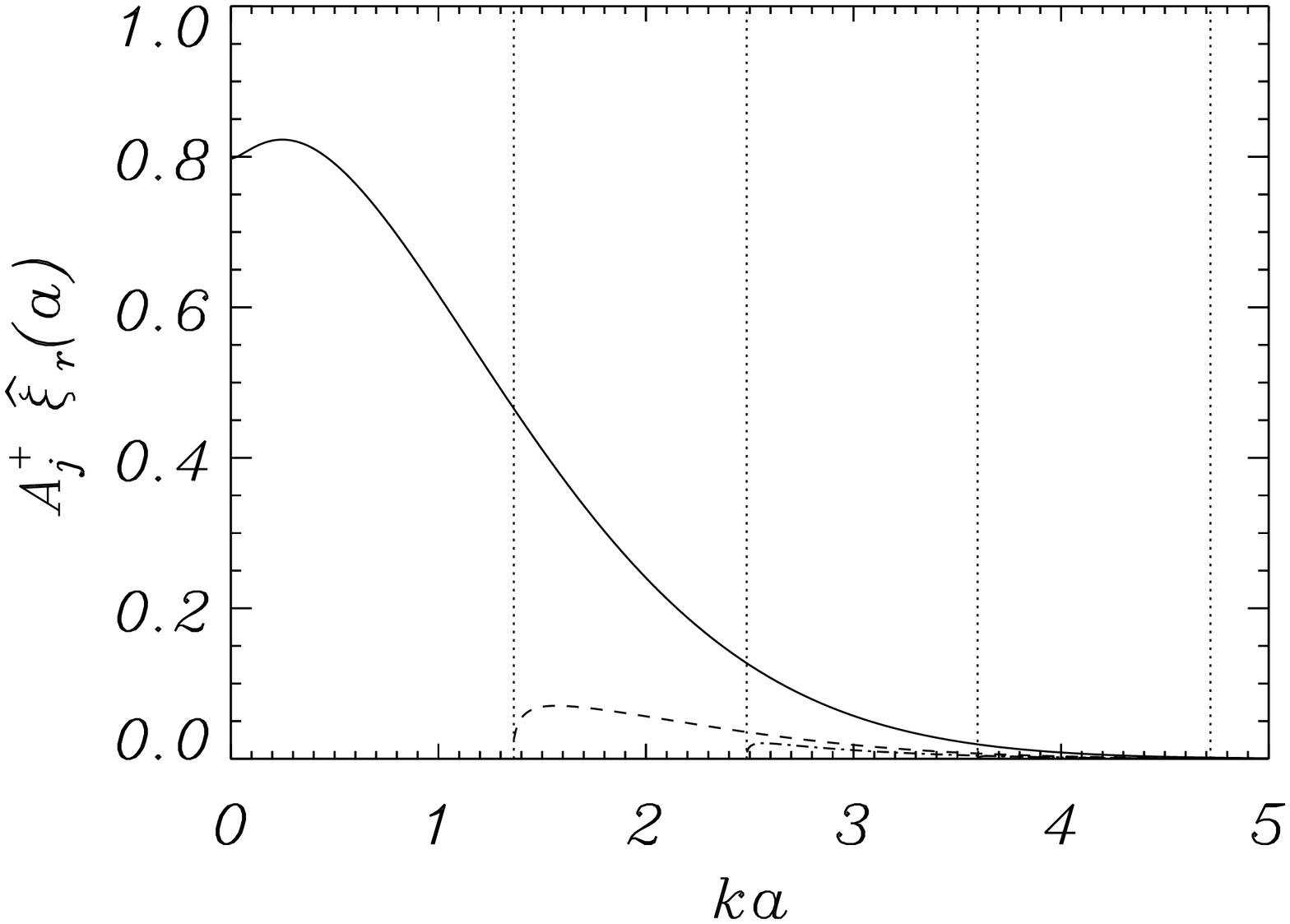} \\
  }
  \caption{Proper mode contribution to the radial displacement as a function of the longitudinal wavenumber for (a) the density ratio $\rhoi/\rhoe=4$ and $\Delta=a$; (b) $\rhoi/\rhoe=4$ and $\Delta=4a$; and (c) $\rhoi/\rhoe=9$ and $\Delta=a$. Solid, dashed, and dash-dotted lines correspond to the global kink mode and its first two overtones. The third and fourth overtones in panel (c) cannot be discerned with this vertical scale. Vertical dotted lines denote the cut-off wavenumbers of the overtones.}
  \label{fig_amplr}
\end{figure}

Regarding the importance of proper and improper modes, Figures~\ref{fig_t=0}(a) and (b) show that the first ones contribute much more than the second ones to the tube displacement at $t=0$ and for this reason proper modes are expected to determine the main features of the wave train propagation. Moreover, we have already proven that improper modes contribution tends to zero as $t$ tends to infinity (Section~\ref{sect_asymptotic}). Then, we next concentrate on the contribution of the various proper modes (i.e. the global kink mode and its radial overtones) to the tube transverse displacement. We restrict ourselves to the wave packet propagating in the positive $z$-direction, that comes from the terms with amplitude $A_j ^+(k)$ (with $j=0$ for the global kink mode and $j=1,2,\ldots$ for its overtones) in Equation~(\ref{expand_eigen-main}). Each of these terms is multiplied by the factor $\hat{\boldsymbol{\xi}}_j(r,k)$. Hence, the contribution of each proper mode to the radial displacement for a fixed $k$ is given by the product $A_j ^+(k)\hat\xi_r(a,k)$, where the second factor is computed with Equation~(\ref{xir_of_r}). Note that the radial eigenfunction, $\hat\xi_r$, is calculated at $r=a$ because we are only interested in the radial displacement of the tube boundary. Figure~\ref{fig_amplr}(a) presents the product $A_j ^+(k)\hat\xi_r(a,k)$ for the density ratio $\rhoi/\rhoe=4$ and $\Delta=a$; these parameter values are those of Figure~\ref{fig_t=0}(a). The global kink mode contribution is much larger than that of its overtones and is concentrated in a restricted range of wavenumbers, namely $ka\lesssim 4$. The first overtone, that only exists for wavenumbers larger than its cut-off, has a much smaller contribution that also decreases with increasing $k$. And, finally, the second overtone has an even smaller contribution and so its importance in the wave train propagating along the magnetic tube must be very weak. Next we consider the same density ratio, $\rhoi/\rhoe=4$, and a longer initial perturbation with $\Delta=4a$ (Figure~\ref{fig_amplr}(b)); these parameter values are those of Figure~\ref{fig_t=0}(b). Now only the global kink mode has a significant amplitude, the first overtone having a maximum value $\max_kA_1^+(k)\hat\xi_r(a,k) = 1.1\times 10^{-9}$. The global kink mode amplitude is now concentrated in a narrower range of $k$ because a longer initial disturbance ($f_r$) has a more concentrated Fourier transform ($\tilde f_r$); cf. Equations~(\ref{f_xi_x0}) and (\ref{hat_f_xi_x0}). Finally, increasing the density ratio while keeping the shape of the initial perturbation unchanged has very little effect in the global mode contribution: compare the solid lines of Figures~\ref{fig_amplr}(a) and (c). Nevertheless, an increase of $\rhoi/\rhoe$ lowers the cut-off wavenumbers of the radial overtones and so the first overtone is present for smaller $k$, although its amplitude is roughly the same: compare the dashed lines of Figures~\ref{fig_amplr}(a) and (c). On the other hand, decreasing the density ratio with respect to the value of Figure~\ref{fig_amplr}(a) has the opposite effect and the first overtone becomes even less important. Hence, we conclude that the shape of the wave packet is mainly defined by the global kink mode.

\subsection{Evolution of the Wave Train Shape}\label{subsubsect_spatial}

\begin{figure}[ht!]
  \centerline{
    \scriptsize{(a)}
    \includegraphics[width=0.33\textwidth,angle=-90]{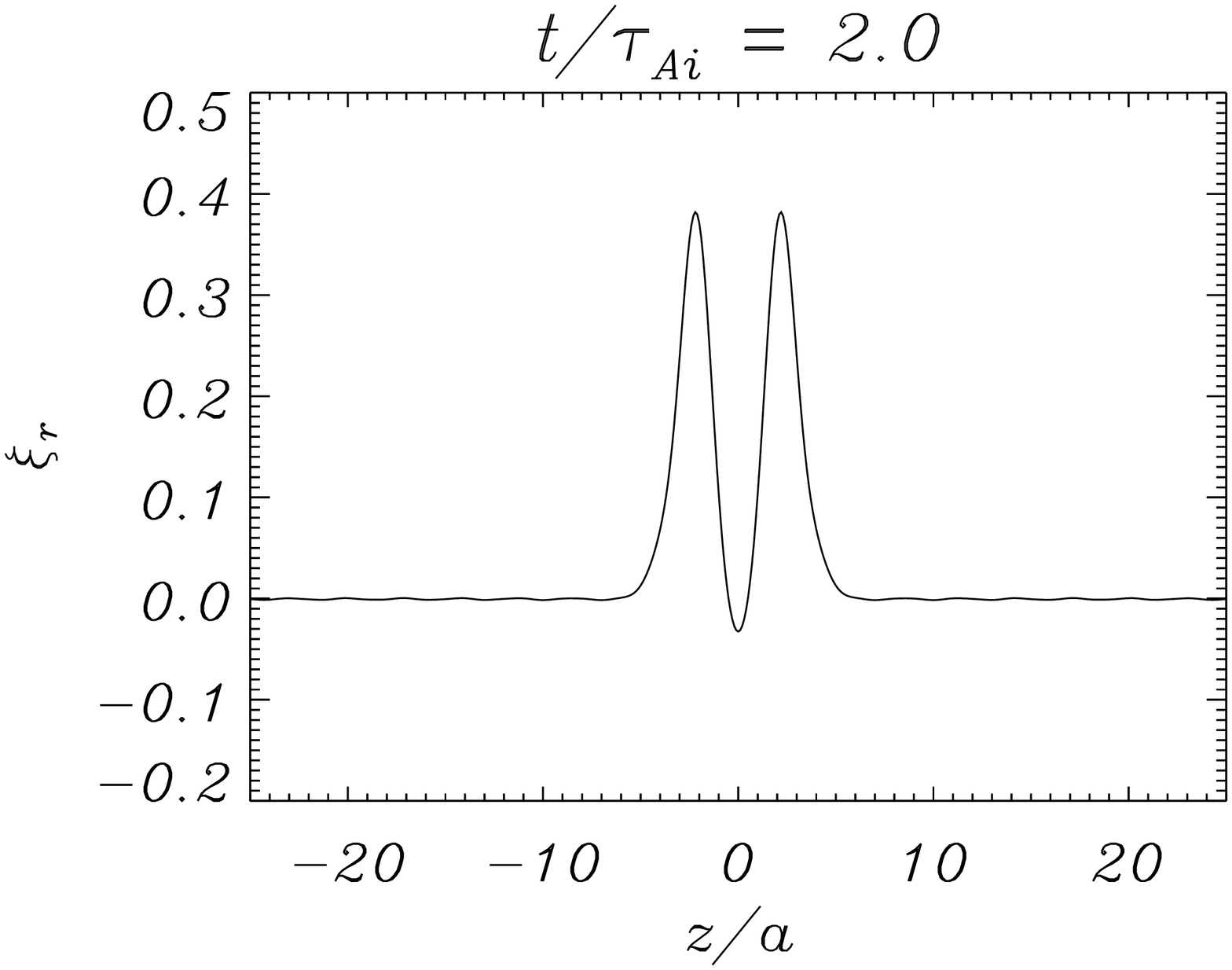} \\
    \scriptsize{(b)}
    \includegraphics[width=0.33\textwidth,angle=-90]{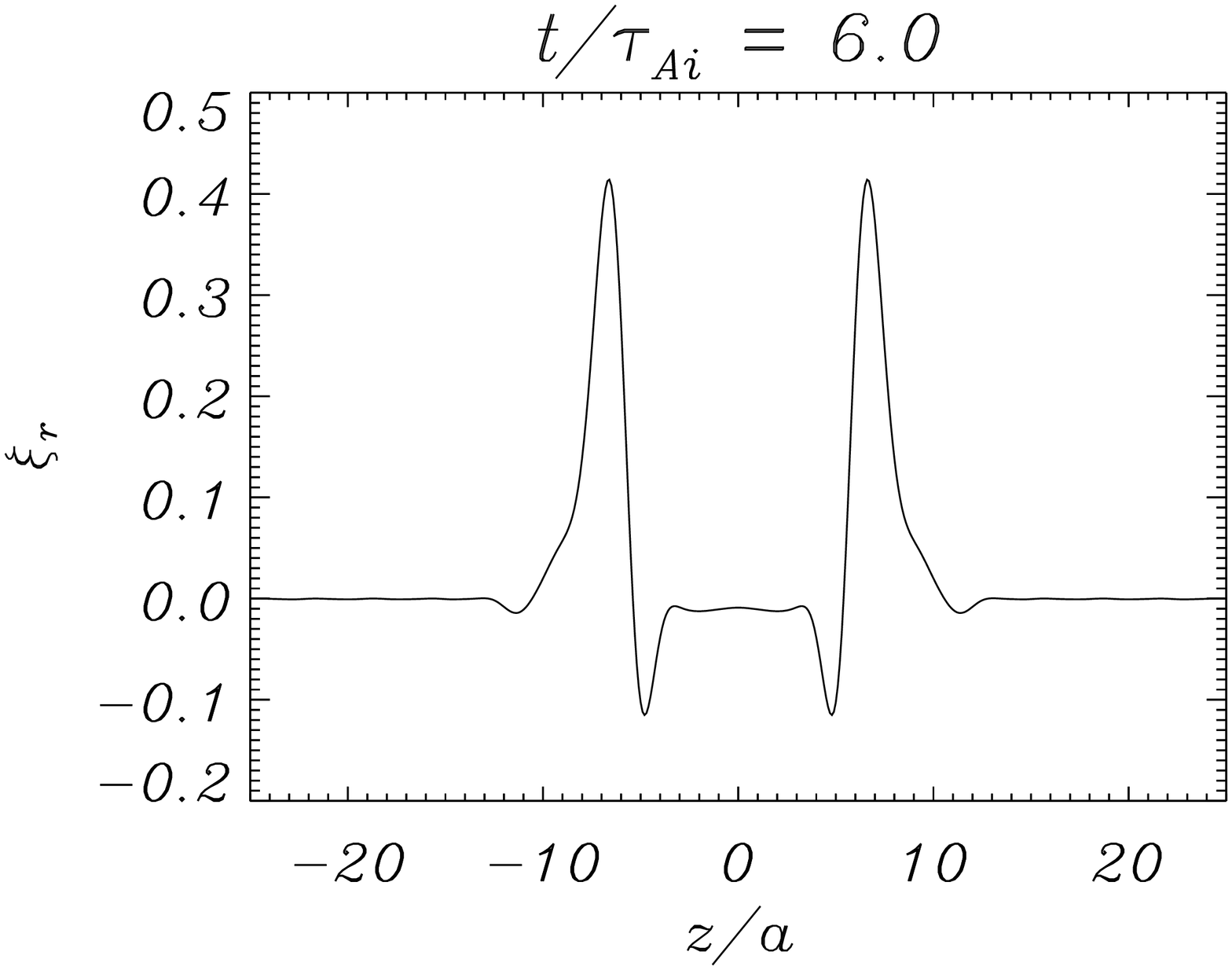} \\
  }
  \centerline{
    \scriptsize{(c)}
    \includegraphics[width=0.33\textwidth,angle=-90]{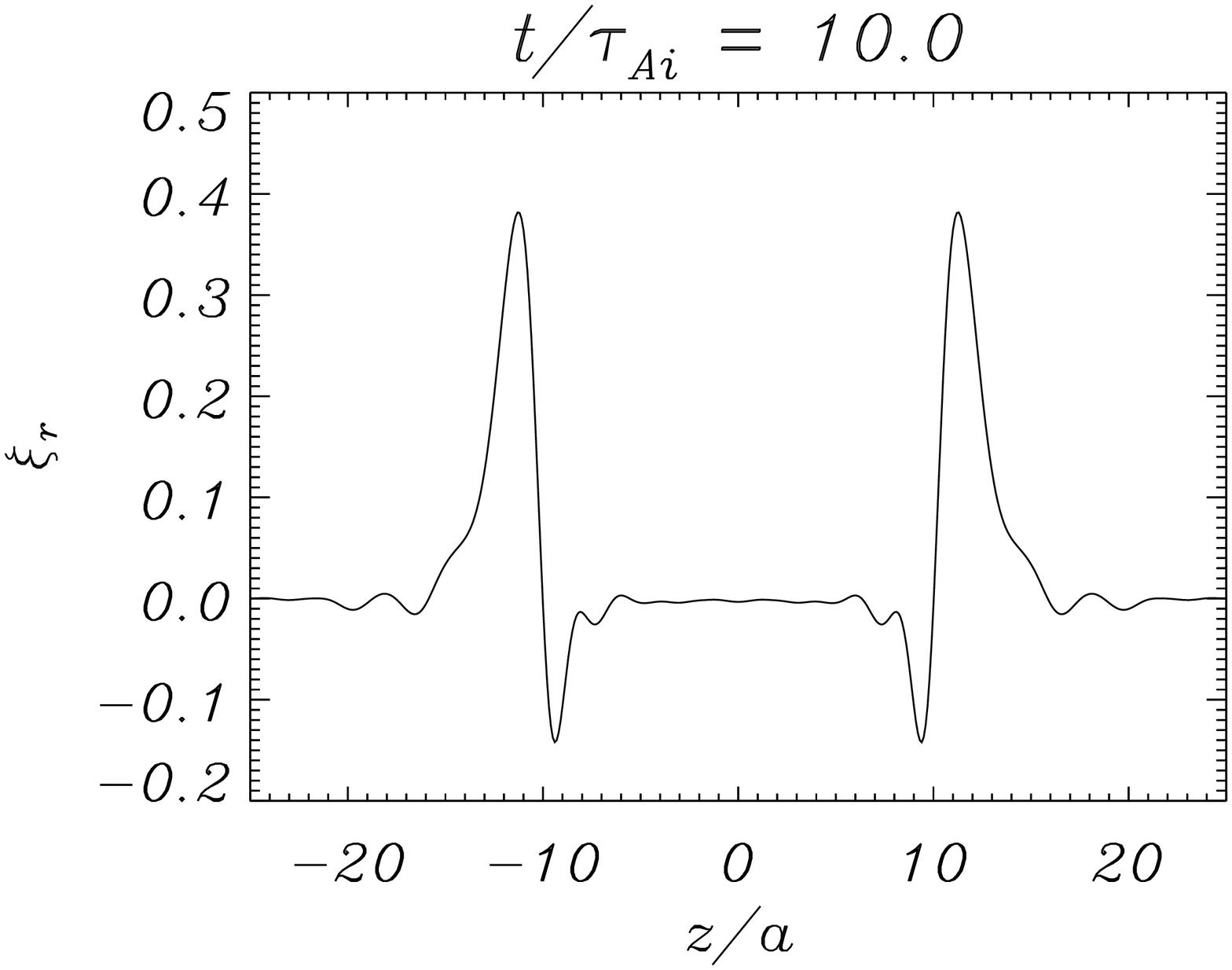} \\
  }
  \caption{Radial displacement of the magnetic tube boundary as a function of $z$ for several times (shown at the top of each frame). This figure shows the early evolution of the initial pulse of Figure~\ref{fig_t=0}. The density ratio is $\rhoi/\rhoe=4$ and $\Delta = a$, while the length of the initial disturbance is of the order of $4\Delta$.}
  \label{fig_vae2_delta1_short_t}
\end{figure}

In Figure~\ref{fig_vae2_delta1_short_t} we consider the initial solution of Figure~\ref{fig_t=0}(a) and show how the initial disturbance splits into two symmetric wave packets travelling at the same speed in opposite directions along the magnetic tube (Figure~\ref{fig_vae2_delta1_short_t}(a)). After a very short time (Figure~\ref{fig_vae2_delta1_short_t}(b)) both packets become distorted because of dispersion and display a strong oscillation behind them. This effect becomes more pronounced as time passes (Figure~\ref{fig_vae2_delta1_short_t}(c)) and more oscillations at the front and back of the wave packet develop. Note that Figure~\ref{fig_vae2_delta1_short_t} corresponds to particular values of the density ratio and length of the initial perturbation. The results are similar for other values of $\rhoi/\rhoe$, but changing $\Delta$ has strong influence on the evolution of the wave train shape, as we will show later.

\begin{figure}[ht!]
  \centerline{
    \scriptsize{(a)}
    \includegraphics[width=0.33\textwidth,angle=-90]{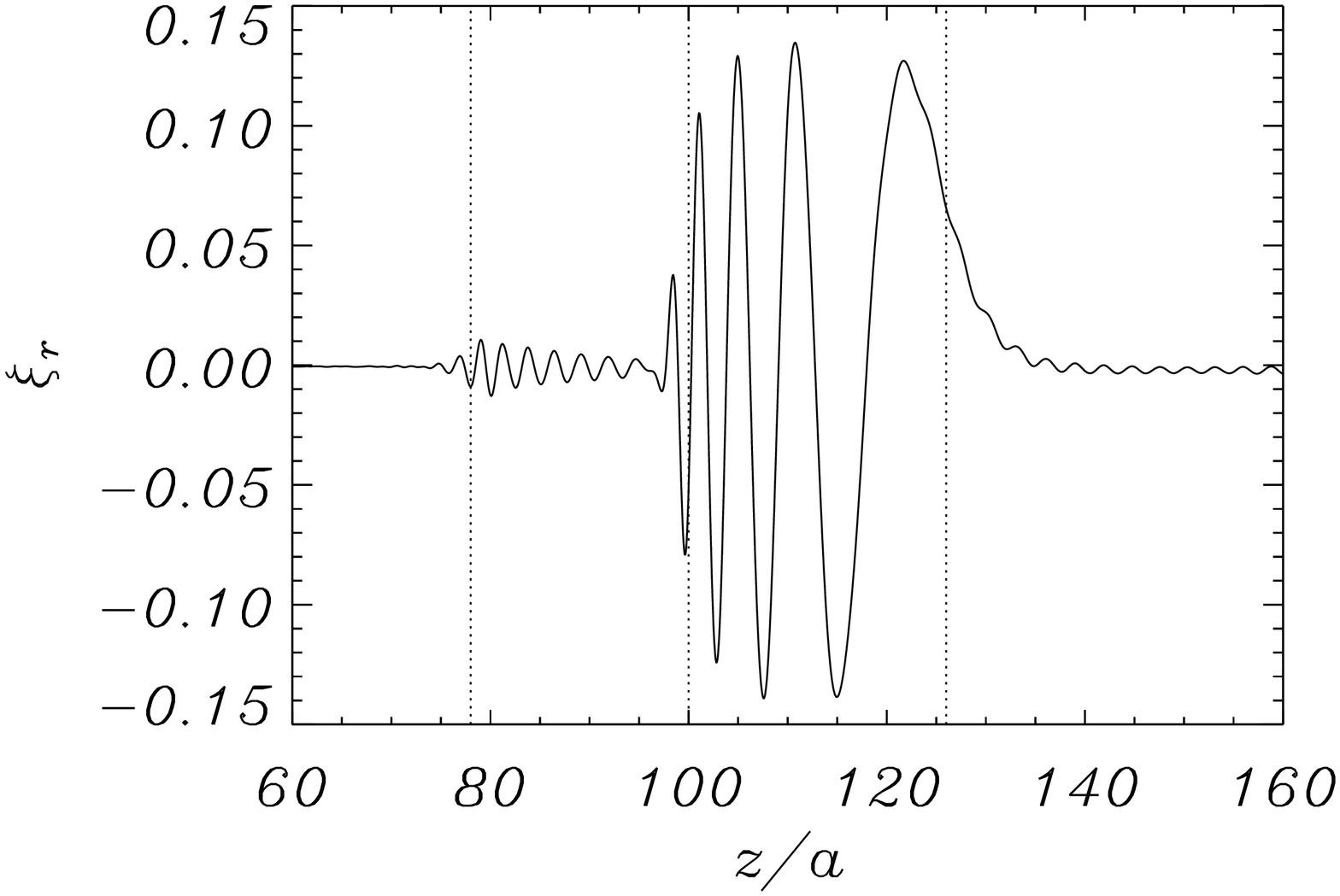} \\
    \scriptsize{(b)}
    \includegraphics[width=0.33\textwidth,angle=-90]{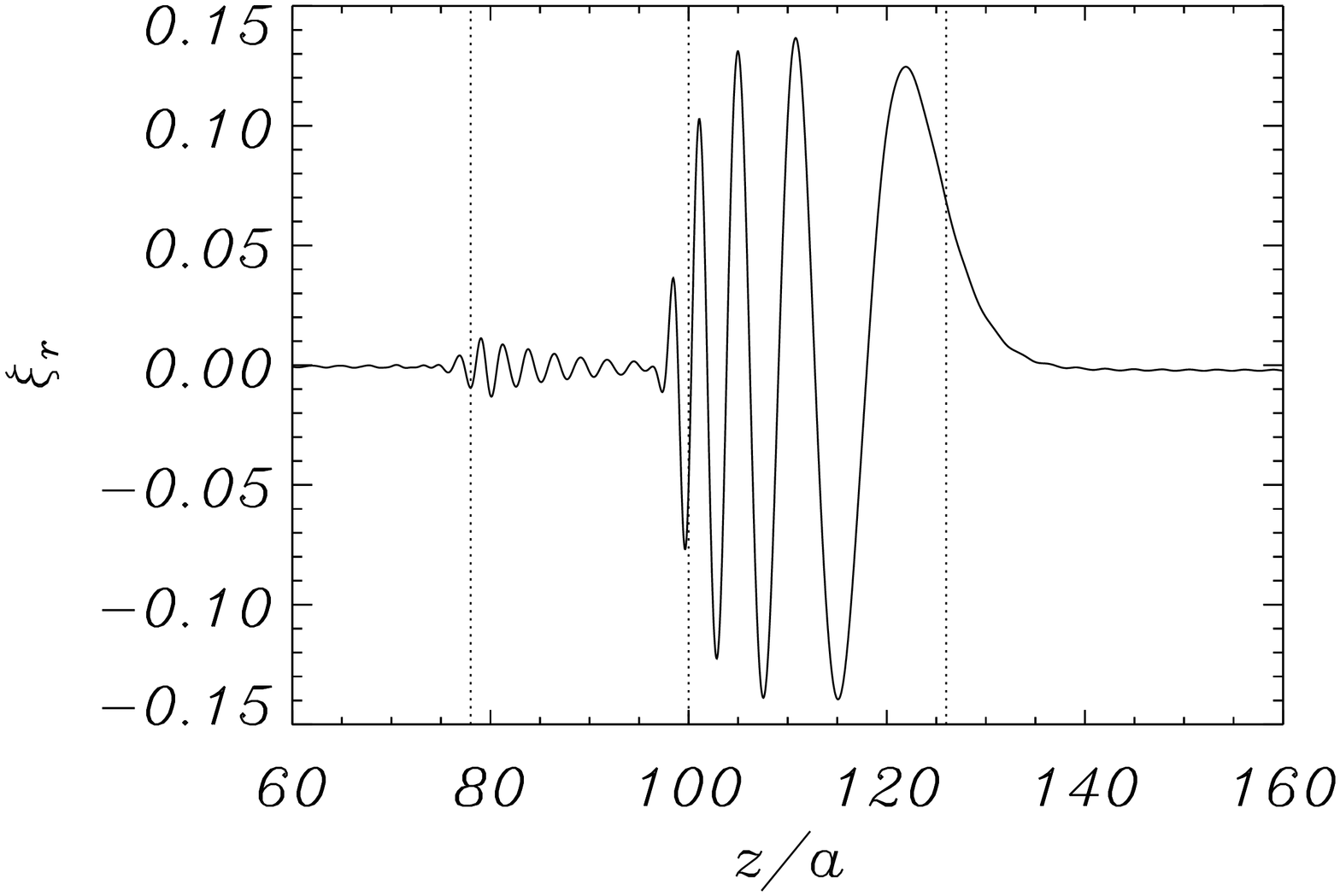} \\
  }
  \centerline{
    \scriptsize{(c)}
    \includegraphics[width=0.33\textwidth,angle=-90]{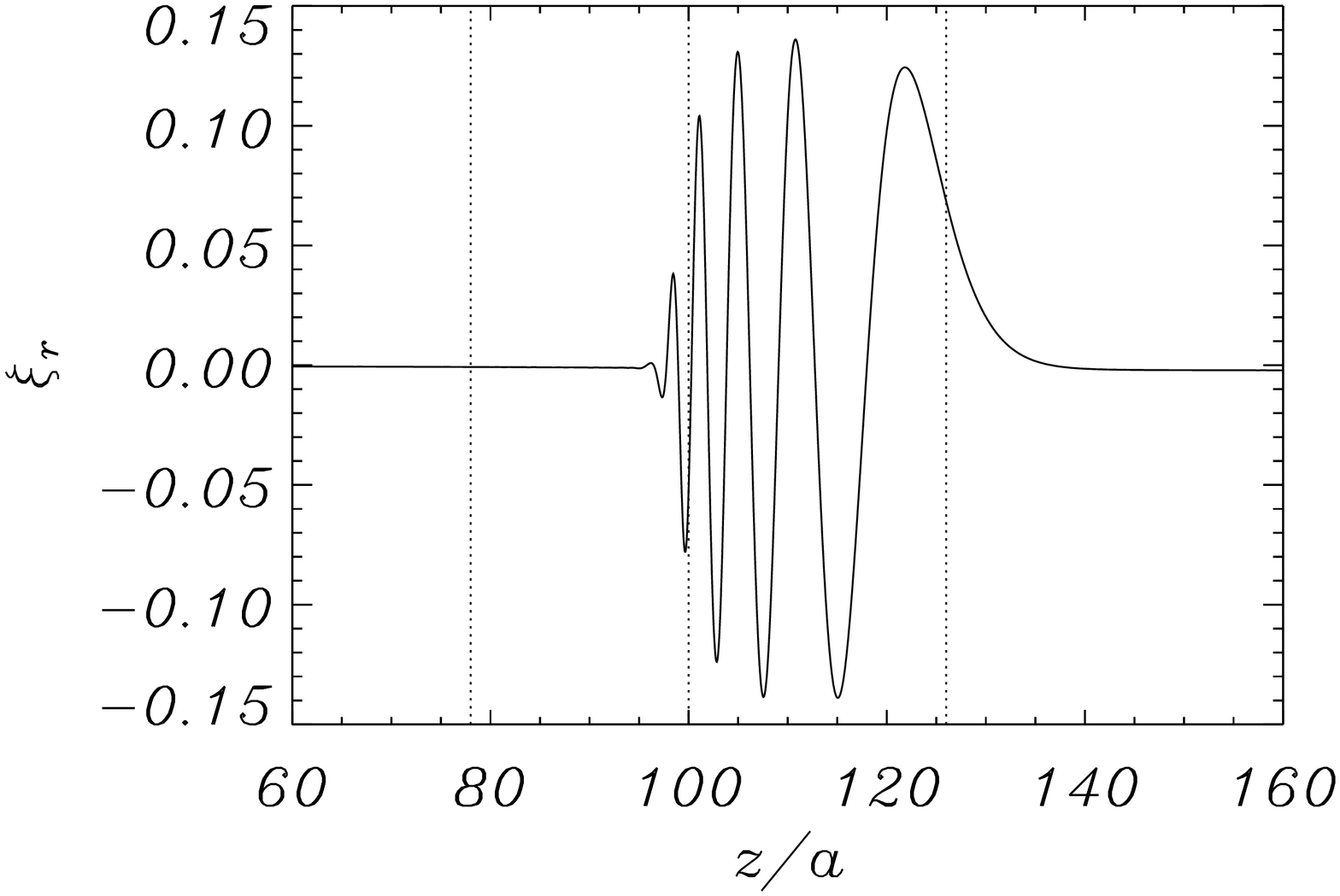} \\
    \scriptsize{(d)}
    \includegraphics[width=0.33\textwidth,angle=-90]{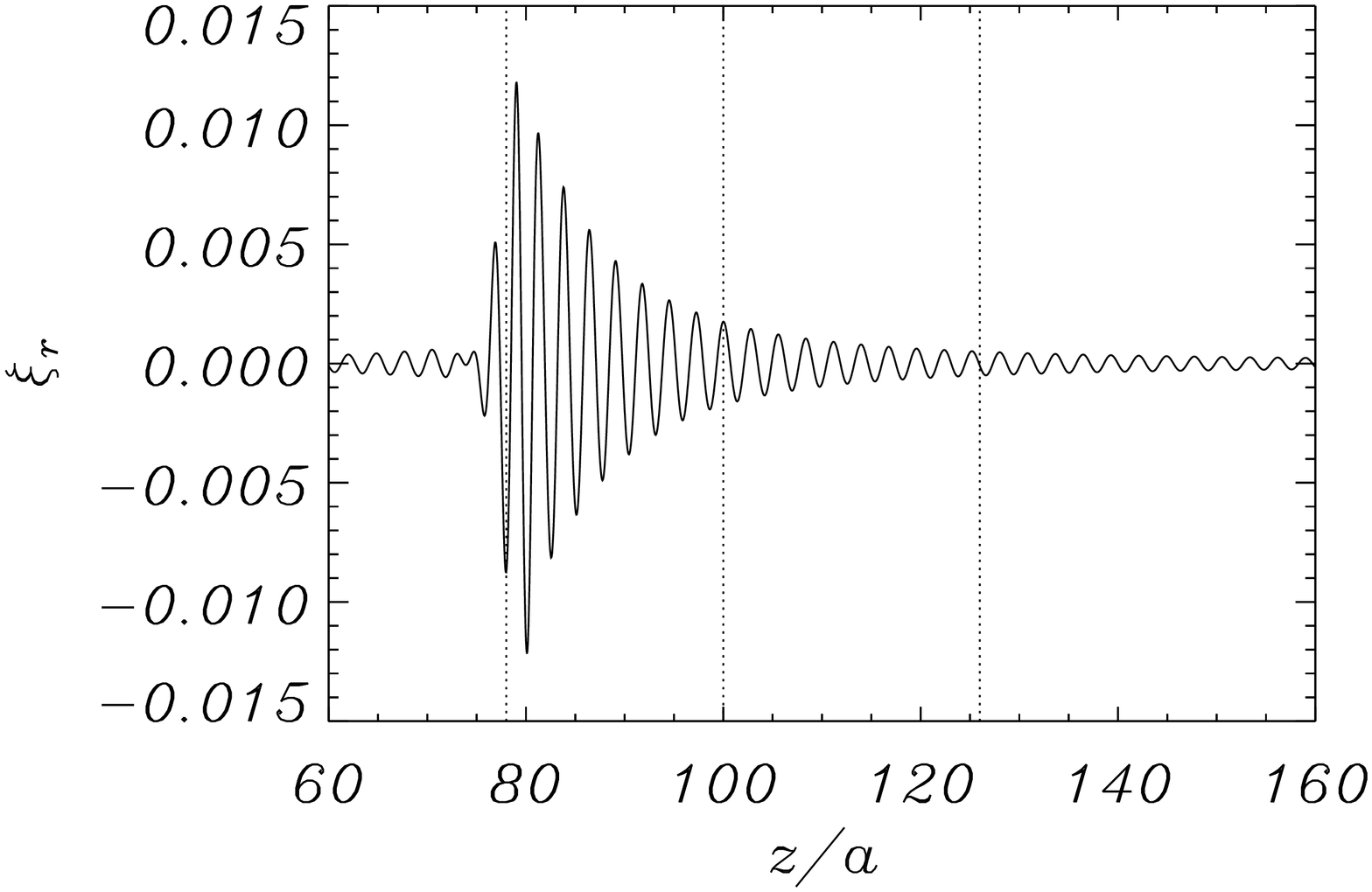} \\
  }
  \caption{(a) Radial displacement of the magnetic tube boundary as a function of $z$ for $t=100\tauAi$. (b) Contribution to the signal in panel (a) coming from all proper modes. (c) and (d) Contributions coming from the global kink mode and its first overtone, respectively. The vertical lines correspond to positions along the magnetic tube $z=0.78\vai t$, $z=\vai t$, and $z=1.26\vai t$ (see text). The density ratio is $\rhoi/\rhoe=4$ and $\Delta=a$.}
  \label{fig_vae2_delta1_t}
\end{figure}

Figure~\ref{fig_vae2_delta1_t} illustrates the effect of wave dispersion on the wave train at later time ($t=100\tauAi$). A comparison of Figure~\ref{fig_vae2_delta1_t}(a) with Figure~\ref{fig_vae2_delta1_short_t}(c) shows that the wave train has become very structured because of wave dispersion and that a considerable length of the magnetic tube has become affected by the initial disturbance: while the initial wave pulse has length $\simeq 4\Delta=4a$, at $t=100\tauAi$ the magnetic tube section in the range $100a\lesssim z\lesssim 125a$ displays transverse displacements, and so the wave train is about 6 times longer. Obviously, as the wave train disperses, its amplitude decreases and the amplitude of the radial motion in Figure~\ref{fig_vae2_delta1_t}(a) is much smaller than that of the initial perturbation. Figure~\ref{fig_vae2_delta1_t}(a) displays several distinct features: first, the maximum wave packet amplitude is attained at its central part, where long wavelengths are dominant. Second, the trailing part of the packet is made of a smaller amplitude subpacket with shorter wavelengths. And third, the leading part of the wave train displays small amplitude, small wavelength oscillations. The origin of these features is clear if the contributions of different modes are separated. Figure~\ref{fig_vae2_delta1_t}(b) shows the part of the signal that comes from proper modes. This signal is almost identical to that of panel (a), except for the leading oscillations, that therefore come from improper modes. We go one step further and plot separately the contributions of the global kink mode (Figure~\ref{fig_vae2_delta1_t}(c)) and its first overtone (Figure~\ref{fig_vae2_delta1_t}(d)). Now it is evident that the main part of the wave train is caused by the first of these two modes and that the trailing subpacket arises because of the first overtone. Let us first concentrate on the first of these two subpackets. Its properties can be fully comprehended with the help of Figures~\ref{fig_dr_over}(c) and \ref{fig_amplr}(a). The solid line of Figure~\ref{fig_dr_over}(c) tells us that the global kink mode has a group speed that monotonically decreases from $\ck\simeq 1.26\vai$ for $k=0$ to $\vai$ as $k\rightarrow\infty$, with a very shallow minimum at an intermediate wavenumber. Hence, in Figure~\ref{fig_vae2_delta1_t}(c) (see the two rightmost vertical lines) the leading part of the subpacket travels at a velocity $\simeq 1.26\vai$ and contains long wavelengths, whereas its trailing part travels at a speed $\simeq\vai$ and is composed of shorter wavelengths. While Figure~\ref{fig_dr_over}(c) explains the propagation speed of different wavenumbers, the solid line of Figure~\ref{fig_amplr}(a) explains their amplitude: wavenumbers around $ka=0.5$, travelling at the middle of the wave train, attain the maximum amplitude, while those with smaller and higher $k$ have smaller contribution and so the transverse displacement becomes smaller as one moves from the wave train center to its sides. We now turn our attention to the first overtone subpacket (Figure~\ref{fig_vae2_delta1_t}(d)), that displays its own distinctive features: its amplitude is some 10 times smaller than that of the global mode subpacket, it does not contain long wavelengths, and its main contribution travels at a speed around $0.75\vai$. All these properties are well explained by Figures~\ref{fig_dr_over}(c) and \ref{fig_amplr}(a). First, the first overtone has a wavenumber cut-off and it only exists for wavenumbers $k\geq k_{c1}\simeq 2.2/a$, so that its contribution to the radial displacement cannot contain long wavelengths. Second, the different amplitude of the subpackets in Figures~\ref{fig_vae2_delta1_t}(c) and (d) is a consequence of the amplitude of these two modes, that from Figure~\ref{fig_amplr}(a) differ by an order of magnitude more or less. And third, Figure~\ref{fig_dr_over}(c) indicates that the first overtone has a maximum group velocity $\simeq 0.98\vai$ and a minimum group velocity $\simeq 0.78\vai$ for $k\simeq 3.2/a$, and these two speeds very precisely limit the propagation speed of the first overtone subpacket (see the two leftmost vertical lines in Figure~\ref{fig_vae2_delta1_t}(d)).

\begin{figure}[ht!]
  \centerline{
    \scriptsize{(a)}
    \includegraphics[width=0.33\textwidth,angle=-90]{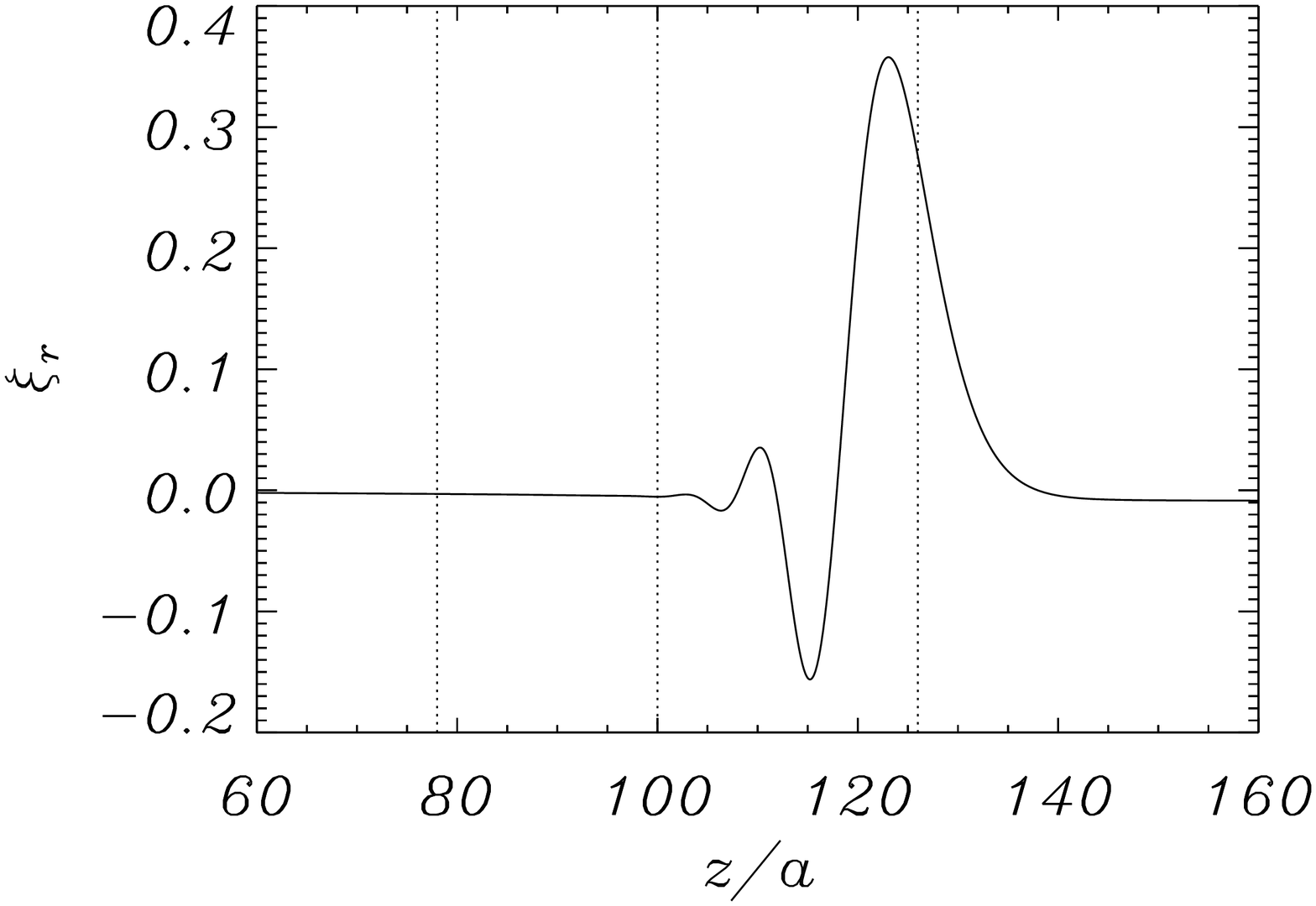}
    \scriptsize{(b)}
    \includegraphics[width=0.33\textwidth,angle=-90]{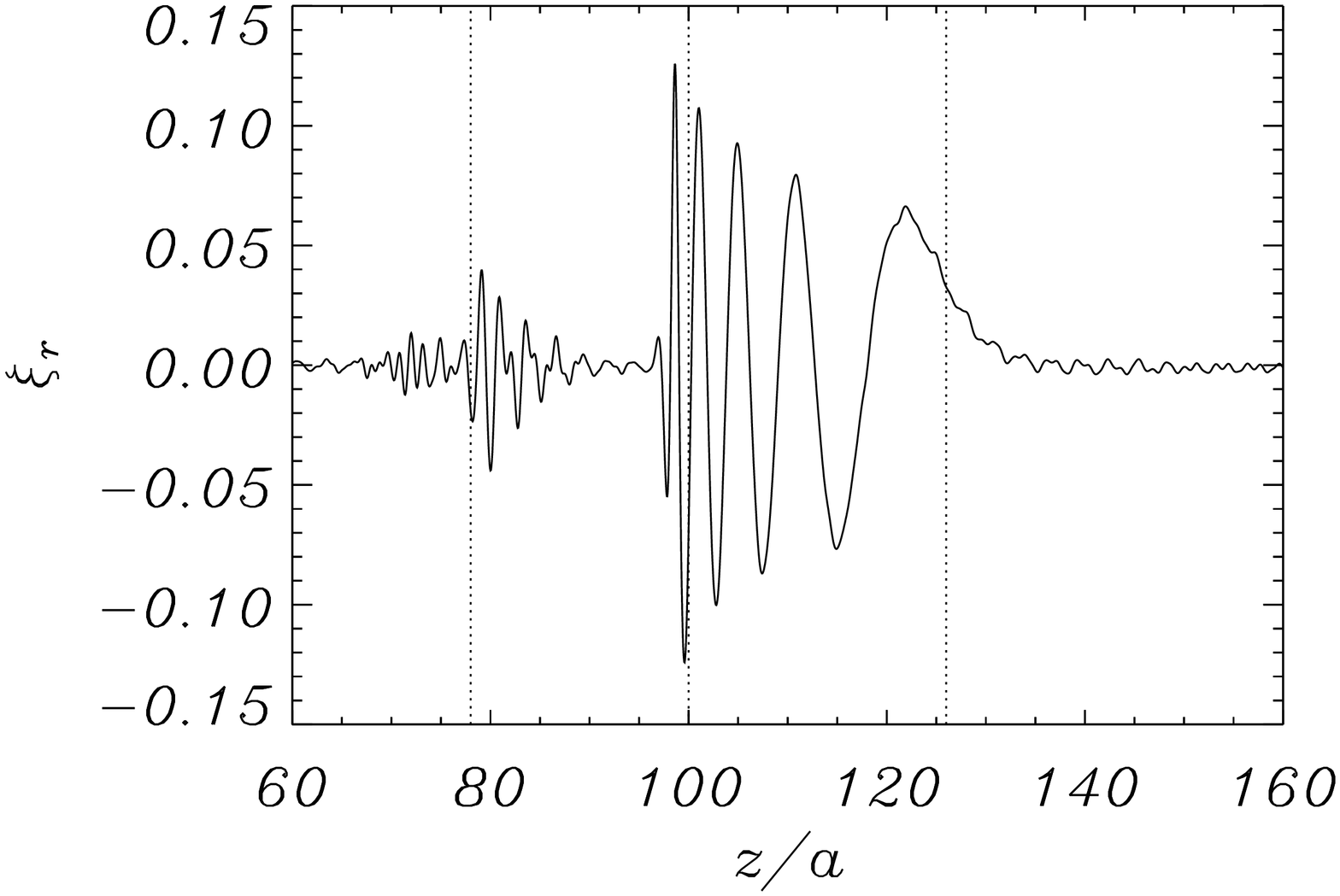}
  }
  \caption{Same as Figure~\ref{fig_vae2_delta1_t}(a) for an initial perturbation of different lengths: (a) $\Delta=4a$, (b) $\Delta=0.5a$.}
  \label{fig_vae2_delta_t}
\end{figure}

We now show that the length of the initial pulse strongly determines its dispersion. Figures~\ref{fig_vae2_delta1_short_t} and \ref{fig_vae2_delta1_t} have been obtained with $\Delta=a$. Increasing this parameter leads to a longer initial disturbance for which the Fourier transform of the initial conditions (functions $\tilde f$ and $\tilde g$ in Equation~(\ref{hat_f_xi_x0})) becomes more concentrated around small wavenumbers. This implies that, for the particular density ratio $\rhoi/\rhoe=4$ used so far, an increase in $\Delta$ can lead to the first overtone not being excited by the initial perturbation. This has been shown to be the case for $\Delta=4a$ (see Figure~\ref{fig_amplr}(b)). Figure~\ref{fig_vae2_delta_t}(a) presents the wave train for $t=100\tauAi$ and for an initial perturbation with $\Delta=4a$. Thus, except for the value of $\Delta$, all parameters are those of Figure~\ref{fig_vae2_delta1_t}. The wave packet for $\Delta=4a$ shows no signs of the trailing oscillations (caused by the slower propagation of the first overtone) and the leading oscillations (ascribed to improper modes) that were present in Figure~\ref{fig_vae2_delta1_t}(a). In fact, by plotting separately these two contributions we have verified that they are negligible, so that the wave train at $t=100\tauAi$ only contains discernible contributions from the global kink mode. There are some similarities between the global kink mode contribution for $\Delta=a$ and the wave train for $\Delta=4a$ (Figures~\ref{fig_vae2_delta1_t}(c) and \ref{fig_vae2_delta_t}(a)), namely the presence of long wavelengths at the front of the wave train and shorter wavelengths at its back, and the propagation of the wave train front at a speed $\ck$. But there are some important differences too. The wave train in Figure~\ref{fig_vae2_delta_t}(a) lacks the short wavelengths of Figure~\ref{fig_vae2_delta1_t}(c), the reason being that short wavelengths have not been excited because their amplitude is negligible (see Figure~\ref{fig_amplr}(b)). And since these missing wavelengths are the ones that propagate at a smaller velocity, the wave packet is shorter compared to that of Figure~\ref{fig_vae2_delta1_t}(c) and undergoes less dispersion, which in turn implies that it contains less extrema and has a larger amplitude, by a factor $\simeq 3$. On the other hand, a shorter initial displacement with $\Delta=0.5a$ imparts energy to the global kink mode and its first two overtones, whose signature is present in the wave train of Figure~\ref{fig_vae2_delta_t}(b). The global kink mode and its first overtone are responsible for the oscillations between the two rightmost and the two leftmost vertical lines, such as previously found in Figure~\ref{fig_vae2_delta1_t}. Note, however, that the relative amplitude of the first overtone to that of the global kink mode increases when $\Delta/a$ is decreased. In addition, the second overtone is not excited for $\Delta=a$, but in Figure~\ref{fig_vae2_delta_t}(b) it causes the transverse magnetic tube displacement that can be seen for $70\lesssim z/a\lesssim 80$.

\begin{figure}[ht!]
  \centerline{
    \scriptsize{(a)}
    \includegraphics[width=0.33\textwidth,angle=-90]{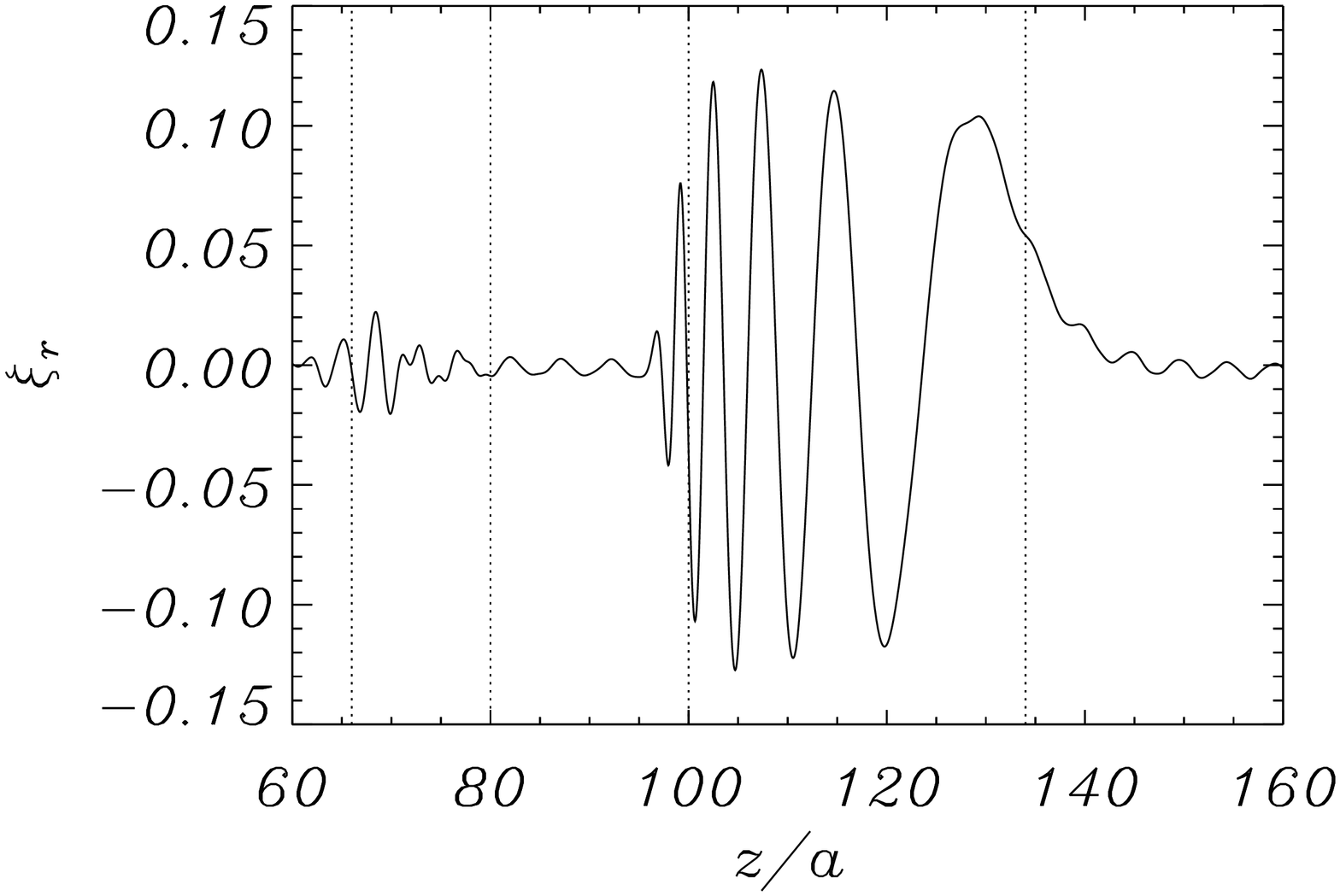}
    \scriptsize{(b)}
    \includegraphics[width=0.33\textwidth,angle=-90]{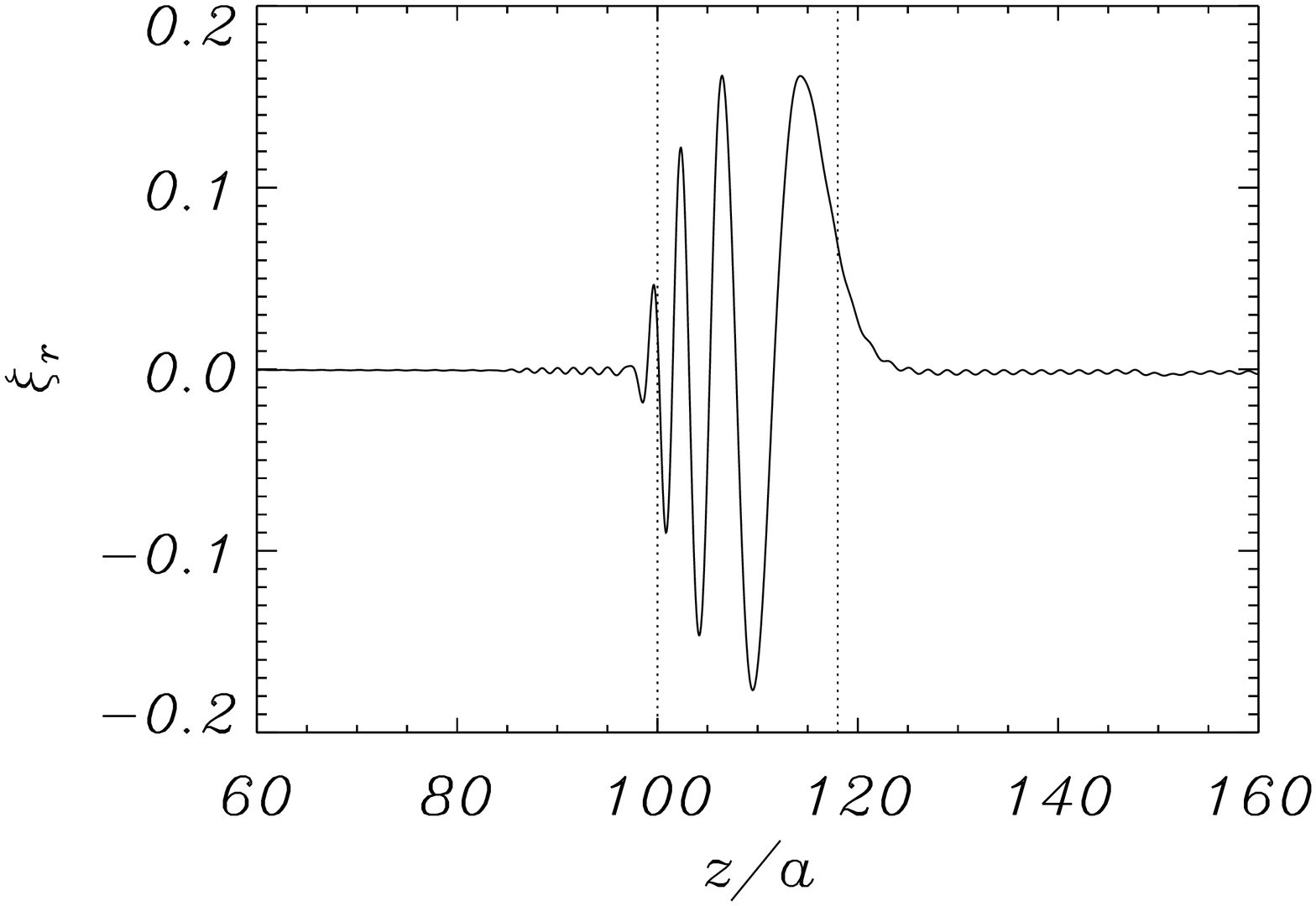}
  }
  \caption{Same as Figure~\ref{fig_vae2_delta1_t}(a) for different density ratios: (a) $\rhoi/\rhoe=9$; here vertical lines correspond to positions along the magnetic tube $z=0.66\vai t$, $z=0.8\vai t$, $z=\vai t$, and $z=1.34\vai t$. (b) $\rhoi/\rhoe=2.25$; here vertical lines correspond to positions along the magnetic tube $z=\vai t$ and $z=1.18\vai t$.}
  \label{fig_vae_delta1_t}
\end{figure}

Finally, the relevance of the density ratio on the dispersive spreading of the initial perturbation is analysed. Figure~\ref{fig_vae_delta1_t}(a) shows the magnetic tube radial displacement for $\rhoi/\rhoe=9$ and for the range of positions of Figure~\ref{fig_vae2_delta1_t}(a), so that a comparison of the two figures can be done. The main part of the wave train, i.e. that in the range $90\leq z/a\leq 140$, corresponds to the global kink mode, that propagates with group speeds between $\simeq\vai$ and $\cg\simeq 1.34\vai$ (see two rightmost dotted lines of Figure~\ref{fig_vae_delta1_t}(a)). Since the kink speed increases with the density ratio, the contribution of the global kink mode affects a slightly wider range of $z$ in Figure~\ref{fig_vae_delta1_t}(a) than in Figure~\ref{fig_vae2_delta1_t}(a). Regarding the kink mode overtones, for $\rhoi/\rhoe=9$ the first one has $0.66\vai\leq\cg\leq 0.80\vai$ and so its contribution lies roughly between the two leftmost vertical lines in Figure~\ref{fig_vae_delta1_t}(a); higher overtones have even smaller amplitude and so their contribution is negligible. Thus, depending on the value of $\rhoi/\rhoe$ the global kink mode and its first overtones can contribute to the wave train with packets separated by some space (as in Figure~\ref{fig_vae_delta1_t}(a)) or with packets that have almost no space between them (as in Figure~\ref{fig_vae2_delta1_t}(a)). While in Figure~\ref{fig_vae_delta1_t}(a) the density ratio is larger than the one used in Figure~\ref{fig_vae2_delta1_t}, in Figure~\ref{fig_vae_delta1_t}(b) it is smaller. Now the wave train is almost free from short waves and is more spatially concentrated. The first issue is caused by the dependence of the wavenumber cut-offs with the density ratio: for $\rhoi/\rhoe=2.25$ we have $k_{c1}a\simeq 3.5$ and this implies that the first overtone receives little energy from the initial lateral displacement. Thus, the wave train of Figure~\ref{fig_vae_delta1_t}(b) is dominated by the global kink mode, whose range of propagation speeds decreases for smaller values of the density ratio and for this reason the wave packet length is smaller than that of Figures~\ref{fig_vae2_delta1_t}(a) and \ref{fig_vae_delta1_t}(a).

\section{APPLICATION TO CORONAL LOOP OSCILLATIONS}\label{sect_loop_application}

Assuming that a coronal loop suffers a sudden lateral displacement analogous to that of Figure~\ref{fig_cylinder_kink}(b), a wave train will then propagate along the tube and will be dispersed as it travels. To apply our results to coronal loops, we consider a loop radius $a=250$~km and a loop Alfv\'en velocity $\vai=500$~km~s$^{-1}$. Then, the internal travel time is $\tauAi=0.5$~s. Regarding the initial disturbance, it has a Gaussian shape along the loop (Equation~(\ref{f_xi_x0})) and we impose an amplitude $\xi_0=a/10=25$~km and a length given by $\Delta=a$ (recall that the initial transverse displacement has a length $\simeq 4\Delta$ along the loop). The initial disturbance is, therefore, twice longer than that depicted in Figure~\ref{fig_cylinder_kink}(b). With these parameter values, Figure~\ref{fig_vae2_delta1_t} gives the shape of the loop boundary after $t=50$~s have elapsed from the initial transverse displacement and covers the range of distances 15~Mm~$\leq z\leq$~40~Mm along the tube from the position of the initial disturbance. For the adopted value of $\xi_0$, the vertical scale in this figure transforms into maximum transverse displacements of the order of $3.5$~km. This is a very small figure, so one may wonder whether the events studied in this work can be detected with present day instruments.

\subsection{Perturbed Velocity}\label{sect_velocity}

\begin{figure}[ht!]
  \centerline{
    \scriptsize{(a)}
    \includegraphics[width=0.33\textwidth,angle=-90]{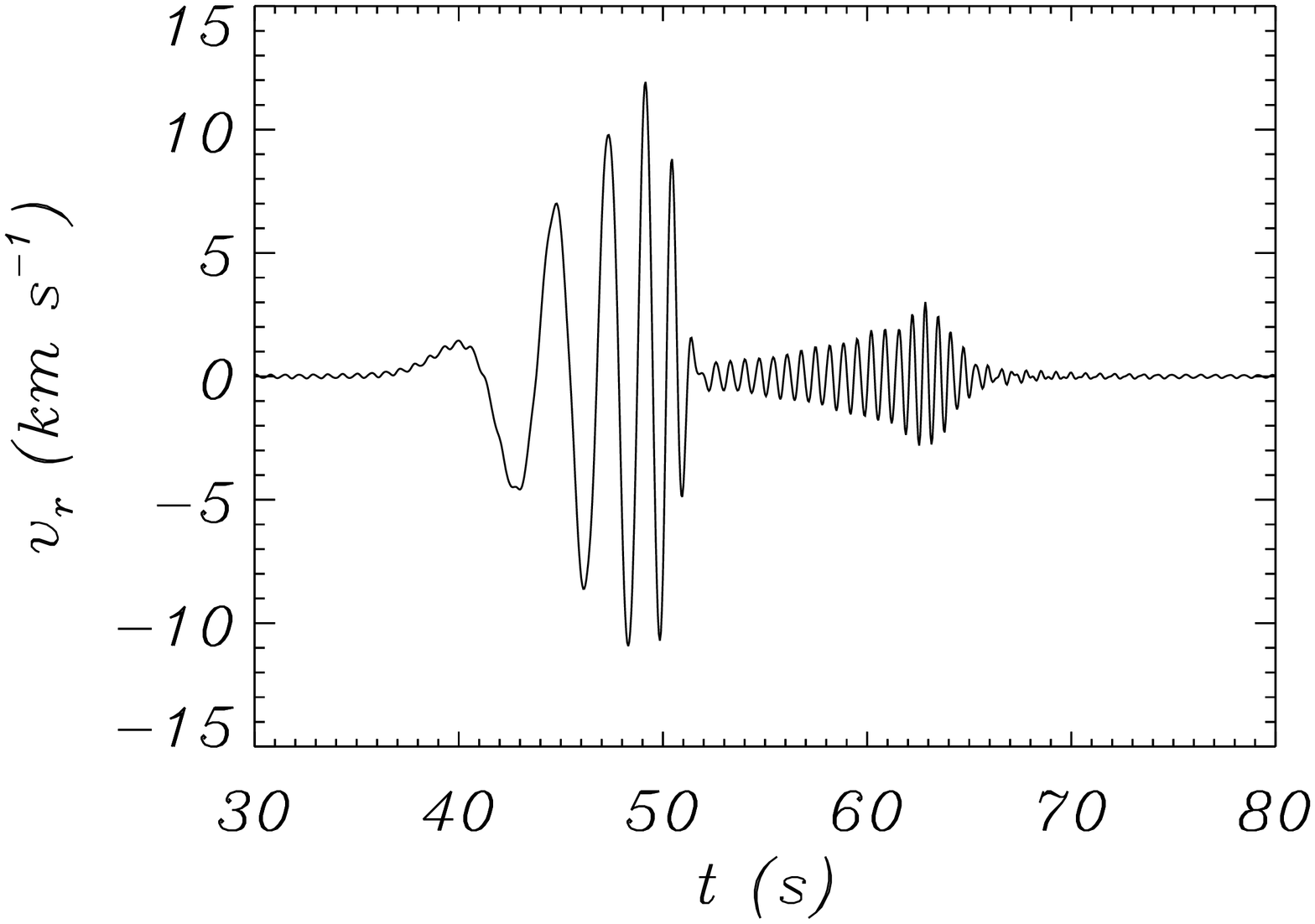} \\
    \scriptsize{(b)}
    \includegraphics[width=0.33\textwidth,angle=-90]{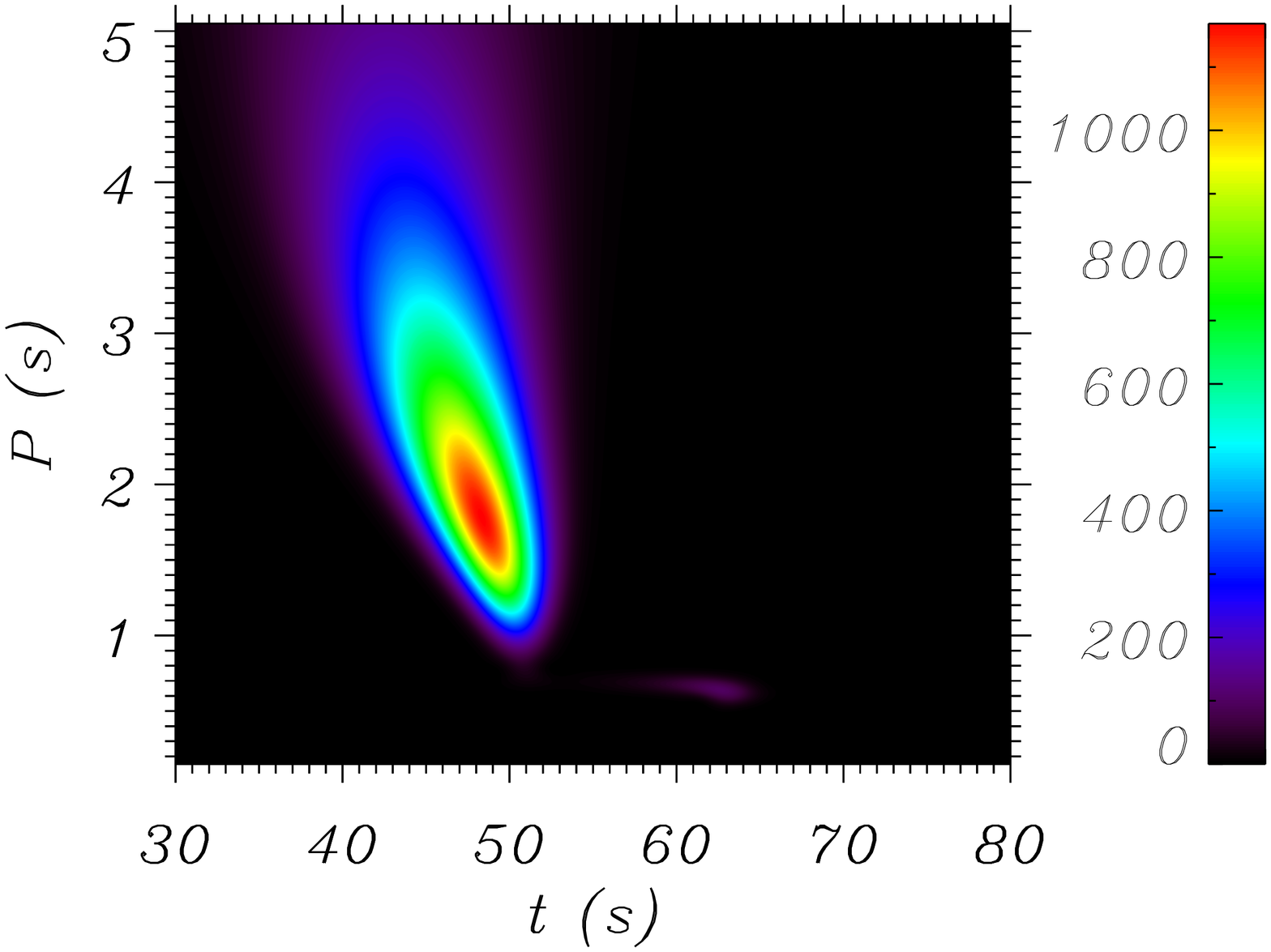} \\
  }
  \centerline{
    \scriptsize{(c)}
    \includegraphics[width=0.33\textwidth,angle=-90]{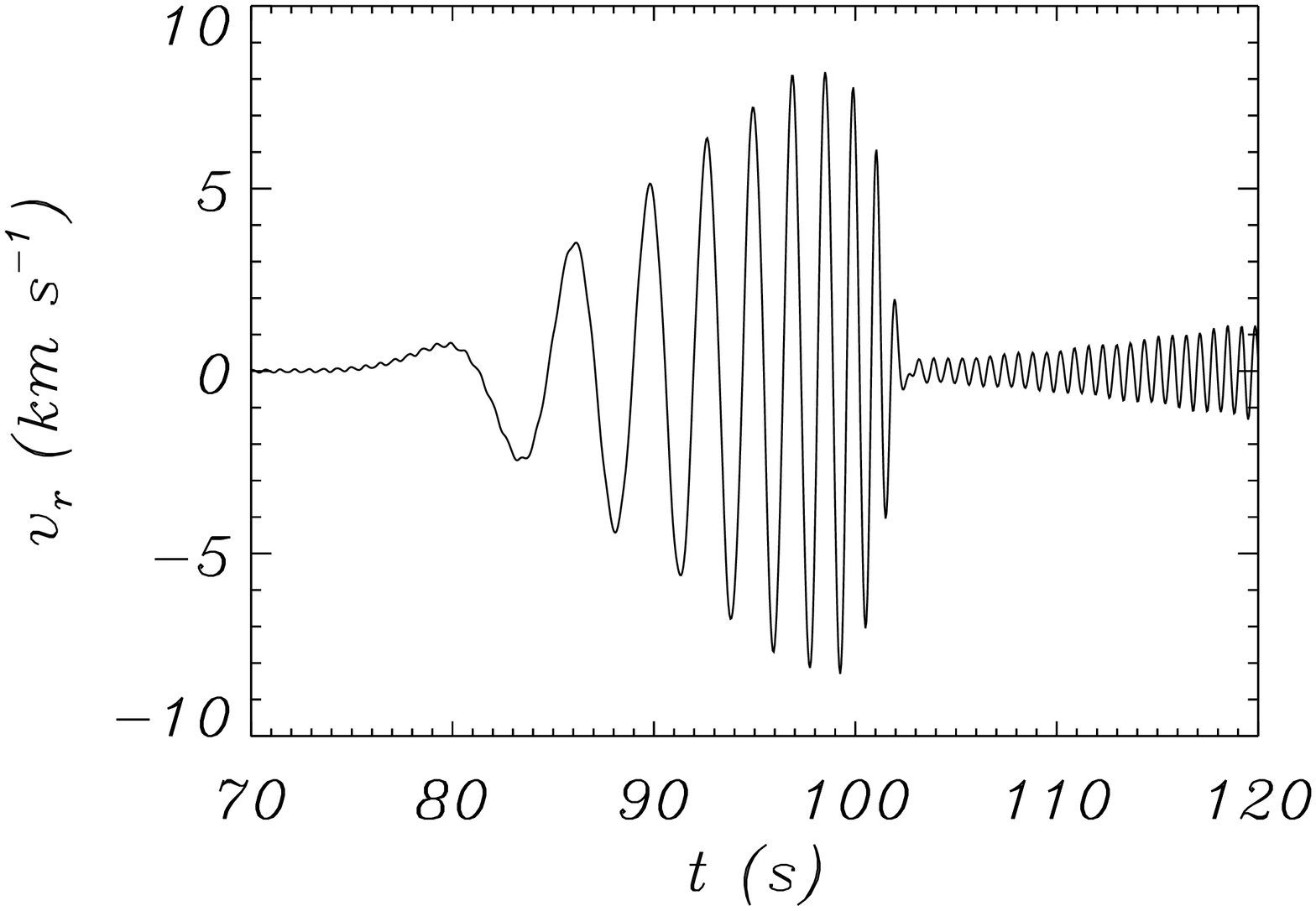} \\
    \scriptsize{(d)}
    \includegraphics[width=0.33\textwidth,angle=-90]{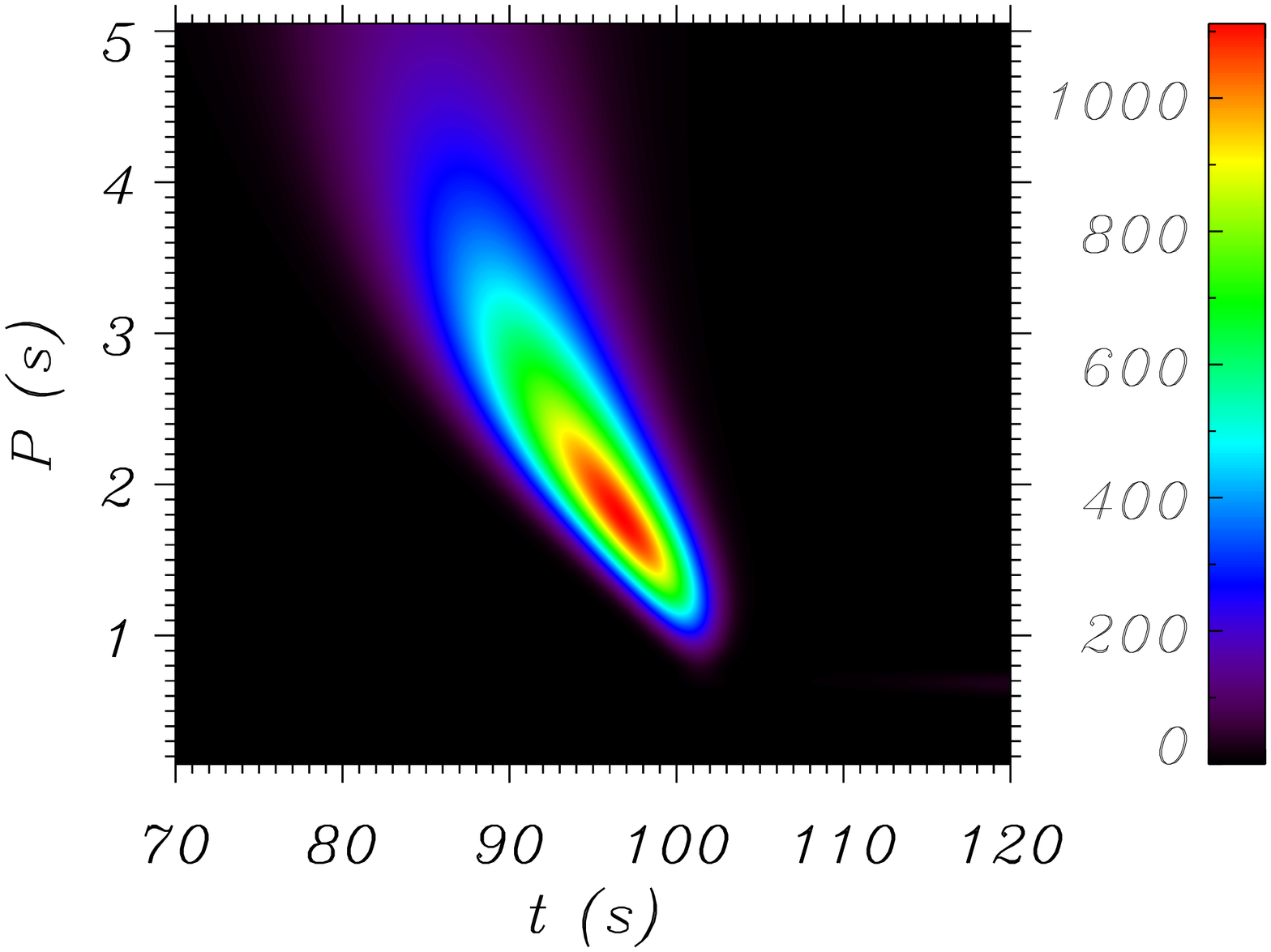} \\
  }
  \caption{Radial velocity of the loop boundary as a function of time at a distance (a) $25$~Mm and (c) $50$~Mm from the position of the initial disturbance. (b) and (d) Wavelet diagrams of the signals in panels (a) and (c), respectively. The density ratio is $\rhoi/\rhoe=4$, the tube radius and Alfv\'en speed are 250~km and 500~km~s$^{-1}$, and the initial disturbance has a length of the order of 1~Mm ($\Delta=a$ in Equation~(\ref{init_xi_x})). In this figure only the contribution of proper modes has been taken into account.}
  \label{fig_vae2_delta1_z}
\end{figure}

To address this question, we consider the radial velocity, $v_r$, on the loop surface at a fixed position along the loop. Temporal variations of this quantity can in principle be observed as Doppler shift variations if the line-of-sight is contained in the polarization plane of the oscillations or makes a small angle with it. The radial velocity for the density ratio $\rhoi/\rhoe=4$ is presented in Figure~\ref{fig_vae2_delta1_z}(a), where the detection point is located 25~Mm along the loop from the initial perturbation site. The temporal variation of $v_r$ is easy to understand from Figure~\ref{fig_vae2_delta1_t}: the contribution of the global kink mode arrives first to the detection point (at $t\simeq 40$~s) and this is followed by the contribution from the first overtone (starting after $t=50$~s). The global kink mode subpacket contains a range of wavenumbers, with long wavelengths travelling faster than shorter ones, and so the former arrive before the later. We note that each of these two subpackets only lasts about 10~s, so we conclude that the whole event has very short duration. The wavelet diagram of $v_r$ is shown in Figure~\ref{fig_vae2_delta1_z}(b). Most of the power in this plot originates from the global kink mode and has periods between 1 and 4~s, with long periods being the first to arrive at the detection point, as described before. The first overtone also leaves its contribution in the wavelet diagram, in the form of a small power hump around $t=60$~s with period smaller than 1~s. Having a detection point at a larger distance (50~Mm) from the wave packet source implies that the wave train has more time to disperse because of the different group speeds of its constituent wavelengths (see Figure~\ref{fig_vae2_delta1_z}(c)). Therefore, the wave train takes longer to transit at the detection point and the large power blob in the wavelet diagram is more inclined (Figure~\ref{fig_vae2_delta1_z}(d)). Nevertheless, the wave train contains the same wavenumbers (and so the same periods) and this means that the power level at a fixed period does not change when moving the detection point along the loop.

In the numerical results of Figure~\ref{fig_vae2_delta1_z} only the global kink mode has significant power. From the results of Figures~\ref{fig_vae2_delta_t}(b) and \ref{fig_vae_delta1_t}(a) we know that the global kink mode becomes more dispersive if shorter initial disturbances or larger density ratios are considered. It is worth mentioning again that the signature of higher wave dispersion is a more inclined power blob in the wavelet diagram of the line-of-sight velocity. Moreover, kink mode overtones can be excited by shorter initial disturbances and, to a lesser extent, by larger density ratios. In particular, for the values $\Delta=0.5a$ and $\rhoi/\rhoe=4$ the first overtone leaves a clear signature in the wavelet spectrum (similar to that present in Figure~\ref{fig_vae2_delta1_z}(b) at $t\simeq 60$~s) with power half that of the global kink mode.

At this point, we test the consistency of these results. Figure~\ref{fig_amplr}(a) points out that, when the global kink mode is excited, then a train of waves with $0\leq k\lesssim 4a$ will be generated. Using the shortest wavelength in this range and the approximation $\omega \approx k\vai$ (cf. Figure~\ref{fig_dr_over}(a)), we get the period $P\simeq 1$~s; longer wavelengths result in larger periods. All this is in good agreement with Figures~\ref{fig_vae2_delta1_z}(b) and (d). A second property that can be tested is the amplitude of the $v_r$ oscillations. Such as described above, the maximum radial displacement at $t=50$~s is close to $3.5$~km. Let us assume that this maximum displacement corresponds to the wavenumber with maximum amplitude, namely $ka\simeq 0.4$; see Figure~\ref{fig_amplr}(a). We further assume that the maximum oscillatory displacement and speed follow the relation of a harmonic oscillator, i.e. $\max(v_r)=\omega\max (\xi_r)$. Using the approximation $\omega=k\vai$, we obtain $\max (v_r)\simeq 0.056\vai=28$~km~s$^{-1}$. This is consistent with the velocity amplitudes of Figures~\ref{fig_vae2_delta1_z}(a) and (c).

\subsection{Perturbed Density}\label{sect_density}

\begin{figure}[ht!]
  \centerline{
    \scriptsize{(a)}
    \includegraphics[width=0.33\textwidth,angle=-90]{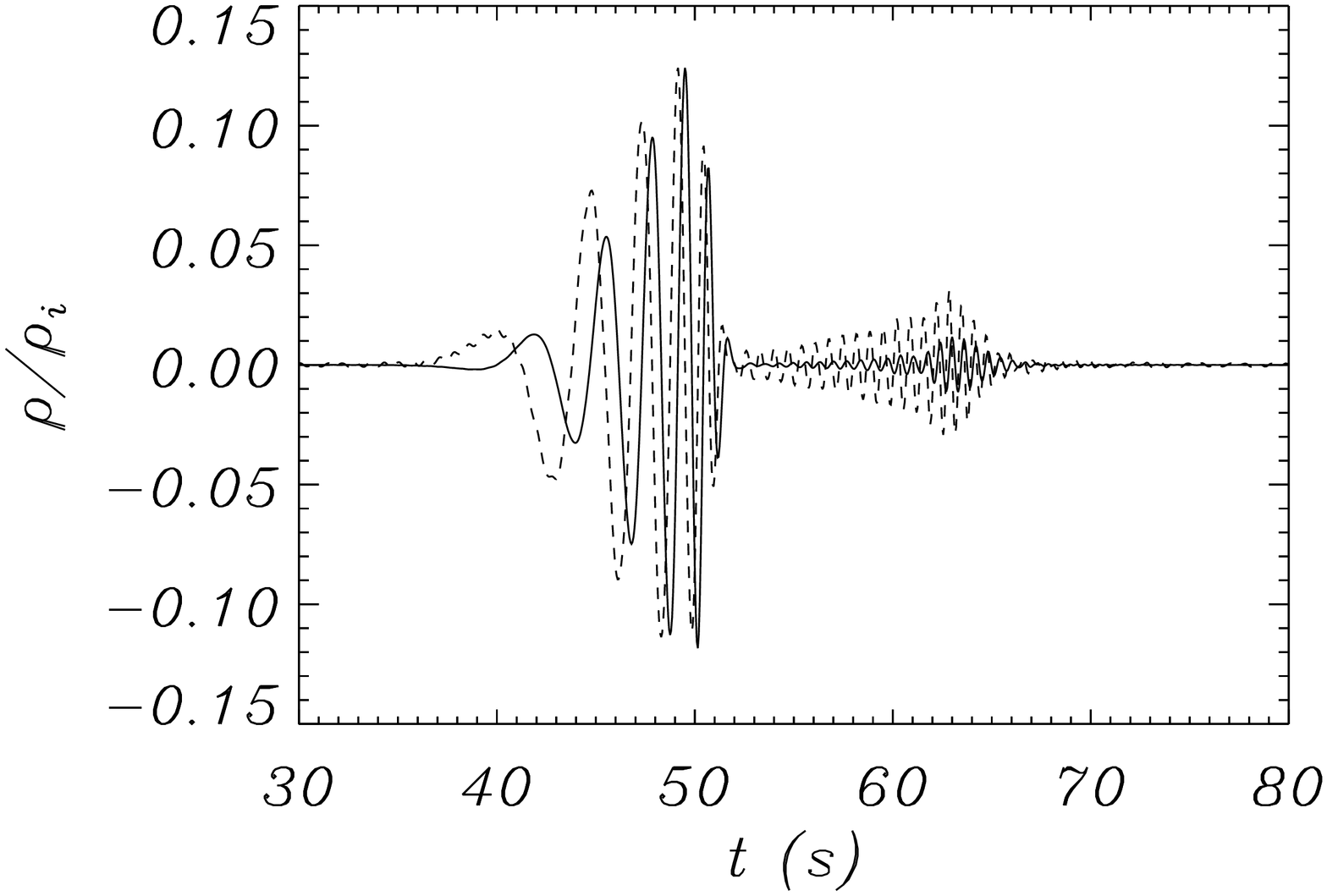} \\
    \scriptsize{(b)}
    \includegraphics[width=0.33\textwidth,angle=-90]{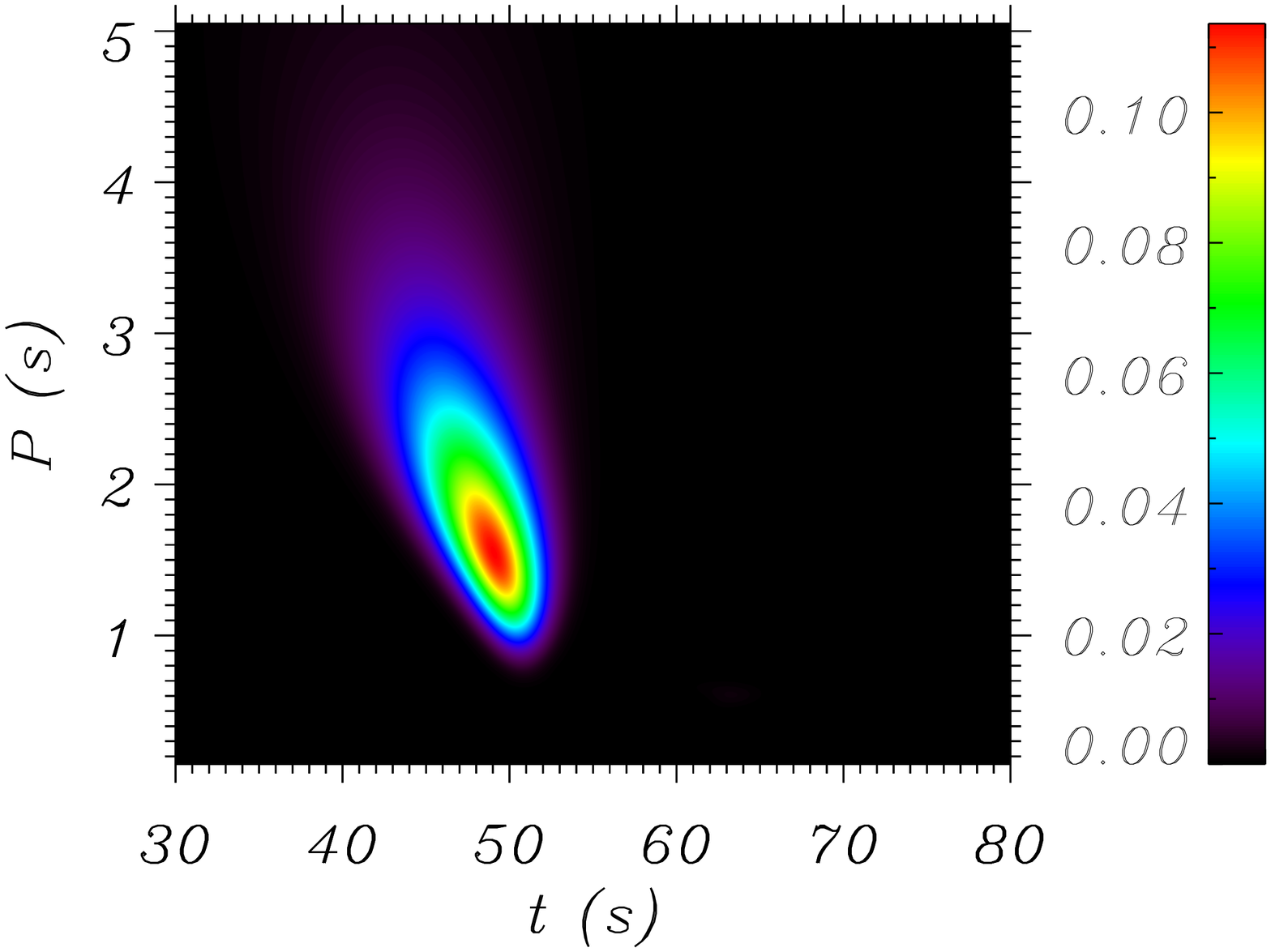} \\
  }
  \caption{(a) The solid line shows the ratio of the perturbed density to the loop density at the loop boundary as a function of time at a distance $25$~Mm from the position of the initial perturbation.  The dashed line is the radial velocity of Figure~\ref{fig_vae2_delta1_z}(a) rescaled by an appropriate amount. (b) Wavelet diagram of the solid line in panel (a). All parameter values are those of Figure~\ref{fig_vae2_delta1_z}.}
  \label{fig_vae2_delta1_rho_z}
\end{figure}

The global kink mode is responsible for the most important part of the propagating wave train. For very long wavelengths, this mode is characterized by transverse motions of the magnetic tube with very small density variations. For the shorter wavelengths that are excited by a localized disturbance, however, the global kink mode produces periodic variations of the loop radius and so the transverse motions are accompanied by appreciable density changes. Here we consider, as an example, the density associated to the radial velocity variations of Figure~\ref{fig_vae2_delta1_z}(a). The ratio of perturbed to unperturbed density is plotted in Figure~\ref{fig_vae2_delta1_rho_z}(a) together with the radial velocity itself, rescaled to fit in the same vertical scale. The density changes are quite appreciable and reach 10--15\% the equilibrium loop density. In addition, both subpackets of Figure~\ref{fig_vae2_delta1_z}(a) are present in Figure~\ref{fig_vae2_delta1_rho_z}(a) and from the variations they cause we conclude that the global kink mode produces more important density changes at the loop boundary than its first overtone. In fact, although the later oscillations can be discerned in the signal of Figure~\ref{fig_vae2_delta1_rho_z}(a), their power is so small that they are not visible in the wavelet diagram of Figure~\ref{fig_vae2_delta1_rho_z}(b), where the power of global kink mode oscillations is dominant. One can also see that the velocity and density oscillations are slightly out of phase, with the changes of $v_r$ preceding those of $\rho$ by about 1~s. On the other hand, the two signals have the same distribution of power versus time (compare the wavelet diagrams of Figures~\ref{fig_vae2_delta1_z}(b) and \ref{fig_vae2_delta1_rho_z}(b)).

\section{DISCUSSION}\label{sect_conclusions}

We have analyzed the propagation of impulsively excited, linear fast kink wave packets in cylindrical magnetic tubes. The method we use is based in decomposing the initial perturbation in a sum of eigenmodes. We have imposed a concentrated transverse disturbance of the magnetic tube and have found that most of its energy is imparted to the global kink mode. The concentrated transverse impulse is the sum of many different global fast kink modes with their own amplitude and that travel at their own group velocity in the packet. For this reason, the wave train dispersion is determined by two key ingredients: on one hand, the dependence of the group velocity on the longitudinal wavenumber, that is represented in Figures~\ref{fig_dr_over}(c) and \ref{fig_cg_vae}. On the other hand, the amplitude of the global kink mode as a function of the longitudinal wavenumber (Figure~\ref{fig_amplr}). We have found that large internal to external density ratios and especially shorter initial perturbations enhance wave dispersion.

The signature of a wave train at a fixed point in a coronal loop has been investigated using the radial velocity and the density perturbation as measurable quantities. We have shown that both variables display very fast oscillations with periods varying between 1 and 4~s and with detectable amplitudes. These values have been obtained for a loop radius $a=250$~km and an internal Alfv\'en speed $\vai=500$~km~s$^{-1}$, although these are just illustrative values. If the loop radius is doubled or its internal Alfv\'en speed is halved, then the obtained periods and time scales are doubled. That is, both the horizontal and vertical scales in Figures~\ref{fig_vae2_delta1_z}(b), (d) and \ref{fig_vae2_delta1_rho_z}(b) are doubled. In view of the diversity of both $a$ and $\vai$ in coronal loops, we conclude that the dispersion of a transverse coronal loop perturbation can be detected as periodic variations of the velocity and density with periods ranging from a fraction of a second to tens of seconds.

\citet{murawski1998} performed a numerical simulation of impulsively generated, small-amplitude waves in a slab coronal loop model. Regarding transverse oscillations, these authors concluded that they could be detected as changes in the loop position through difference images of time series taken with a cadence of seconds or better. Moreover, \citet{selwa2004} solved the same problem for a cylindrical loop geometry and obtained that time scales of the fast kink wave are of the order of a dozen of seconds, in agreement with the results presented here. \citet{selwa2004} did not describe the dispersive properties of localized transverse loop disturbances and, for a perturbation centered in the loop axis, did not obtain large density variations. This contradicts our conclusions (see Figure~\ref{fig_vae2_delta1_rho_z}), probably because the initial disturbance of \citet{selwa2004} has a very small amplitude, of the order of 0.1~km. Impulsively generated transverse waves in coronal loops modeled as straight or curved slabs \citep{selwa2006,selwa2007,rial2013} or as curved cylinders \citep{selwa2011a,selwa2011b} have also been investigated, although the main aim of these studies is the analysis of transverse oscillations in line-tied loops, rather than the propagation of localized disturbances.

A localized axisymmetric compression (rather than the localized transverse excitation investigated here) is the sum of fast sausage eigenmodes, which in a uniform cylinder are highly dispersive. Studies of propagating fast sausage wave trains have been performed for example by \citep{roberts1984}. These authors showed that an impulsive axisymmetric excitation of a magnetic cylinder generates a clear time signature when the wave packet arrives at a point along the tube away from the initial disturbance. This time signature consists of a periodic phase, followed by a quasi-periodic phase, and a decaying phase \citep[see also][]{murawski1993b,murawski1993d,nakariakov2004,nakariakov2009}. This complex temporal behavior is produced by the dispersive character of the fast sausage mode, that results in different frequencies arriving at different times at the detection position. In the study of \citet{nakariakov2004}, the detected density perturbation contains a continuous variation of the period and its wavelet diagram shows large power distributed between 3 and 10~s, with a total duration of the oscillatory event of about 20~s. In fact, the wavelet spectra of Figure~\ref{fig_vae2_delta1_z}(d) and Figures~3 to 5 of \citet{nakariakov2004} are remarkably similar in spite of the different nature of the propagating disturbances. Fast sausage wave trains propagating along coronal loops have been invoked by \citet{nakariakov2004,nakariakov2005} as the cause of the compressible waves analyzed by \citet{williams2001,williams2002} and \citet{katsiyannis2003}. These waves have also been put forward as the cause of the ``wavelet tadpoles'' found in solar decimetric radio bursts \citep{meszarosova2009,meszarosova2011,karlicky2013}. But in view of the similar density perturbations produced by a transverse excitation and an axisymmetric compression, these conclusions may need additional work to be confirmed or refuted.

Our method for solving Equations~(\ref{eq_mass})--(\ref{eq_induc}) in a magnetic cylinder has no underlying assumptions, except that the magnetic tube and the environment are uniform and that the internal to external density ratio is larger than one. It has several advantages over the direct numerical solution of Equations~(\ref{eq_mass})--(\ref{eq_induc}): the numerical integration of this set of partial differential equations requires using a two-dimensional mesh of points in the $r$- and $z$-directions. Moreover, to obtain the solution at a given time, $t$, the temporal evolution of the five perturbed variables $\xi_r$, $\xi_\varphi$, $b_r$, $b_\varphi$, and $P$ must be computed from $t=0$ to $t$. To make things worse, to produce the results of Figure~\ref{fig_vae2_delta1_t}, for example, one needs to consider, at least, the large range of distances $0\leq z\leq 160a$, with $a$ the magnetic tube radius. The method we use is a numerical one and so it also suffers from numerical inaccuracy. It is nevertheless optimised to study wave propagation along a uniform magnetic tube. Imposing the initial conditions allows to obtain the amplitudes of the eigenmodes ($A_j^\pm(k)$ from Equation~(\ref{A_pm-main}) and $A_\omega^\pm(k)$ from Equation~(\ref{A_pm_om-main})). Once these amplitudes are known, any perturbed variable can be computed at any position and time with Equation~(\ref{Four_trans_inv-main}). A similar approach was used by \citet{terradas2007}, based on the work of \citet{ruderman2006}, to estimate the energy deposited in the normal modes given an external perturbation located in the solar corona.

The method can be extended to study the propagation of, for example, an axisymmetric compression of the magnetic tube. Such a disturbance can be expressed as the sum of fast sausage eigenmodes. This study will be almost a complete repetition of the analysis in this paper. The only difference is that, in the case of fast sausage waves, there is a cut-off wavenumber even for the mode fundamental in the radial direction. This implies that we can expect more substantial contribution from improper eigenmodes.

In this paper we have considered a magnetic tube with a sharp boundary. If, instead, we consider a tube with the density continuously varying from its value inside the tube to a lower value outside, then there will be wave damping due to resonant absorption. Even in this case the initial value problem can be solved using the same method as the one used in this paper. The only difference is that, in the case of a tube inhomogeneous in the radial direction, it is usually not possible to calculate the proper and improper eigenfunctions analytically, so they have to be calculated numerically.

\section*{Acknowledgements}

R.O. and J.T. acknowledge the support from the Spanish MINECO and FEDER funds through project AYA201-22846 and from CAIB through the ``Grups Competitius'' program and FEDER funds. J.T. acknowledges support from MINECO through a Ram\'on y Cajal grant. R.O. also thanks R. Soler and J.L. Ballester for useful discussions. Wavelet software was provided by C. Torrence and G. Compo, and is available at URL: http://atoc.colorado.edu/research/wavelets/. R.O. also thanks D. W. Fanning for making available the Coyote Library of IDL programs (http://www.idlcoyote.com/). A part of this work was carried out when M.S.R. was a guest of Departament de F\'isica of Universitat de les Illes Balears. He acknowledges the financial support received from the Universitat de les Illes Balears and the warm hospitality of the Department. He also acknowledges the support by the STFC grant.

\appendix

\section{General Expressions for Traveling Disturbances}\label{app-fourier}

Here the method for obtaining solutions to Equations~(\ref{eq_xi_r})--(\ref{init_cond}) is presented in detail. Since the initial conditions~(\ref{init_cond}) are given in terms of the radial and azimuthal components and their time derivative, solutions for $\xi_r(t,r,z)$ and $\xi_\varphi(t,r,z)$ are sought. The obtained expressions are written in a more compact form if the displacement vector $\boldsymbol{\xi}(t,r,z)=(\xi_r,\xi_\varphi)$ is used. We introduce the Fourier transform of this vector with respect to $z$ defined by

\begin{equation}\label{Four_trans}
\tilde{\boldsymbol{\xi}}(t,r,k) = \int_{-\infty}^\infty
   \boldsymbol{\xi}(t,r,z) e^{-ikz}\,dz,
\end{equation}

\noindent
with inverse transform,

\begin{equation}\label{Four_trans_inv}
\boldsymbol{\xi}(t,r,z) = \frac1{2\pi}\int_{-\infty}^\infty
   \tilde{\boldsymbol{\xi}}(t,r,k) e^{ikz}\,dk.
\end{equation}

\noindent
Substituting the expression for $P$ in terms of $\boldsymbol{\xi}$ given by Equations~(\ref{eq_press}) in Equations~(\ref{eq_xi_r}) and (\ref{eq_xi_phi}), we obtain a system of two equations for the two components of $\boldsymbol{\xi}$\/, which can be also considered as one vector equation for vector $\boldsymbol{\xi}$\/. We apply the Fourier transform to this equation and obtain an equation for $\tilde{\boldsymbol{\xi}}$\/. The solution to this system has to be subject to two boundary conditions at $r = a$\/: one is the continuity of $\xi_r$\/, and the second one is obtained using the continuity of $P$\/. 

We define the Hilbert space of two-dimensional vector-functions on the interval $[0,\infty)$ by the scalar product

\begin{equation}\label{scalar_prod}
\langle\boldsymbol{\eta},\boldsymbol{\zeta}\rangle = \int_0^\infty \rho_0(r) 
   (\boldsymbol{\eta}\cdot\boldsymbol{\zeta}^*) r\,dr .
\end{equation}

\noindent
where the asterisk indicates complex conjugate. Recall that $\rho_0(r)$ is a piecewise-constant function equal to $\rhoi$ for $0<r<a$ and $\rhoe$ for $r > a$\/. The norm $\|\boldsymbol{\eta}\|$ of a vector-function $\boldsymbol{\eta}$ is defined by $\|\boldsymbol{\eta}\|^2 = \langle\boldsymbol{\eta},\boldsymbol{\eta}\rangle$\/. Now we take $\tilde{\boldsymbol{\xi}}$ proportional to $e^{-i\omega t}$ in the equation for $\tilde{\boldsymbol{\xi}}$\/. As a result, we obtain an eigenvalue problem for an ordinary linear differential operator defined on the Hilbert space \citep[e.g.][]{Goedbloed2004}. As we have already mentioned, the spectrum of this operator is the union of the point spectrum and the continuous spectrum. The point spectrum consists of the eigenfrequencies defined by the dispersion Equation~(\ref{dr_mod}). The corresponding eigenfunctions (cf. Equations~(\ref{xir_of_r}) and (\ref{xiphi_of_r})) are square integrable with respect to $r$\/, i.e.\ their norms are finite and they belong to the Hilbert space. When $|k| < k_{c1}$\/, the point spectrum consists of two points, $\omega_0$ and $-\omega_0$\/, where $\omega_0$ is the eigenfrequency of the global fast kink mode. For larger $|k|$ it also contains points $\pm\omega_j$\/, ($j = 1,2,\dots,N$), where $\omega_j$ is the eigenfrequency of the $j$\/th overtone with respect to the radial variable, and $N$ is the number of such overtones. In what follows we use the agreement that $\omega_j/k > 0$, so $\omega_j$ corresponds to the wave propagating in the positive $z$\/-direction, and $-\omega_j$ to the wave propagating in the negative $z$\/-direction. The eigenfunction is the same for $\omega_j$ and $-\omega_j$\/. We denote this vector eigenfunction as $\hat{\boldsymbol{\xi}}_j(r,k)$\/. Its components are obtained by substituting $\omega_j$ for $\omega$ in Equations~(\ref{xir_of_r}) and (\ref{xiphi_of_r}).   

As it has been already stated in Section~\ref{sect_cont_modes}, the continuous spectrum is the union of two intervals, $(-\infty,|k|\vae]$ and $[|k|\vae,\infty)$. The improper eigenfunction $\hat{\boldsymbol{\xi}}_\omega(r,k)$ corresponding to point $\omega$ from the continuous spectrum has an oscillatory behavior for large $r$ with amplitude approximately proportional to $r^{-1/2}$ (cf. Equations~(\ref{xir_of_r_im}) and (\ref{xiphi_of_r_im})). As a result, it is not square-integrable with respect to $r$\/, i.e.\ $\|\hat{\boldsymbol{\xi}}_\omega\| = \infty$ and it does not belong to the Hilbert space. Note that $\hat{\boldsymbol{\xi}}_{-\omega}(r,k) = \hat{\boldsymbol{\xi}}_\omega(r,k)$. Its components are given by Equations~(\ref{xir_of_r_im}) and (\ref{xiphi_of_r_im}). 

The function $\tilde{\boldsymbol{\xi}}$ can be expanded with respect to the proper and improper eigenfunctions as    

\begin{align}\label{expand_eigen}
\tilde{\boldsymbol{\xi}}(t,r,k) &= \left[A_0^+(k)e^{-i\omega_0(k)t} + 
   A_0^-(k)e^{i\omega_0(k)t}\right]\hat{\boldsymbol{\xi}}_0(r,k) \nonumber\\
&+ \sum_{j=1}^N \left[A_j^+(k)e^{-i\omega_j(k)t} + 
   A_j^-(k)e^{i\omega_j(k)t}\right]\hat{\boldsymbol{\xi}}_j(r,k) \nonumber\\
&+ \int_{|k|\vae}^\infty\left[A_\omega^+(k)e^{-i\omega t} + 
   A_\omega^-(k)e^{i\omega t}\right]\hat{\boldsymbol{\xi}}_\omega(r,k)\,d\omega .  
\end{align}

\noindent
To determine the functions $A_j^\pm(k)$ ($j = 0,1,\dots, N$) and $A_\omega^\pm(k)$ we use the initial conditions. Taking $t = 0$ in Equation~(\ref{expand_eigen}) we obtain 

\begin{align}\label{init_expand_xi}
\tilde{\boldsymbol{f}}(r,k) &= \left[A_0^+(k) + 
   A_0^-(k)\right]\hat{\boldsymbol{\xi}}_0(r,k) \nonumber\\
&+ \sum_{j=1}^N \left[A_j^+(k) + 
   A_j^-(k)\right]\hat{\boldsymbol{\xi}}_j(r,k) \nonumber\\
&+ \int_{|k|\vae}^\infty\left[A_\omega^+(k) + 
   A_\omega^-(k)\right]\hat{\boldsymbol{\xi}}_\omega(r,k)\,d\omega ,  
\end{align}

\noindent
where $\tilde{\boldsymbol{f}}(r,k)$ is the Fourier transform of the vector function $\boldsymbol{f}(r,k) = (f_r,f_\varphi)$. Differentiating Equation~(\ref{expand_eigen}) with respect to $t$ and, once again, taking $t = 0$ yields

\begin{align}\label{init_expand_der}
\tilde{\boldsymbol{g}}(r,k) &= -i\omega_0(k)\left[A_0^+(k) - 
   A_0^-(k)\right]\hat{\boldsymbol{\xi}}_0(r,k) \nonumber\\
&- i\sum_{j=1}^N \omega_j(k)\left[A_j^+(k) - 
   A_j^-(k)\right]\hat{\boldsymbol{\xi}}_j(r,k) \nonumber\\
&- i\int_{|k|\vae}^\infty \omega\left[A_\omega^+(k) - 
   A_\omega^-(k)\right]\hat{\boldsymbol{\xi}}_\omega(r,k)\,d\omega ,  
\end{align}

\noindent
where $\tilde{\boldsymbol{g}}(r,k)$ is the Fourier transform of the vector function $\boldsymbol{g}(r,k) = (g_r,g_\varphi)$. The eigenfunctions satisfy the orthogonality conditions

\begin{equation}\label{scalar}
\begin{array}{l}
\langle\hat{\boldsymbol{\xi}}_j,\hat{\boldsymbol{\xi}}_l\rangle = 0 \quad (j \neq l) ,
    \vspace*{2mm}\\
\langle\hat{\boldsymbol{\xi}}_j,\hat{\boldsymbol{\xi}}_\omega\rangle = 0 , \vspace*{2mm}\\
\langle\hat{\boldsymbol{\xi}}_\omega,\hat{\boldsymbol{\xi}}_{\omega'}\rangle = 
   q(\omega)\delta(\omega-\omega'), 
\end{array} 
\end{equation}

\noindent
where $\omega > 0$ and $\omega' > 0$. The function $q(\omega)$ is calculated in Appendix~\ref{app_q}. It reads 

\begin{equation}\label{q_omega}
q(\omega) = \frac{\rho_e\vae^2\ke^2}\omega\left(C_J^2 + C_Y^2\right).
\end{equation}

Taking the scalar product of Equations~(\ref{init_expand_xi}) and (\ref{init_expand_der}) with $\boldsymbol{\xi}_j$ we obtain 

\begin{equation}\label{A_pm_eq}
\begin{array}{l}
\left[A_j^+(k) + A_j^-(k)\right]\|\hat{\boldsymbol{\xi}}_j\|^2 = 
   \langle\tilde{\boldsymbol{f}},\hat{\boldsymbol{\xi}}_j\rangle, \vspace*{2mm}\\
\omega_j(k)\left[A_j^+(k) - A_j^-(k)\right]\|\hat{\boldsymbol{\xi}}_j\|^2 = 
   i\langle\tilde{\boldsymbol{g}},\hat{\boldsymbol{\xi}}_j\rangle,
\end{array} 
\end{equation}

\noindent
where $j = 0,1,\dots,N$\/. It follows from these equations that

\begin{equation}\label{A_pm}
A_j^\pm(k) = \frac{\langle\omega_j\tilde{\boldsymbol{f}} \pm i\tilde{\boldsymbol{g}},
   \hat{\boldsymbol{\xi}}_j\rangle}{2\omega_j\|\hat{\boldsymbol{\xi}}_j\|^2} .
\end{equation}

\noindent
With the aid of Equations~(\ref{xir_of_r}) and (\ref{xiphi_of_r}) we obtain

\begin{align}\label{A_numer}
\langle\omega_j\tilde{\boldsymbol{f}} \pm 
   i\tilde{\boldsymbol{g}},\hat{\boldsymbol{\xi}}_j\rangle &=
   \frac{\rhoi K_1(\kappa_e a)}{\ki^2} \int_0^a \left[(\omega_j \tilde{f}_r \pm 
   i\tilde{g}_r) \ki r J'_1(\ki r) \right. \nonumber\\
&\hspace{18ex}-\left. (i\omega_j \tilde{f}_\varphi \mp \tilde{g}_\varphi)J_1(\ki r)\right]\,dr \nonumber\\ 
&- \frac{\rhoe J_1(\ki a)}{\kappa_e^2} \int_a^\infty \left[(\omega_j \tilde{f}_r\pm
   i\tilde{g}_r) \kappa_e r K'_1 (\kappa_e r) \right. \nonumber\\ 
&\hspace{18ex}-\left. (i\omega_j \tilde{f}_\varphi \mp \tilde{g}_\varphi) K_1(\kappa_e r)\right]\,dr ,
\end{align}

\begin{align}\label{A_denom}
\|\hat{\boldsymbol{\xi}}_j\|^2 &= \frac{\rhoi K_1^2(\kappa_e a)}{\ki^4} 
   \int_0^a \left[(r\ki)^2 {J'_1}^2(\ki r) + J_1^2(\ki r)\right]\frac{dr}r \nonumber\\
&+ \frac{\rhoe J_1^2(\ki a)}{\kappa_e^4} \int_a^\infty 
   \left[(r\kappa_e)^2 {K'_1}^2(\kappa_e r) + K_1^2(\kappa_e r)\right]\frac{dr}r  .
\end{align}

\noindent
Equations~(\ref{A_pm}), (\ref{A_numer}), and (\ref{A_denom}) correspond to Equations~(\ref{A_pm-main}), (\ref{A_numer-main}), and (\ref{A_denom-main}).

Now we obtain the expressions for $A_\omega^+(k)$ and $A_\omega^-(k)$. We substitute $\omega'$ for $\omega$ in Equation~(\ref{expand_eigen}) and then take the scalar product of this equation with $\hat{\boldsymbol{\xi}}_\omega(r,k)$. As a result we obtain

\begin{eqnarray}\label{A_omega_f}
\langle\tilde{\boldsymbol{f}},\hat{\boldsymbol{\xi}}_\omega\rangle &=& 
    \int_{|k|\vae}^\infty\left[A_{\omega'}^+(k) + A_{\omega'}^-(k)\right]
    q(\omega)\delta(\omega - \omega')\,d\omega \nonumber\\
&=& q(\omega)\left[A_\omega^+(k) + A_\omega^-(k)\right] .
\end{eqnarray}

\noindent
In a similar way we obtain

\begin{equation}\label{A_omega_h}
\langle\tilde{\boldsymbol{g}},\hat{\boldsymbol{\xi}}_\omega\rangle =
   -i\omega q(\omega)\left[A_\omega^+(k) - A_\omega^-(k)\right] .
\end{equation}

\noindent
It follows from Equations~(\ref{A_omega_f}) and (\ref{A_omega_h}) that

\begin{equation}\label{A_pm_om}
A_\omega^\pm(k) = \frac{\langle\omega\tilde{\boldsymbol{f}} \pm i\tilde{\boldsymbol{g}},
   \hat{\boldsymbol{\xi}}_\omega\rangle}{2\omega q(\omega)} .
\end{equation}

\noindent
With the aid of Equations~(\ref{xir_of_r_im}) and (\ref{xiphi_of_r_im}) we obtain

\begin{align}\label{A_numer_om}
\langle\omega\tilde{\boldsymbol{f}} \pm 
   i\tilde{\boldsymbol{g}},\hat{\boldsymbol{\xi}}_\omega\rangle &=
   \frac{\rhoi a^2k_e^2}{\ki^2} \int_0^a \left[(\omega \tilde{f}_r \pm 
   i\tilde{g}_r) \ki r J'_1(\ki r) - (i\omega\tilde{f}_\varphi \mp \tilde{g}_\varphi)J_1(\ki r)\right]\,dr  \nonumber\\
&+ \frac{\rhoe a^2k_e^2}{\ke^2} \int_a^\infty \left\{(\omega\tilde{f}_r \pm i\tilde{g}_r) \ke r 
    \left[C_J J'_1(\ke r) + C_Y Y'_1(\ke r)\right] \right. \nonumber\\
&\hspace{12ex} -\left. (i\omega\tilde{f}_\varphi \mp \tilde{g}_\varphi)
   \left[C_J J_1(\ke r) + C_Y Y_1(\ke r)\right]\right\}\,dr ,
\end{align}

\noindent
Equations~(\ref{A_pm_om}) and (\ref{A_numer_om}) correspond to Equations~(\ref{A_pm_om-main}) and (\ref{A_numer_om-main}).

\section{Calculation of $q(\omega)$}\label{app_q}

In this appendix we calculate the function $q(\omega)$ that appears in Equation~(\ref{scalar}). Using Equations~(\ref{xir_of_r_im}) and (\ref{xiphi_of_r_im}) we write the expression for the scalar product of two improper eigenfunctions as

\begin{eqnarray}\label{appen1}
\langle\hat{\boldsymbol{\xi}}_\omega,\hat{\boldsymbol{\xi}}_{\omega'}\rangle &=& 
   \rho_i\int_0^a\left[\frac{J'_1(\ki r) J'_1(\ki' r)}{\ki\ki'} + 
   \frac{J_1(\ki r) J_1(\ki' r)}{r^2\ki^2{\ki'}^2}\right]r\,dr \nonumber\\
&+& \rho_e\int_a^\infty\left[\frac{F'_1(\ke r) F'_1(\ke' r)}{\ke\ke'} + 
   \frac{F_1(\ke r) F_1(\ke' r)}{r^2\ke^2{\ke'}^2}\right]r\,dr ,
\end{eqnarray}

\noindent
where $\ki'$ and $\ke'$ are given by Equations~(\ref{ks}) and (\ref{ke}) with $\omega'$ substituted for $\omega$\/, and we have introduced the notation

\begin{equation}\label{appen2}
F_1(x) = C_J J_1(x) + C_Y Y_1(x) .
\end{equation}

\noindent
In what follows the prime is used to indicate that a quantity that depends on $\omega$ is calculated with $\omega$ substituted by $\omega'$\/. Although the prime is also used to indicate the derivative, this double use of the same symbol cannot cause any confusion.

Functions $J_1(z)$ and $Y_1(z)$ can be written as \citep{abramowitz1964}

\begin{equation}\label{appen3} 
\begin{array}{l}\displaystyle
J_1(z) = \sqrt{\frac2{\pi z}}\cos\left(z - \mbox{$\frac34$}\pi\right)\left[1 + j_1(z)\right] ,
   \vspace*{2mm}\\ \displaystyle
Y_1(z) = \sqrt{\frac2{\pi z}}\sin\left(z - \mbox{$\frac34$}\pi\right)\left[1 + y_1(z)\right] ,
\end{array}
\end{equation}

\noindent
where $j_1(z)$ and $y_1(z)$ decay as $z^{-1}$ when $z \to \infty$\/. Similar expressions are valid for the Bessel functions of order zero, $J_0(z)$ and $Y_0(z)$, although with $\frac14\pi$ substituted for $\frac34\pi$\/. Then, using the relations \citep{abramowitz1964}

\begin{equation}\label{appen4}
J'_1(z) = J_0(z) - \frac1z J_1(z), \quad Y'_1(z) = Y_0(z) - \frac1z Y_1(z),
\end{equation}

\noindent
we obtain expressions similar to those given by Equation~(\ref{appen3}) for the derivatives of Bessel functions,

\begin{equation}\label{appen5} 
\begin{array}{l}\displaystyle
J'_1(z) = \sqrt{\frac2{\pi z}}\cos\left(z - \mbox{$\frac14$}\pi\right)\left[1 + j_2(z)\right] ,
   \vspace*{2mm}\\ \displaystyle
Y'_1(z) = \sqrt{\frac2{\pi z}}\sin\left(z - \mbox{$\frac14$}\pi\right)\left[1 + y_2(z)\right] ,
\end{array}
\end{equation}

\noindent
where once again $j_2(z)$ and $y_2(z)$ decay as $z^{-1}$ when $z \to \infty$\/.

With the aid of Equations~(\ref{appen3}) and (\ref{appen5}), after some algebra, we rewrite Equation~(\ref{appen1}) as

\begin{equation}\label{appen6}
\langle\hat{\boldsymbol{\xi}}_\omega,\hat{\boldsymbol{\xi}}_{\omega'}\rangle =
   \Gamma(\omega,\omega') + \Upsilon(\omega,\omega') ,
\end{equation}

\noindent
where
 
\begin{eqnarray}\label{appen7}
\Gamma(\omega,\omega') &=& \rho_i\ke^4\int_0^a\left[\frac{J'_1(\ki r) J'_1(\ki' r)}{\ki\ki'} + \frac{J_1(\ki r) J_1(\ki' r)}{r^2\ki^2{\ki'}^2}\right]r\,dr \nonumber\\
&+& \frac{\rho_e\ke^{5/2}}{\pi(\ke')^{3/2}}\int_a^\infty\left\{U_c^-(r)
   \cos\left[r(\ke - \ke')\right] \right.\nonumber\\
&-&  U_c^+(r)\cos\left[r(\ke - \ke')\right] - U_s^-(r)\sin\left[r(\ke - \ke')\right] \nonumber\\
&+& \left. U_s^+(r)\sin\left[r(\ke - \ke')\right]\right\} dr , 
\end{eqnarray}

\begin{eqnarray}\label{appen8}
\Upsilon(\omega,\omega') &=& \frac{\rho_e\ke^{5/2}}{\pi(\ke')^{3/2}}\int_a^\infty\left\{(C_J C'_J + C_Y C'_Y)\cos\left[r(\ke - \ke')\right]\right.  \nonumber\\
&-& (C_J C'_Y + C_Y C'_J)\cos\left[r(\ke + \ke')\right] \nonumber\\
&-& (C_J C'_Y - C_Y C'_J)\sin\left[r(\ke - \ke')\right] \nonumber\\
&+& \left. (C_J C'_J - C_Y C'_Y)\sin\left[r(\ke + \ke')\right]\right\}dr ,
\end{eqnarray}

\noindent
and the functions $U_c^\pm(r)$ and $U_s^\pm(r)$ are given by

\begin{eqnarray}\label{appen9}
U_c^-(r) &=& C_J C'_J[j_2(\ke r) + j_2(\ke' r) + j_2(\ke r) j_2(\ke' r)] \nonumber\\
&+& C_Y C'_Y[y_2(\ke r) + y_2(\ke' r) + y_2(\ke r) y_2(\ke' r)] \nonumber\\
&+& \frac{C_J C'_J}{r^2\ke\ke'}[1 + j_1(\ke r) + j_1(\ke' r) + j_1(\ke r) j_1(\ke' r)]  \nonumber\\
&+& \frac{C_Y C'_Y}{r^2\ke\ke'}[1 + y_1(\ke r) + y_1(\ke' r) + y_1(\ke r) y_1(\ke' r)] ,
\end{eqnarray}

\begin{eqnarray}\label{appen10}
U_c^+(r) &=& C_J C'_Y[j_2(\ke r) + y_2(\ke' r) + j_2(\ke r) y_2(\ke' r)] \nonumber\\
&+& C_Y C'_J[y_2(\ke r) + j_2(\ke' r) + y_2(\ke r) j_2(\ke' r)] \nonumber\\
&-& \frac{C_J C'_Y}{r^2\ke\ke'}[1 + j_1(\ke r) + y_1(\ke' r) + j_1(\ke r) y_1(\ke' r)]  \nonumber\\
&-& \frac{C_Y C'_J}{r^2\ke\ke'}[1 + y_1(\ke r) + j_1(\ke' r) + y_1(\ke r) j_1(\ke' r)] ,
\end{eqnarray}

\begin{eqnarray}\label{appen11}
U_s^-(r) &=& C_J C'_Y[j_2(\ke r) + y_2(\ke' r) + j_2(\ke r) y_2(\ke' r)] \nonumber\\
&-& C_Y C'_J[y_2(\ke r) + j_2(\ke' r) + y_2(\ke r) j_2(\ke' r)] \nonumber\\
&+& \frac{C_J C'_Y}{r^2\ke\ke'}[1 + j_1(\ke r) + y_1(\ke' r) + j_1(\ke r) y_1(\ke' r)]  \nonumber\\
&-& \frac{C_Y C'_J}{r^2\ke\ke'}[1 + y_1(\ke r) + j_1(\ke' r) + y_1(\ke r) j_1(\ke' r)] ,
\end{eqnarray}

\begin{eqnarray}\label{appen12}
U_s^+(r) &=& C_J C'_J[j_2(\ke r) + j_2(\ke' r) + j_2(\ke r) j_2(\ke' r)] \nonumber\\
&-& C_Y C'_Y[y_2(\ke r) + y_2(\ke' r) + y_2(\ke r) y_2(\ke' r)] \nonumber\\
&-& \frac{C_J C'_J}{r^2\ke\ke'}[1 + j_1(\ke r) + j_1(\ke' r) + j_1(\ke r) j_1(\ke' r)]  \nonumber\\
&+& \frac{C_Y C'_Y}{r^2\ke\ke'}[1 + y_1(\ke r) + y_1(\ke' r) + y_1(\ke r) y_1(\ke' r)] .
\end{eqnarray}

\noindent
The fact that the functions $j_1(r)$, $y_1(r)$, $j_2(r)$, and $y_2(r)$ decay as $r^{-1}$ as $r \to \infty$ implies that the functions $U_c^\pm(r)$ and $U_s^\pm(r)$ have the same property. Then the integrand in the second integral on the right-hand side of Equation~(\ref{appen7}) is an oscillatory function for large $r$ with the oscillation amplitude tending to zero as $r^{-1}$\/, so the integral is convergent. Then it follows that $\Gamma(\omega,\omega')$ is a regular continuous function. In contrast, the integrand in the integral on the right-hand side of Equation~(\ref{appen8}) oscillates, but the oscillation amplitude does not tend to zero as $r \to \infty$\/. Hence, this integral does not exist in the classical sense and has to be treated as a generalized function. In what follows we consider $\Upsilon(\omega,\omega')$ as a generalized function of $\omega'$\/. 

Using the variable substitution $r = a + x$ we transform Equation~(\ref{appen8})

\begin{eqnarray}\label{appen13}
\Upsilon(\omega,\omega') &=& \frac{\rho_e\ke^{5/2}}{\pi(\ke')^{3/2}}\int_0^\infty\left\{W_c^- \cos\left[x(\ke - \ke')\right] \right.\nonumber\\
&-& W_c^+ \cos\left[x(\ke + \ke')\right] - W_s^- \sin\left[x(\ke - \ke')\right] \nonumber\\
&+& \left. W_s^+\sin\left[x(\ke + \ke')\right]\right\}dx, 
\end{eqnarray}

\noindent
where

\begin{equation}\label{appen14}
\begin{array}{l}
W_c^- = (C_J C'_J + C_Y C'_Y)\cos\left[a(\ke - \ke')\right] \\
\hphantom{xxx} - (C_J C'_Y - C_Y C'_J)\sin\left[a(\ke - \ke')\right], \vspace*{1mm}\\
W_c^+ = (C_J C'_Y + C_Y C'_J)\cos\left[a(\ke + \ke')\right] \\
\hphantom{xxx} - (C_J C'_J - C_Y C'_Y)\sin\left[a(\ke + \ke')\right], \vspace*{1mm}\\
W_s^- = (C_J C'_Y - C_Y C'_J)\cos\left[a(\ke - \ke')\right] \\
\hphantom{xxx} + (C_J C'_J + C_Y C'_Y)\sin\left[a(\ke - \ke')\right], \vspace*{1mm}\\
W_s^+ = (C_J C'_J - C_Y C'_Y)\cos\left[a(\ke + \ke')\right] \\
\hphantom{xxx} + (C_J C'_Y + C_Y C'_J)\sin\left[a(\ke + \ke')\right] .
\end{array}
\end{equation}

\noindent
Now we evaluate the generalized functions $\int_0^\infty\cos(ux)\,dx$ and $\int_0^\infty\sin(ux)\,dx$\/. For this we apply these functions to a trial function $\phi(u)$, which is an infinitely differentiable function tending to zero as $|u| \to \infty$ faster than any negative power of $|x|$. We have

\begin{eqnarray}
&& \int_{-\infty}^\infty \phi(u)\,du \int_0^\infty\cos(ux)\,dx \nonumber\\ 
&& \hphantom{xx} = \frac12\int_{-\infty}^\infty \phi(u)\,du 
   \int_{-\infty}^\infty\cos(ux)\,dx \nonumber\\
&& \hphantom{xx} = \frac14\int_{-\infty}^\infty dx \int_{-\infty}^\infty \phi(u)
   \left(e^{iux} + e^{-iux}\right)\,du \nonumber\\
&& \hphantom{xx} = \frac14\int_{-\infty}^\infty\big[\tilde{\phi}(-x) + 
   \tilde{\phi}(x)\big]dx = \pi\phi(0), \nonumber
\end{eqnarray} 

\noindent
where, as before, the tilde denotes the Fourier transform. It follows from this result that \citep[e.g.][]{Richtmyer1978}

\begin{equation}\label{appen15}
\int_0^\infty\cos(ux)\,dx = \pi\delta(u),
\end{equation}

\noindent
where $\delta(u)$ is the Dirac delta-function. To evaluate the second generalized function we write

$$ 
\int_{-\infty}^\infty \phi(u)\,du \int_0^\infty\sin(ux)\,dx
= \int_0^\infty dx \int_{-\infty}^\infty \phi(u)\sin(ux)\,du
$$
$$
= \lim_{\alpha \to +0}\int_0^\infty e^{-\alpha x} dx \int_{-\infty}^\infty \phi(u)\sin(ux)\,du
$$
$$
= \lim_{\alpha \to +0}\int_{-\infty}^\infty \phi(u)\,du
   \int_0^\infty e^{-\alpha x}\sin(ux)\,dx 
$$
$$
= \lim_{\alpha \to +0}\Im\left(\int_{-\infty}^\infty \phi(u)\,du
   \int_0^\infty e^{-\alpha x}\sin(ux)\,dx\right) 
$$
$$
= \lim_{\alpha \to +0}\Im\left(\int_{-\infty}^\infty \frac{\phi(u)}{\alpha - iu}du\right)
\hphantom{xxxxxxxxxx}
$$
$$
= \lim_{\alpha \to +0}\int_{-\infty}^\infty \frac{u\phi(u)}{\alpha^2 + u^2}du
= {\cal P}\int_{-\infty}^\infty \frac{\phi(u)}{u}du ,
$$
where $\Im$ indicates the imaginary part of a quantity, and $\cal P$ denotes the principal Cauchy part of an integral. This result implies that

\begin{equation}\label{appen16}
\int_0^\infty\sin(ux)\,dx = {\cal P}\frac1u .
\end{equation} 

Since $\ke$ is a monotonic function of $\omega$\/, we can consider $\Upsilon(\omega,\omega')$ as a generalized function of $\ke'$\/. Then it follows from Equations~(\ref{appen13}), (\ref{appen15}), and (\ref{appen16}) that

\begin{equation}\label{appen17}
\Upsilon(\omega,\omega') = \Upsilon_1(\omega,\omega') + \Upsilon_2(\omega,\omega'),
\end{equation} 

\noindent
where

\begin{equation}\label{appen18}
\Upsilon_1(\omega,\omega') = \frac{\rho_e\ke^{5/2}}{(\ke')^{3/2}}
   \left[W_c^-\delta(\ke - \ke') - W_c^+\delta(\ke + \ke')\right], 
\end{equation} 

\begin{equation}\label{appen19}
\Upsilon_2(\omega,\omega') = \frac{\rho_e\ke^{5/2}}{\pi(\ke')^{3/2}}
   \left(W_s^+{\cal P}\frac1{\ke + \ke'} - W_s^-{\cal P}\frac1{\ke - \ke'}\right). 
\end{equation}

\noindent
First of all, we note that $\ke > 0$ and $\ke' > 0$ when $\omega > |k|\vae$ and $\omega' > |k|\vae$\/. Hence, the second term in the square brackets on the right-hand side of Equation~(\ref{appen18}) is zero in the whole domain of variation of $\omega$ and $\omega'$ and can be dropped. Also the first term in the brackets on the right-hand side of Equation~(\ref{appen19}) is a regular continuous function. Now we notice that $W_s^- = 0$ when
$\omega' = \omega$\/, so the second term in the brackets also has no singularity when $\omega' = \omega$\/, so the simbol $\cal P$ in Equation~(\ref{appen19}) can be dropped. Hence, $\Upsilon_2(\omega,\omega')$ is a regular continuous function.
Using the identity

\begin{equation}\label{appen20}
\ke - \ke' = \frac{(\omega + \omega')(\omega - \omega')}{\vae^2(\ke + \ke')},
\end{equation}

\noindent
and the formulae \citep[e.g.][]{Richtmyer1978}

\begin{equation}\label{appen21}
f(x)\delta(x) = f(0)\delta(x), \quad \delta(\phi(x)) = \frac1{|\phi'(0)|}\delta(x),
\end{equation}

\noindent
where $f(x)$ and $\phi(x)$ are infinitely differentiable functions and, in addition, $\phi(x)$ is monotonic, we rewrite Equation~(\ref{appen18}) as

\begin{equation}\label{appen22}
\Upsilon_1(\omega,\omega') = 
   \frac{\rho_e\vae^2\ke^2}\omega W_c^-\delta(\omega - \omega'),
\end{equation} 

\noindent
where $W_c^-$ is calculated at $\omega' = \omega$\/. Now, using Equations~(\ref{appen1}), (\ref{appen6}), (\ref{appen17}), and (\ref{appen22}), we obtain

\begin{equation}\label{appen23}
\langle\hat{\boldsymbol{\xi}}_\omega,\hat{\boldsymbol{\xi}}_{\omega'}\rangle = 
    Q(\omega,\omega') + \frac{\rho_e\vae^2\ke^2}\omega W_c^-\delta(\omega - \omega'),
\end{equation} 

\noindent
where $Q(\omega,\omega')$ is a regular continuous function of $\omega'$\/. In accordance with 
Equation~(\ref{scalar}) the support of the generalized function $\langle\hat{\boldsymbol{\xi}}_\omega,\hat{\boldsymbol{\xi}}_{\omega'}\rangle$ considered as a function of $\omega'$ consists of exactly one point $\omega' = \omega$\/. The support of any regular continuous function that is not identically equal to zero contains at least one interval. Hence, we conclude that $Q(\omega,\omega') \equiv 0$. The function $q(\omega)$ is equal to the coefficient at $\delta(\omega - \omega')$ in Equation~(\ref{appen1}). Then, using Equations~(\ref{appen14}) we arrive at

\begin{equation}\label{appen23-2}
q(\omega) = \frac{\rho_e\vae^2\ke^2}\omega\left(C_J^2 + C_Y^2\right) .
\end{equation} 

\noindent
This is Equation~(\ref{q_omega-main}).


\end{document}